\pdfoutput=1 % arXiv: compile with pdflatex
\documentclass[sigconf,nonacm]{acmart}

\usepackage{amsmath}

\usepackage{amssymb}
\usepackage{array}
\usepackage{booktabs}
\usepackage{tabularx}
\usepackage{multirow}
\usepackage{placeins}
\usepackage{float}
\usepackage{graphicx}
\usepackage{algorithm}
\usepackage{algpseudocode}
\algrenewcommand{\alglinenumber}[1]{\small\texttt{>}}
\usepackage{enumitem}
\usepackage[most]{tcolorbox}
\tcbset{
  mathbox/.style={
    colback=gray!5, colframe=gray!40,
    coltitle=black,
    boxrule=0.4pt, arc=2pt,
    left=4pt, right=4pt, top=4pt, bottom=4pt,
    before skip=6pt, after skip=6pt,
    fonttitle=\footnotesize, fontupper=\footnotesize
  }
}

\newcommand{\qual}[1]{^{{\scriptscriptstyle #1}}}

\setcopyright{none}
\renewcommand\footnotetextcopyrightpermission[1]{}
\AtBeginDocument{\newgeometry{top=0.5in, bottom=0.5in, left=0.5in, right=0.5in, columnsep=18pt}}
\AddToHook{shipout/before}{\ifnum\value{page}=1\relax\setbox\ShipoutBox=\hbox{\kern7pt\box\ShipoutBox}\fi}
\makeatletter\renewcommand*{\enlargethispage}{\@ifstar\@gobble\@gobble}\makeatother

\title{GRCA: Fast Geometrically Modeled Emitter-Centric Ray-Casting Acceleration in Highly Dynamic LiDAR Simulation}

\author{Rabin Gajmer}
\authornote{Primary contributor.}
\affiliation{
  \institution{Woven by Toyota}
  \city{Tokyo}
  \country{Japan}
}
\email{rabin.gajmer@woven.toyota}

\author{Joonas Haapala}
\affiliation{
  \institution{Woven by Toyota}
  \city{Tokyo}
  \country{Japan}
}
\email{joonas.haapala@woven.toyota}

\author{Zoltan Beck}
\affiliation{
  \institution{Woven by Toyota}
  \city{Tokyo}
  \country{Japan}
}
\email{zoltan.beck@woven.toyota}



\newcommand{\grca}{GRCA}
\newcommand{\grcaname}{Gajmer Ray-Casting Acceleration}
\newcommand{\grcagpu}{GRCA\textsubscript{GPU}}
\newcommand{\grcacpu}{GRCA\textsubscript{CPU}}
\def\originscalingfigname{origin_scaling}
\def\satthresholdfigname{sat_threshold}
\def\scalevsembrefigname{scale_vs_bvh}
\def\triareacombinedfigname{tri_area_combined}
\def\mainsuitefigname{main_suite}
\def\costcontributorsfigname{cost_contributors}
\def\multiplatformfigname{multi_platform}
\def\rangevariationfigname{range_variation}
\def\scalebatsubdivfigname{scale_bat_subdiv}
\def\scenesizestaticfigname{scene_size_static}
\def\hybridpipelinefigname{hybrid_pipeline_benchmark}
\newcommand{\figpagesinlayoutorder}{}
\begin{document}

\begin{teaserfigure}
  %% Teaser content, shared by body.tex (teaserfigure) and the IEEE wrappers (figure*).
%% One LiDAR per static scene (columns, in Table tab:meshes order), four views from the same camera (rows); all but the first
%% rendered with GRCA-CUDA by GRCA/presets/benchmark_config_vis_teaser_<Scene>.json. The 1200x900 renders are shown with \tztrim px cut from
%% top and bottom (LiDAR stays centred) to fit the 7-page body; the supplementary shows them uncropped.
\centering
%% Six panels per row, exactly one sixth of the text width each, with no gaps between panels or rows.
\newcommand{\tzw}{\dimexpr\textwidth/6\relax}%
%% Largest panels that keep the body on 7 pages with the one-line caption (tested 2026-11-24): trim 80 with the named
%% acmtog author block (75 overflows), 5 for the anonymous review (0 overflows).
%% A wrapper may \newcommand{\tztrim}{..} before \input{body}.
\providecommand{\tztrim}{80}%
\newcommand{\tzimg}[1]{\edef\tzopts{width=\noexpand\tzw,trim=0 \tztrim\space 0 \tztrim,clip}%
  \expandafter\includegraphics\expandafter[\tzopts]{#1}}%
%% TODO: realistic renders. Each slot shows a placeholder until Teaser_<Scene>_render.png (1200x900, same camera) exists.
\newcommand{\tzrender}[1]{\IfFileExists{#1}{\tzimg{#1}}{{\setlength{\fboxsep}{0pt}\colorbox{black!12}{\parbox[b][%
  \dimexpr\tzw*\numexpr900-2*\tztrim\relax/1200\relax][c]{\tzw}{\centering\scriptsize TODO: render}}}}}%
\newcommand{\tzrow}[2][\tzimg]{%
  #1{Teaser_Vokselia_#2.png}%
  #1{Teaser_Bistro_#2.png}%
  #1{Teaser_SanMiguel_#2.png}%
  #1{Teaser_Rungholt_#2.png}%
  #1{Teaser_EmeraldSquare_#2.png}%
  #1{Teaser_PowerPlant_#2.png}}%
{\offinterlineskip
\tzrow[\tzrender]{render}\par
\tzrow{plain}\par
\tzrow{heat}\par
\tzrow{rainbow}\par}
\caption{Scenes as in Table~\ref{tab:meshes}. Rows: render, mesh, rays tested per triangle (gray = culled), hits by LiDAR
  range (red = near; enlarged: Figs.~\ref{fig:supp_scene_views}--\ref{fig:supp_scene_views2}).}
\Description{A grid of six columns, one per static scene, and four rows: a rendering of the scene, its triangle mesh in
  white with black edges, a heat map of rays tested per triangle with culled triangles in gray, and the LiDAR hit points
  coloured red to blue by distance.}
\label{fig:teaser}

\end{teaserfigure}

%%
%% Abstract
%%
\begin{abstract}
Real-time Light Detection And Ranging (LiDAR) simulation must find, per emitted ray, the closest intersecting triangle
even in dynamic scenes containing large numbers of moving and deformable objects.
Dominant acceleration-structure approaches require rebuilding each frame for dynamic geometry---a cost that
compounds directly with scene dynamics and cannot be amortized regardless of how little actually changed.

This paper presents the \grcaname{} (\grca{}) algorithm, which inverts the question: instead of asking
\emph{what does each ray hit?} it asks \emph{which rays can each triangle possibly hit?}
\grca{} geometrically models spinning LiDAR emitters as rotation-traced cones or planes and uses each
triangle's emitter-centric apparent area to cull, per triangle, which channels and the rays within those channels
can possibly reach it---without any acceleration structure.
\grca{} is compute-based and vendor-agnostic by design, targeting highly dynamic, high-resolution simultaneous multi-sensor simulation.
Beyond LiDAR, it applies to any setting that casts large numbers of rays against complex geometry.

Benchmarks evaluate 2--8 simultaneous $128\times4096$-ray LiDARs ($360^\circ$/$180^\circ$) over complex dynamic scenes, across five hardware platforms.
At 1000\,m sensor range: standalone \grca{} reaches up to \textbf{14.55$\times$} over Embree~(CPU) and \textbf{7.97$\times$} over OptiX~(GPU); adding a static bounding volume hierarchy (BVH) for the unchanging geometry (hybrid pipeline) pushes this to \textbf{22.7$\times$} CPU and \textbf{34.9$\times$} GPU.
At 100\,m range: standalone reaches up to \textbf{14.66$\times$} CPU and \textbf{8.33$\times$} GPU; combined with the hybrid pipeline, \textbf{17.9$\times$} CPU and \textbf{36.5$\times$} GPU.

\medskip
%% Code link: review builds ([I]RCA.tex) define \coderepo as the Anonymous GitHub mirror of the GRCA review branch
\providecommand{\coderepo}{\url{https://github.com/woven-by-toyota/GRCA}}
{\centering\small\makebox[\linewidth]{Code \& supplementary: \textcolor{blue}{\coderepo}}\par}

\end{abstract}

\maketitle

%% =============================================================================

\section{Introduction}
%% =============================================================================

\subsection{Problem Statement}

Real-time Light Detection And Ranging (LiDAR) simulation over dynamic geometry is central to closed-loop autonomy training, robotic manipulation, and multi-sensor perception evaluation, where scenes contain large numbers of moving and deformable objects.
A LiDAR device fires a large number of rays into a scene composed of triangles;
for each ray the simulation must find the closest triangle it intersects and
its distance, every frame at real-time rates.

The brute-force cost per frame, in terms of Ray-Triangle Intersection Checks (RTIC; notation: Table~\ref{tab:notation}), is:
\begin{equation}
  \text{Total RTIC} = \sum_{n=1}^{\Omega} \gamma_n \times \chi_n \times \tau
  \label{eq:brute}
\end{equation}
where $\Omega$ is the ray origin count (independent LiDAR scan instances),
$\gamma_n$ the total vertical channels for origin $n$, $\chi_n$ the horizontal rays per
channel for origin $n$, and $\tau$ the triangle count in the scene.

For a representative scene ($\tau = 10\,\text{M}$, $\Omega = 1$, $\gamma_n = 128$, $\chi_n = 4096$) this gives ${\approx}5.2{\times}10^{12}$ tests per frame.

$\gamma_n$, $\chi_n$, and $\tau$ can all be cut dramatically by identifying which ray--triangle
pairs are actually worth testing; at brute-force scale, per-ray atomic contention
alone becomes a significant bottleneck.

\label{sec:bvh_limits}
\textbf{Ray-tracing-based solutions} are the state-of-the-art approach for LiDAR simulation:
OptiX~\cite{parker2010optix} and Embree~\cite{wald2014embree} are the leading
bounding volume hierarchy (BVH)~\cite{wald2007bvh,karras2012bvh} libraries, and modern GPUs add dedicated RT cores for BVH traversal and RTIC in hardware.
For LiDAR, BVH-based methods impose three structural shortcomings:
\begin{enumerate}[label=\textbf{S\arabic*.},nosep,wide,leftmargin=0pt,labelsep=0.3em]
  \item \label{sc:dynamic} \textbf{Dynamic scenes.}
        LiDAR simulations inherently involve fast-moving objects that move, rotate, and deform every frame;
        BVH-based methods require a full per-frame rebuild (refitting
        is only valid for small deformations~\cite{wald2007dynamic,kopta2012bvhrefit}; see \S\ref{sec:bvh_mode_selection})---at cost proportional
        to total triangle count regardless of how little actually moved~\cite{karras2012bvh}.
  \item \label{sc:360} \textbf{360\textdegree{} LiDARs.}
        Efficient BVH traversal relies on ray coherence~\cite{aila2009};
        a 360\textdegree{} spinning LiDAR emits rays in all directions at once, removing directional culling and causing incoherent traversal.
  \item \label{sc:nofilter} \textbf{No upfront triangle filtering.}
        In standard BVH APIs, all culling requires an active ray; no triangle is removed before traversal; out-of-range or back-facing triangles
        that share a bounding volume with contributing geometry are still tested per ray.
\end{enumerate}

\subsection{The \grca{} Approach}

\grca{} addresses \hyperref[sc:dynamic]{S1}--\hyperref[sc:nofilter]{S3} by asking
\emph{which rays can each triangle possibly hit?}---no ray is fired first, no acceleration structure is built.
It is built on three observations about spinning LiDAR geometry:

\textbf{Observation 1:} \textit{the ideal solution tests only the channels and rays that intersect each triangle.}
\label{obs:paradigm}
For a given triangle $T$, the minimum necessary work is to predict the small
subset of vertical channels that could possibly contain a ray hitting $T$, and
within those only the subset of rays whose directions pass through $T$.
These form a rectangular index span $[C_{\mathrm{from}},C_{\mathrm{to}}]\times[R_{\mathrm{from}},R_{\mathrm{to}}]$ in the $\gamma_n \times \chi_n$ ray grid, giving channel span $C_{\mathrm{span}}$ and ray span $R_{\mathrm{span}}$.
Running intersection tests solely on those pairs eliminates all guaranteed
misses without checking them.
\begin{figure}[htb]
  \centering
  % left panel gets the larger share (it sets the figure height); right panel vertically centred
  \raisebox{-0.5\height}{\includegraphics[width=0.60\columnwidth]{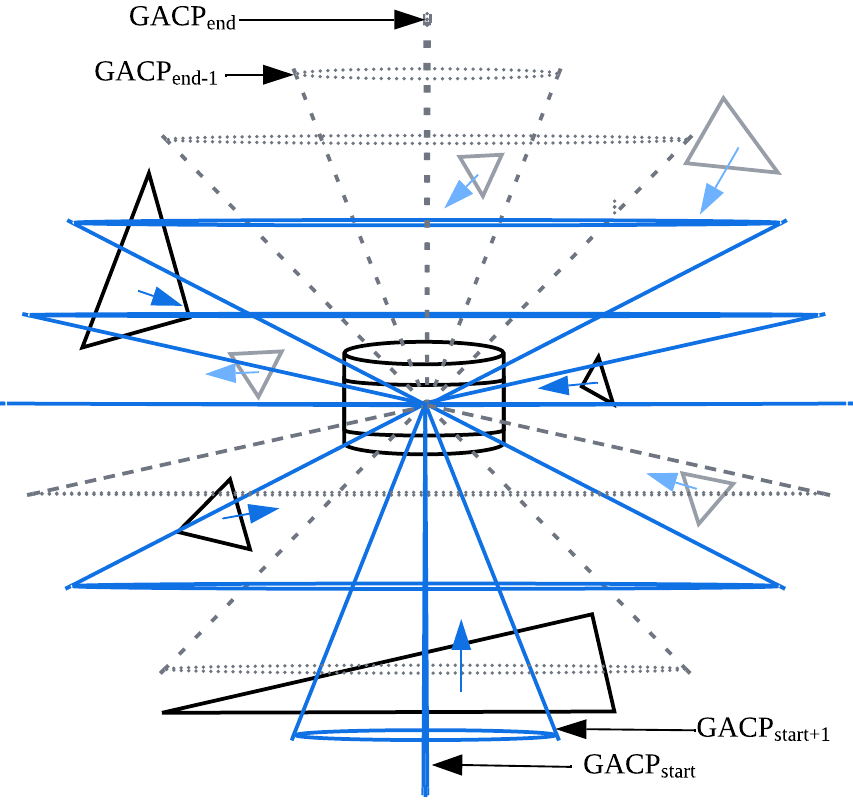}}\hspace{0.01\columnwidth}%
  \raisebox{-0.5\height}{\includegraphics[width=0.38\columnwidth]{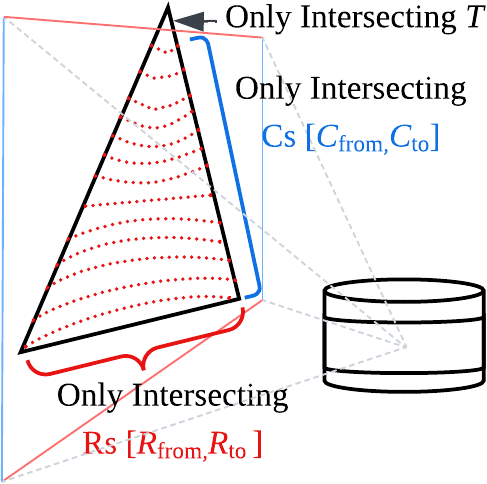}}
  \caption{Left: \grca{} approximates emitted rays as cone or plane surfaces (blue). Right: per-triangle ray filtering: only the span $[C_{\mathrm{from}},C_{\mathrm{to}}]\times[R_{\mathrm{from}},R_{\mathrm{to}}]$ is tested.}
  \Description{Left: diagram of \grca{} approximating emitted LiDAR rays as cones or planes. Right: diagram showing per-triangle ray filtering where only a subset of channels and rays are tested.}
  \label{fig:ideal-solution}
\end{figure}

\textbf{Observation 2:} \textit{each vertical channel traces a conic surface.}
\label{obs:gacp}
As the LiDAR rotates, all rays in a channel collectively form a cone surface
around the spin axis---or a horizontal plane at zero elevation (Figure~\ref{fig:ideal-solution}, left).  This is
distinct from cone tracing~\cite{amanatides1984} and beam
tracing~\cite{heckbert1984}, which widen individual rays into volumetric
primitives---\grca{} keeps each ray a directed half-line and instead tests the Geometrically
Approximated Cone or Plane (GACP) surface against the triangle.  Where this surface intersects $T$, the
candidate rays trace conic sections across the triangle face
(Figure~\ref{fig:ideal-solution}, right; red dots).  Each channel therefore reduces to
a single GACP-Triangle (GACP-T) surface test that predicts the intersecting subset.

\textbf{Observation 3:} \textit{the apparent-area law partitions the scene into two
processing classes with different costs and contributions.}
\label{obs:apparent_area}
The \emph{origin-centric apparent area} is a triangle's footprint seen from the ray origin:
\begin{equation}
  A_{\mathrm{origin}} \propto A_T\cos\phi_{\mathrm{view}} / \delta^2
  \label{eq:aorigin}
\end{equation}
where $A_T$ is the triangle area, $\phi_{\mathrm{view}}$ the angle between the triangle normal and the line of sight, and $\delta$ the sensor-to-triangle distance.
In the \emph{emitter-centric} view, this maps to the span of Observation~1:
\begin{equation}
  \gamma_{\mathrm{span}} = C_{\mathrm{to}} - C_{\mathrm{from}} + 1,
  \qquad
  \chi_{\mathrm{span}} = R_{\mathrm{to}} - R_{\mathrm{from}} + 1
\end{equation}
\begin{equation}
  A_{\mathrm{emitter}} \approx
    \underbrace{\gamma_{\mathrm{span}}}_{\text{height}}
    \;\times\;
    \underbrace{\chi_{\mathrm{span}}\qual{\mathrm{widest}}}_{\text{width}}
  \label{eq:aemitter}
\end{equation}
where $\chi_{\mathrm{span}}\qual{\mathrm{widest}}$ is the maximum $\chi_{\mathrm{span}}$ across all channels in the triangle's span.
Triangles whose $A_{\mathrm{emitter}}$ fits within classification thresholds $\gamma_T$ (channels) $\times\,\chi_T$ (rays) are classified as
\emph{Smaller-Appearing Triangles} (SAT); all others are \emph{Bigger-Appearing Triangles} (BAT).
Two implications follow:
\begin{itemize}[nosep,wide,leftmargin=0pt,labelsep=0.3em]
  \item \textbf{SAT vastly outnumber BAT yet contribute few hits.}
        Footprint falls as $1/\delta^2$; mixed-distance geometry produces far more SAT than BAT, but triangles below the angular sample spacing are unlikely to be the closest hit---a conservative prediction suffices.
  \item \textbf{BAT dominate scan output and require precise prediction.}
        Large or close-up surfaces are the closest hit for most rays; per-channel prediction is most critical for BAT.
        Classification is per origin: a triangle can be SAT for one origin and BAT for another.
\end{itemize}

Together these observations naturally yield a two-pass design addressing \hyperref[sc:dynamic]{S1}--\hyperref[sc:nofilter]{S3}: an \emph{early pass} processes all triangles with a
conservative approximate span prediction, completing SAT intersections inline;
BAT triangles are deferred to a \emph{late pass} that applies the full precise GACP-T intersection per channel;
per-triangle geometry precomputed by the early pass is carried in the deferred list and reused by the late pass,
avoiding redundant calculation.

\subsection*{The contributions of this paper are:}
\begin{itemize}[nosep,wide,leftmargin=0pt,labelsep=0.3em]
  \item \textbf{Paradigm inversion}: We propose the \grcaname{} (\grca{}) algorithm, which inverts the conventional
        question---instead of \emph{what does each ray hit?}, it asks
        \emph{which rays can each triangle possibly hit?}---yielding a
        per-triangle process whose cost scales with scene geometry and each triangle's predicted ray span, not the full ray count.
  \item \textbf{Emitter-centric model using GACP} (\S\ref{sec:gacp_model}--\ref{sec:pipeline}):
        a closed-form GACP surface approximates all rays in a spinning LiDAR channel,
        producing per-triangle integer index spans
        $[C_{\mathrm{from}},C_{\mathrm{to}}]\!\times\![R_{\mathrm{from}},R_{\mathrm{to}}]$
        without scene partitioning.
        Apparent-area equations (Eqs.~\ref{eq:aorigin}--\ref{eq:aemitter}) then classify
        every triangle into SAT or BAT, driving a two-pass pipeline.
        No prior system (spatial discretization, directional binning, or coarse-to-fine hierarchies) interprets rays as a
        single geometric surface for per-triangle intersection prediction while keeping
        each ray an individual directed half-line, nor applies emitter-centric
        apparent area as a per-triangle culling primitive (\S\ref{sec:related}).
  \item \textbf{Open-source and comprehensive benchmark study}
        (\S\ref{sec:benchmarks}--\ref{sec:accuracy}):
        a fully reproducible evaluation---source code, benchmark harness, scenes, and pre-recorded pose/scale streams are public---spanning \textbf{216 configurations} across six scenes and three deformation regimes on one machine, plus cross-platform runs on four more,
        in which open-source CUDA and AVX2 \grca{} implementations outperform OptiX by up to $7.97\times$ and Embree by up to $14.55\times$.
\end{itemize}
\enlargethispage{2\baselineskip}
%% =============================================================================

\section{Related Work}
\label{sec:related}
%% =============================================================================

We discuss the most relevant related work here; additional discussion and comparisons are in Suppl.~\ref{sec:related_full}.

\begin{itemize}[nosep,wide,leftmargin=0pt,labelsep=0.3em]
  \item \textbf{BVH-based LiDAR simulation.}
        L\'{o}pez et al.~\cite{lopez2022lidar} and Mock et al.~\cite{mock2025radarays} apply hardware-accelerated RT to spinning LiDAR and FMCW radar, inheriting the per-frame rebuild cost of \hyperref[sc:dynamic]{S1}.
        Spatialized normal-cone hierarchies~\cite{johnson2001snch} augment a BVH with per-node normal cones for closest-geometry queries but discard per-ray directional identity and still require per-frame rebuild under deformation (extended discussion in Suppl.~\ref{sec:related_snch}).
        \grca{} maintains no BVH.
  \item \textbf{Simulator-based approaches.}
        CARLA~\cite{dosovitskiy2017carla}, LGSVL~\cite{rong2020lgsvl}, AirSim~\cite{shah2017airsim}, and Gazebo~\cite{koenig2004gazebo} issue rays through single-ray CPU collision backends with no GPU parallelism, making ${{\sim}}4$\,M rays per frame infeasible.
        GPU-backed production simulators such as Isaac~Sim~\cite{isaacsim} and AWSIM~\cite{awsim} (via RobotecGPULidar~\cite{robotecgpulidar}) route LiDAR rays through OptiX and thus inherit the same BVH-rebuild bottleneck.
        Rasterization back-projection~\cite{denis2023rasterization} requires $\ge$8 cameras per 360$^\circ$ origin; offline and neural methods~\cite{manivasagam2020lidarsim,fang2020alsim,guillard2022lidar,yang2023unisim,wu2024dynlidar,toth2025hybrid} target static ML data generation and cannot generalize to arbitrary dynamic geometry.
        \grca{} needs no scene pre-processing or per-origin cameras.
  \item \textbf{Ray classification and apparent-area methods.}
        Arvo and Kirk~\cite{arvo1987ray} pre-index primitives into 5D angular cells ($O(\tau \cdot \nu)$ re-indexing for $\nu$ angular cells per scene change); Schuetz et al.~\cite{schuetz2026curast} partition geometry by apparent screen coverage for GPU dispatch.
        \grca{} applies the same apparent-area idea as a per-triangle culling predicate at the ray origin with no pre-built index, running in $O(1)$ per triangle and origin.
  \item \textbf{Source-driven light transport.}
        Classical light-transport methods---backward ray tracing~\cite{arvo1986backward}, bidirectional path tracing~\cite{lafortune1993,veach1997}, photon mapping~\cite{jensen1996photon}, radiosity~\cite{goral1984}, and many-lights / instant radiosity~\cite{walter2005lightcuts,keller1997instant}---all trace \emph{from} an emitter but still rely on an acceleration structure for the per-ray intersection query. \grca{} instead bounds, from the emitter's geometry, \emph{which rays can hit a triangle}: a closed-form per-triangle predicate.
\end{itemize}

%% =============================================================================

\section{Emitter-Centric Geometric Modeling}
\label{sec:gacp_model}
%% =============================================================================

This section builds \grca{}'s geometry: rays, their channel cones (GACP) and point-to-index maps (\S\ref{sec:ray_dir}--\ref{sec:ray_indexing}); the cone--triangle test GACP-T (\S\ref{sec:gacp_t}); and each triangle's channel and ray spans (\S\ref{sec:cspan_predict}--\ref{sec:cw_ccw}).
Throughout, \textbf{bold} denotes vectors, $\hat{\cdot}$ denotes unit vectors, and \textit{italics} denote scalars and other variables.
Symbols and abbreviations used throughout the paper are summarized in Table~\ref{tab:notation} (Suppl.~\ref{sec:notation}).

\subsection{Ray Direction Formula}
\label{sec:ray_dir}

Let $\hat{f}$, $\hat{r}$ and $\hat{u}$ be the forward, right and up axes of the left-handed sensor frame.
Rays are indexed horizontally by $R_i$ (clockwise from above) and vertically by $C_i$ (bottom\,=\,0).
The ray direction at azimuth $\theta$ and elevation $\phi$ is:
\begin{equation}
  \hat{d} = \cos\theta\cos\phi\,\hat{f}
           + \sin\theta\cos\phi\,\hat{r}
           + \sin\phi\,\hat{u}
  \label{eq:ray_dir}
\end{equation}
With angular steps $\Delta\theta$, $\Delta\phi$ and center-aligned offsets
$\theta_0 = -\lfloor \chi_n/2 \rfloor\cdot\Delta\theta$,
$\phi_0 = -\lfloor \gamma_n/2 \rfloor\cdot\Delta\phi$:
\begin{equation}
  \theta = \theta_0 + R_i\cdot\Delta\theta,
  \qquad
  \phi  = \phi_0  + C_i\cdot\Delta\phi
\end{equation}
For a $360^\circ$/$180^\circ$ sensor this gives $\theta \in [-\pi,\pi)$ and $\phi \in [-\tfrac{\pi}{2},\tfrac{\pi}{2})$;
the middle-most ray in the middle-most channel points along $\hat{f}$.

\subsection{GACP Definition}

As established in \hyperref[obs:gacp]{Observation~2}, every ray direction in a channel with
elevation $\phi$ satisfies
\begin{equation}
  \hat{d} \cdot \hat{u} = \sin\phi
\end{equation}

\begin{figure}[htb]
  \centering
  \includegraphics[width=0.78\columnwidth]{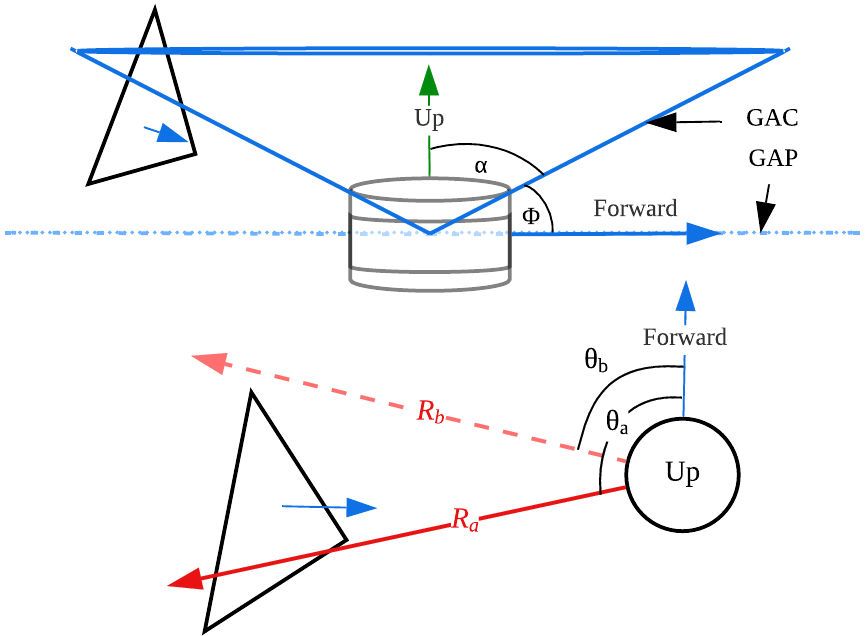}
  \caption{GACP surface with vertical channel and horizontal ray indexing.}
  \Description{Diagram of the GACP cone surface with vertical channel and horizontal ray angular indexing.}
  \label{fig:angular-indexing}
\end{figure}

This places every direction on a cone around $\hat{u}$.
Let $f(\phi,\theta)$ denote the single-ray half-line at azimuth $\theta$ and elevation $\phi$
from origin $\mathbf{o}$:
\begin{equation}
  f(\phi,\theta) \;=\;
    \bigl\{\,\mathbf{o} + \kappa\,\hat{d}(\theta,\phi) : \kappa \ge 0\,\bigr\}
\end{equation}
The GACP surface at elevation $\phi$ is the limit of the discrete ray union as the
azimuthal step vanishes:
\begin{equation}
  \mathrm{GACP}(\phi) \;=\;
    \lim_{\Delta\theta \to 0}\;
    \bigcup_{\theta \in \Theta_{\Delta\theta}} f(\phi,\theta)
  \;=\;
  \bigl\{\,\mathbf{p}
    : (\mathbf{p}-\mathbf{o})\cdot\hat{u}
      = |\mathbf{p}-\mathbf{o}|\sin\phi
  \,\bigr\}
\end{equation}
where $\Theta_{\Delta\theta} = \{\theta_0 + k\,\Delta\theta : k \in \mathbb{Z}\} \cap [-\pi,\pi)$
and $\mathbf{p}$ is any world-space point.
When $\phi = 0$ the cone degenerates
to the horizontal plane through $\mathbf{o}$. This surface is the GACP
(Geometrically Approximated Cone or Plane), illustrated in Figure~\ref{fig:angular-indexing}.
The cone half-angle $\alpha = \tfrac{\pi}{2} - |\phi|$ is the angle from $\hat{u}$ to the cone surface, i.e.\ $\cos\alpha = \sin|\phi|$.

\subsection{Angular Channel Indexing}
\label{sec:channel_indexing}

For a world-space point $\mathbf{p}$, the nearest channel index is:
\begin{equation}
  C_i = \operatorname{clamp}\!\left(
    \operatorname{round}\!\left(
      \frac{\arcsin\!\bigl(\widehat{\mathbf{p}-\mathbf{o}}\cdot\hat{u}\bigr)
            - \phi_0}{\Delta\phi}
    \right),\; 0,\; \gamma_n-1
  \right)
\end{equation}

\subsection{Angular Ray Indexing}
\label{sec:ray_indexing}

The horizontal projection of $\mathbf{p}$ (vertical component removed) is:
\begin{equation}
  \mathbf{q} = (\mathbf{p}-\mathbf{o})
             - \bigl((\mathbf{p}-\mathbf{o})\cdot\hat{u}\bigr)\hat{u}
\end{equation}

The azimuth and its ray index are:
\begin{equation}
  \theta = \operatorname{atan2}\!\bigl(\mathbf{q}\cdot\hat{r},\;
                                        \mathbf{q}\cdot\hat{f}\bigr),
  \qquad
  R_i = \operatorname{clamp}\!\left(
    \operatorname{round}\!\left(
      \frac{\theta - \theta_0}{\Delta\theta}
    \right),\; 0,\; \chi_n-1
  \right)
\end{equation}

$\operatorname{atan2}$ is scale-invariant so normalizing $\mathbf{q}$ is
unnecessary.

\subsection{GACP-T Intersection}
\label{sec:gacp_t}

GACP-T tests whether the GACP surface of channel elevation $\phi$ meets triangle $T$ and returns up to two world-space crossing points, whose azimuths bound that channel's $R_{\mathrm{span}}$.
Its boolean variant serves the early pass and $C_{\mathrm{span}}$ search (Suppl.~\ref{sec:supp_bool}).

\textbf{Early exit.}
Vertex elevations classify $T$ first: all in the wrong half-space or inside the convex cone means no crossing; a mixed triangle must cross (boolean: true at once); an all-outside triangle hit by the apex ray within range spans the cone axis, so the whole channel crosses it (boolean: true). Otherwise both variants solve the edge quadratic (boolean: first root).

\textbf{Cone} ($\phi{\neq}0$).
Squaring the GACP condition, $\bigl((\mathbf{p}{-}\mathbf{o})\cdot\hat{u}\bigr)^2{=}\sin^2\!\phi\,|\mathbf{p}{-}\mathbf{o}|^2$, and substituting a point on each triangle edge for $\mathbf{p}$ gives one quadratic per edge:
\begin{equation}
  c_2\lambda^2 + c_1\lambda + c_0 = 0,
  \qquad \lambda\in[0,1] \text{ along the edge}
  \label{eq:edge_quadratic}
\end{equation}
with $c_2, c_1, c_0$ from its endpoints, $\mathbf{o}$, $\hat{u}$ and $\phi$; roots on the forward half-cone are crossings, \emph{up to two per edge}; pooled over all edges, two bound $R_{\mathrm{span}}$ exactly, one is a tangent touch, and more (degenerate) fall back to the whole channel.

\textbf{Plane} ($\phi{=}0$).
No quadratic: an edge crosses the plane through $\mathbf{o}$ where its endpoint heights change sign, at the interpolated point.

\phantomsection\label{sec:gacp_t_bool}

\subsection{Channel-Span Prediction}
\label{sec:cspan_predict}
\begin{figure}[htb]
  \centering
  \includegraphics[width=0.75\columnwidth]{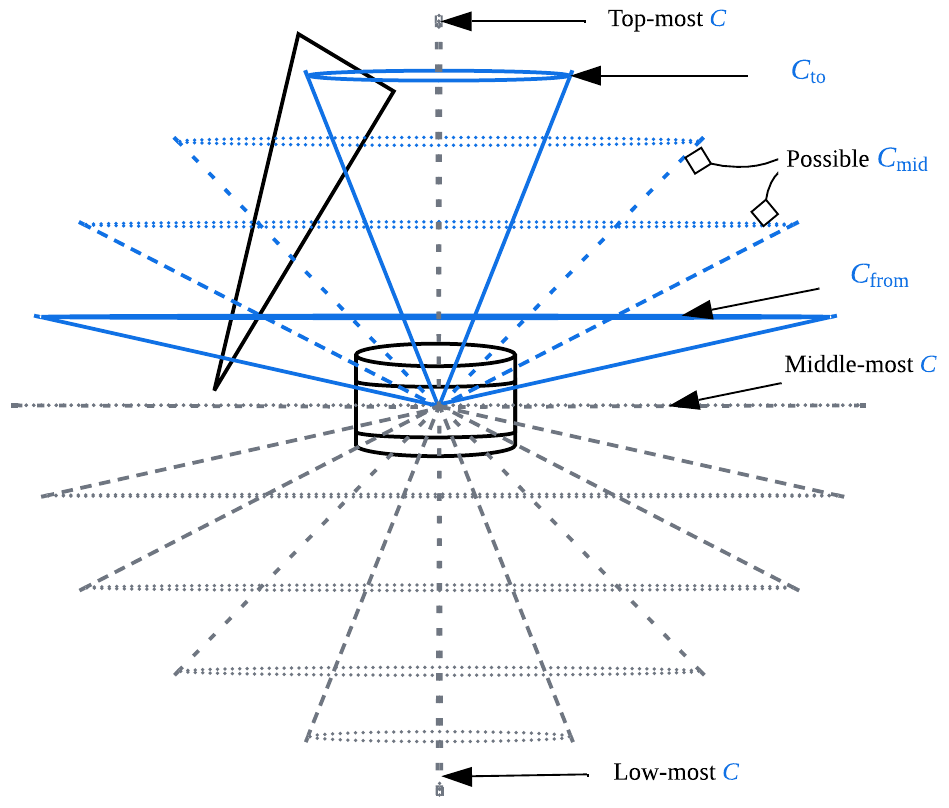}
  \caption{$C_{\mathrm{span}}$ prediction: two binary searches bracket $[C_{\mathrm{from}}, C_{\mathrm{to}}]$ in $O(\log \gamma_n)$.}
  \Description{Diagram showing binary search bracketing of $C_{\mathrm{span}}$ for a triangle.}
  \label{fig:predict-cspan}
\end{figure}

A midpoint channel estimated from the vertex elevations seeds a binary search down to $C_{\mathrm{from}}$, then one up to $C_{\mathrm{to}}$ (Figure~\ref{fig:predict-cspan}, Algorithm~\ref{alg:cspan}); each probe is one $O(1)$ boolean GACP-T call (Algorithm~\ref{alg:gacp_t}),
so the bracket costs $O(\log \gamma_n)$ rather than $O(\gamma_n)$, and a triangle with no valid span is dropped for that origin.

\subsection{Ray-Span Prediction}
\label{sec:rspan_predict}

\begin{figure}[htb]
  \centering
  \includegraphics[width=0.95\columnwidth]{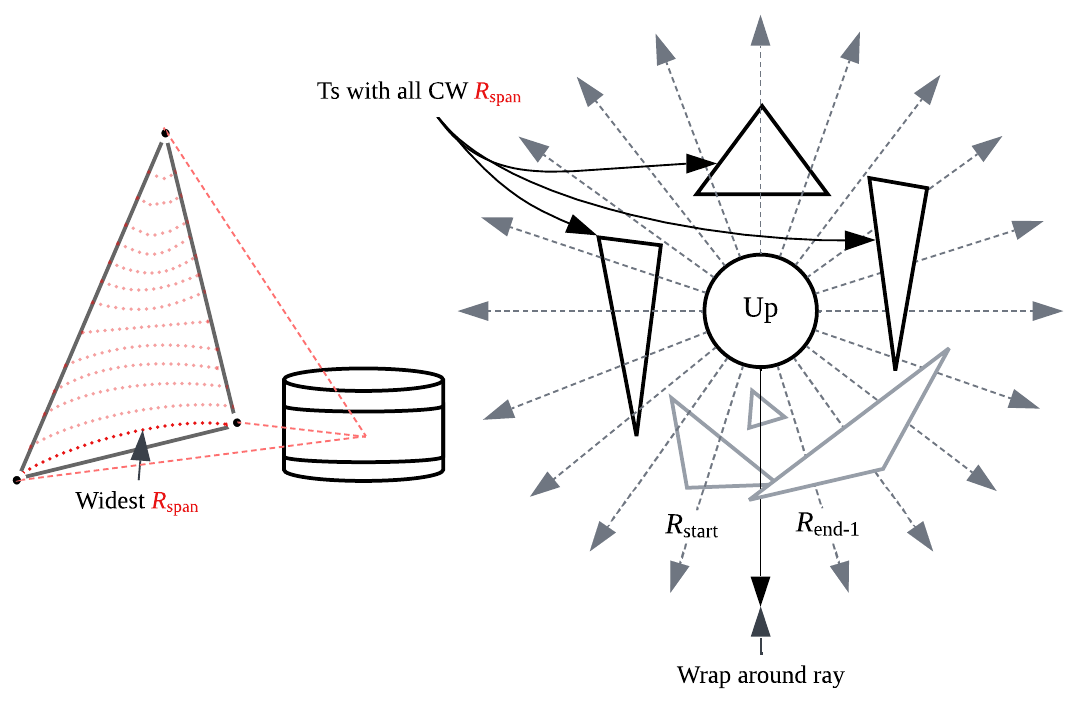}
  \caption{$R_{\mathrm{span}}$ prediction by vertex projection: CW and wrap-around cases.}
  \Description{Diagram showing fast $R_{\mathrm{span}}$ prediction via vertex projection with clockwise and wrap-around cases.}
  \label{fig:fast-rspan}
\end{figure}

Projecting the three vertices to ray indices (\S\ref{sec:ray_indexing}) gives a conservative $R_{\mathrm{span}}$, their minimum to maximum, without any GACP-T test (Figure~\ref{fig:fast-rspan}, Algorithm~\ref{alg:rspan}).
Beyond $180^\circ$ FOV a seam check runs first: the triangle is \emph{all-CW} if any of three dot-product tests in a Gram--Schmidt frame around the center ray (all-front, all-right, all-left) holds; otherwise it may straddle the seam and goes to the late pass as BAT (Suppl.~\ref{sec:supp_rspan}).
\phantomsection\label{eq:all_cw}

\subsection{CW/CCW Arc Disambiguation}
\label{sec:cw_ccw}
\begin{figure}[htb]
  \centering
  \includegraphics[width=0.65\columnwidth]{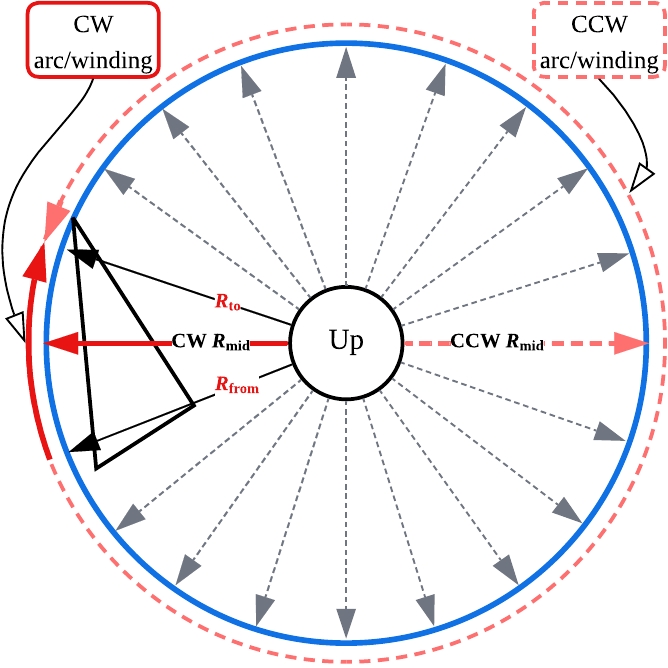}
  \caption{CW vs CCW arc: mid-ray plane test selects the correct arc (bold).}
  \Description{Diagram showing CW versus CCW arc disambiguation via mid-ray plane test.}
  \label{fig:cw-ccw-arc}
\end{figure}

Algorithm~\ref{alg:arc} picks each channel's arc: equal endpoints give one ray, all-CW triangles the CW arc.
For seam-straddling triangles the two GACP-T endpoints are ambiguous (Figure~\ref{fig:cw-ccw-arc}); the CW arc-midpoint ray and its reflection across $\hat{u}$ (the CCW midpoint) are tested against the triangle plane, the hit selects the arc, and M\"oller--Trumbore on the CW mid-ray breaks ties (Suppl.~\ref{sec:supp_cw_ccw}).

%% =============================================================================

\section{The Two-Pass Pipeline}
\label{sec:pipeline}
%% =============================================================================

Two passes turn these predictions into hits in one shared per-ray buffer: the early pass resolves small, well-predicted triangles at once and defers the rest to the late pass (Algorithms~\ref{alg:early_pass}--\ref{alg:late_pass}; Suppl.~\ref{sec:pipeline_detail}--\ref{sec:late_pass}).
\textbf{Early pass.}
Per triangle and origin, it filters (back-face, apparent area, range), predicts both spans and classifies (Suppl.~\ref{sec:supp_filters}--\ref{sec:classification}): all-CW triangles within $(\gamma_T,\chi_T)$ are SAT and run inline RTIC; the rest are BAT (resolved inline on the CPU).
Deferring costly BAT work follows the megakernel-to-wavefront principle~\cite{laine2013megakernels}, keeping occupancy-collapsing branches~\cite{guthmann2023occupancy} out of the hot loop.
\textbf{Late pass.}
It runs precise GACP-T per stored channel for exact $R_{\mathrm{span}}$, takes the CW arc if all-CW (else the mid-ray test, \S\ref{sec:cw_ccw}), then runs RTIC.

\begin{algorithm}[htb]
\caption{GACP-T Intersection (full: crossing points; bool: yes/no)}
\label{alg:gacp_t}
{\footnotesize
\begin{algorithmic}[1]
\If{elevation $= 0$} \Comment{plane case: no quadratic needed}
  \State compute signed axial height for each vertex
  \State \Return sign-changing edge crossings / \textbf{bool:} two edges change sign
\EndIf
\State classify each vertex: inside cone or outside
\If{all inside} \Return no crossings / \textbf{false} \EndIf
\If{mixed} \textbf{bool:} \Return \textbf{true};\quad \textbf{full:} proceed to per-edge quadratic below \EndIf
\If{all outside, apex ray hits within cone reach} \Return whole channel / \textbf{true} \EndIf
\For{each edge~\cite{eberly_geometrictools}} \Comment{quadratic coefficients in Suppl.~\ref{sec:precise_cone}}
  \State solve edge quadratic; accept root in $[0,1]$ on forward half-cone
  \State \textbf{bool:} return \textbf{true} on first root;\quad \textbf{full:} collect crossing point
\EndFor
\State \Return pooled crossings (2: exact span; 1: tangent; ${>}2$: whole channel) / \textbf{false}
\end{algorithmic}
}
\end{algorithm}

\begin{algorithm}[htb]
\caption{Channel-Span Prediction}
\label{alg:cspan}
{\footnotesize
\begin{algorithmic}[1]
\State estimate midpoint channel from vertex elevations \Comment{formula in Suppl.~\ref{sec:supp_cspan}}
\State binary search downward from estimate $\to C_{\mathrm{from}}$ via boolean GACP-T
\State binary search upward from $C_{\mathrm{from}} \to C_{\mathrm{to}}$ via boolean GACP-T
\If{no valid channel span} discard triangle for this origin \EndIf
\State \Return $[C_{\mathrm{from}},C_{\mathrm{to}}]$
\end{algorithmic}
}
\end{algorithm}

\begin{algorithm}[htb]
\caption{Fast R-span Prediction}
\label{alg:rspan}
{\footnotesize
\begin{algorithmic}[1]
\If{FOV $\le 180^\circ$} all-CW $\gets$ true; skip seam check \EndIf
\State build center-ray right vector via Gram--Schmidt
\State all-CW $\gets$ all-front \textbf{or} all-right \textbf{or} all-left \Comment{three dot-product tests}\label{line:allcw}
\If{\textbf{not} all-CW} \Return BAT \Comment{late pass finds the exact span} \EndIf
\State project each vertex to ray index via atan2 (\S\ref{sec:ray_indexing})
\State \Return $R_{\mathrm{from}}={}$min vertex index,\quad $R_{\mathrm{to}}={}$max vertex index
\end{algorithmic}
}
\end{algorithm}

\begin{algorithm}[htb]
\caption{CW/CCW Arc Disambiguation}
\label{alg:arc}
{\footnotesize
\begin{algorithmic}[1]
\If{$R_{\mathrm{from}}=R_{\mathrm{to}}$} \Return clockwise$=$true, sweep$=$0 \EndIf
\If{all-CW} \Return clockwise$=$true, sweep$=R_{\mathrm{to}}-R_{\mathrm{from}}$ \EndIf
\State get ray direction at CW arc midpoint
\State get CCW midpoint ray by reflection across $\hat{u}$ \Comment{azimuthally opposite ray}
\State test both mid-rays against triangle plane
\If{CW hits \textbf{and} CCW misses} \Return clockwise$=$true
\ElsIf{CCW hits \textbf{and} CW misses} \Return clockwise$=$false
\Else{} full M\"oller--Trumbore~\cite{moller1997} on CW mid-ray (tiebreaker) \EndIf
\State compute sweep count from clockwise flag
\end{algorithmic}
}
\end{algorithm}
\begin{algorithm}[htb]
\caption{Early Pass}
\label{alg:early_pass}
{\footnotesize
\begin{algorithmic}[1]
\For{each triangle}
  \For{each ray origin}
    \State \textbf{Filter}: back-face, apparent area~(\hyperref[obs:apparent_area]{Observation~3}), range~\cite{ericson2004}
    \State \textbf{C-span predict}: binary search on GACP-T boolean predicate
    \If{no channel intersects} \textbf{skip} triangle for this origin \EndIf
    \State \textbf{R-span predict}: vertex projection (conservative widest)
    \If{all-CW and spans within thresholds} \Comment{SAT}
      \For{each (channel, ray) in predicted spans}
        \State M\"{o}ller--Trumbore~\cite{moller1997}; record closest hit via atomic min
      \EndFor
    \Else \Comment{BAT}
      \State Store to deferred list with pre-computed flags
    \EndIf
  \EndFor
\EndFor
\end{algorithmic}
}
\end{algorithm}

\begin{algorithm}[htb]
\caption{Late Pass}
\label{alg:late_pass}
{\footnotesize
\begin{algorithmic}[1]
\For{each deferred-list entry (BAT triangles)}
  \For{each origin where origin-active flag is set}
    \For{each channel in stored span}
      \State Precise GACP-T intersection \Comment{exact R-span}
      \If{0 crossings} \textbf{skip} channel \EndIf
      \If{all-CW} arc direction is CW
      \Else{} mid-ray plane test selects CW or CCW arc
      \EndIf
      \For{each ray in R-span (CW or CCW order)}
        \State M\"{o}ller--Trumbore; record closest hit via atomic min
      \EndFor
    \EndFor
  \EndFor
\EndFor
\end{algorithmic}
}
\end{algorithm}

\section{Benchmarks}
\label{sec:benchmarks}
%% =============================================================================

\textbf{Platforms.} \grcacpu{} (AVX2) and \grcagpu{} (CUDA) run on the machines in Table~\ref{tab:hardware} (PC1--PC5) with NVIDIA driver~590, CUDA~13.1, OptiX~9.1.0 and Embree~4.4.0, at a fixed 100\,Hz simulation clock.

\textbf{Scenes and motion.} Six static scenes (Figure~\ref{fig:teaser}) and one dynamic mesh are used (Table~\ref{tab:meshes}), all from McGuire~\cite{mcguire2017} except Emerald Square, from the NVIDIA ORCA archive~\cite{nvidia_orca2017}. The sportscar is instantiated 10, 20 or 30 times (written 10--30~dyn; 3.0/6.0/9.0\,M dynamic triangles) with per-axis scale drawn from $[0.001,30]$ (sparks to vessels, Table~\ref{tab:scalecategories}) unless stated otherwise. All runs use \textbf{random per-frame motion}: every DOF randomizes independently each frame, avoiding trajectory-coherence bias, and all backends replay the same pre-recorded pose and scale stream (fixed seed). Three run conditions vary the deformation (Table~\ref{tab:testsuite}, Figure~\ref{fig:deformation_modes}): none (ND), object-bound (OBD) and scene-wide (SWD); from ND to SWD, \grcacpu{} changes by ${<}1\%$ while Embree slows ${\sim}7\times$ (Suppl.~\ref{sec:pose_sensitivity_detail}).

\textbf{Sensors and parameters.} Each ray origin is one LiDAR ($4096\times128 = 524{,}288$ rays, $360^\circ/180^\circ$ FOV, $0.05$--$1000$\,m), with $\Omega{=}2,4,6,8$ by default. \grca{} uses its defaults (Table~\ref{tab:default_params}, Suppl.~\ref{sec:default_params}): $(\gamma_T,\chi_T){=}(64,64)$, BAT-list capacity $8{\times}10^5$, apparent-area threshold $\varepsilon_A{=}10^{-6}$; $(64,64)$ is within ${\leq}1.1/2.5\%$ (CPU/GPU) of the per-range optimum (Suppl.~\ref{sec:sat_threshold_detail}).

\textbf{Baselines.} Both BVH backends run a flat single BVH (no IAS/instancing), with any-hit and motion blur off and triangle primitives only, isolating BVH traversal cost from scene-graph engineering (Suppl.~\ref{sec:backend_details}). On the CPU, both sides run single-threaded at equal SIMD width. A two-level IAS/BLAS setup~\cite{wald2007dynamic} would gain little here. Under this motion, objects move too far between frames for a refit, so every dynamic BLAS is still rebuilt each frame. Only the static rebuild is saved, and that saving shrinks as dynamic geometry grows.

\textbf{Frame pacing.} ${\sim}57\%$ of \grca{} frames finish below the mean in each condition, so means are upper bounds; fewer fall within ${\pm}20\%$ of the six-scene pooled mean (\grcagpu{} $24$--$26\%$, \grcacpu{} $20$--$21\%$) than OptiX ($36$--$42\%$) and Embree ($28$--$36\%$; Suppl.~\ref{sec:frame_pacing}--\ref{sec:frame_pacing_500}).

\begin{table}[htb]
  \centering
  \caption{Benchmark configuration. Top-left: test hardware. Top-right: scale-to-category map. Middle: benchmark meshes ($^\dagger$avg tri area at scale~1.0; scales as $s^2$ across the dynamic range). Bottom: test suite run conditions.}
  \label{tab:benchmark_config}
  \scriptsize
  \setlength{\tabcolsep}{3pt}
  \begin{minipage}[t]{0.55\columnwidth}
    \centering
    \label{tab:hardware}
    \begin{tabular}{lll}
      \toprule
      PC & CPU & GPU \\
      \midrule
      PC1 & i7-12700H     & RTX 3080 Ti Mobile \\
      PC2 & i9-13900HX    & RTX 4090 Mobile \\
      PC3 & i9-14900KF    & RTX 4090 Desktop \\
      PC4 & i7-14700      & RTX 4060 Desktop \\
      PC5 & Ryzen 9 9950X & RTX 5090 Desktop \\
      \bottomrule
    \end{tabular}
  \end{minipage}%
  \hfill
  \begin{minipage}[t]{0.43\columnwidth}
    \centering
    \label{tab:scalecategories}
    \begin{tabular}{rll}
      \toprule
      Scale & Category & L$\times$W$\times$H \\
      \midrule
      0.001 & sparks   & $4.6{\times}2.3{\times}1.1$\,mm \\
      0.1   & drones   & $46{\times}23{\times}11$\,cm \\
      1.0   & vehicle  & $4.57{\times}2.28{\times}1.08$\,m \\
      10    & structures & $46{\times}23{\times}11$\,m \\
      30    & vessels  & $137{\times}68{\times}32$\,m \\
      \bottomrule
    \end{tabular}
  \end{minipage}

  \vspace{0.6em}
  \label{tab:meshes}
  \resizebox{\columnwidth}{!}{%
  \setlength{\tabcolsep}{4pt}%
  \begin{tabular}{lllllrl}
    \toprule
    Mesh & Scale & Bounds & Diagonal & Scale range & Triangles & Avg tri \\
    \midrule
    Vokselia Spawn           & 100   & 386×66×384\,m      & 549\,m  & static       & 1,875,632  & 5{,}006\,cm$^2$ \\
    Lumberyard Bistro        & 0.02  & 217×64×231\,m      & 323\,m  & static       & 3,858,116  & 577\,cm$^2$ \\
    San~Miguel               & 1     & 69×15×27\,m        & 76\,m   & static       & 5,617,451  & 23\,cm$^2$ \\
    Rungholt                 & 1     & 655×69×550\,m      & 858\,m  & static       & 6,704,264  & 5{,}015\,cm$^2$ \\
    Emerald Square           & 1     & 230×113×230\,m     & 344\,m  & static       & 9,996,068  & 326\,cm$^2$ \\
    Power~Plant              & 0.001 & 611×249×186\,m     & 686\,m  & static       & 12,759,246 & 384\,cm$^2$ \\
    Sportscar               & 1     & 4.6\,mm--137\,m    & \begin{tabular}[b]{@{}l@{}}5.2\,mm--\\157\,m\end{tabular} & $[0.001,30]$ & 300,603    & 3.56\,cm$^2$$^\dagger$ \\
    \bottomrule
  \end{tabular}}

  \vspace{0.6em}
  \label{tab:testsuite}
  \setlength{\tabcolsep}{4pt}
  \begin{tabular}{lll}
    \toprule
    Condition & Topology & Scatter bound \\
    \midrule
    No Deformation (ND)            & fixed      & --- \\
    Object-bound Deformation (OBD) & randomized & mesh object bounding box \\
    Scene-wide Deformation (SWD)   & randomized & world axis-aligned bounding box \\
    \bottomrule
  \end{tabular}
\end{table}

\subsection{BVH Mode Selection}
\label{sec:bvh_mode_selection}

Full rebuild is the baseline reference: of seventeen update policies evaluated across ND, OBD, and SWD (Power~Plant, 30~dyn, $\Omega{=}8$), it is the only one achieving ${\geq}96\%$ pacing within ${\pm}20\%$ across all conditions and both backends.
The others are refit and fifteen periodic rebuilds (alternating rebuild and refit at varying frequencies; per-mode data in Suppl.~\ref{sec:bvh_mode_detail}).
Refit is catastrophically unreliable under \hyperref[sc:dynamic]{S1}: under SWD, Embree refit and a frequent periodic rebuild both exceeded 5\,h.
OptiX's least frequent periodic rebuild achieves the lowest ND mean (17.2\,ms), but its frame pacing collapses to 10\%; the hybrid pipeline (\S\ref{sec:hybrid_pipeline}) supersedes it.

\subsection{Performance Analysis}
\label{sec:performance_analysis}

The following sections analyze \grca{}'s performance across key variables; all use 50~frames (additional analyses in Suppl.~\ref{sec:additional_benchmarks}).

\subsubsection{Scale, Mesh Density, and BAT Growth}
\label{sec:tri_area}
\label{sec:scale_variation}

\grca{}'s advantage grows with mesh density: apparent area bounds the BAT list (\hyperref[obs:apparent_area]{Observation~3}) while BVH construction scales super-linearly in $\tau$~\cite{wald2007bvh}.

\textbf{Mesh density (fixed scale).}
Denser meshes leave \grca{} nearly flat but slow both BVHs: with the sportscar subdivided up to $64\times$ at fixed scale~1.0 (PC1, $\Omega{=}2$--$8$), \grcagpu{} and \grcacpu{} rise only $+20\%$ and $+12\%$, while OptiX degrades $2.6\times$ and Embree $8.1\times$ from denser BVH nodes (Figure~\ref{fig:tri_area_culling}; Table~\ref{tab:subdiv}).
The avg BAT count is \emph{exactly} constant across all subdivision levels---sportscar triangles are predominantly SAT at $1{\times}$ and only shrink further below $\varepsilon_A$, so an increasing fraction is discarded pre-RTIC; the BAT list is driven entirely by the static Power~Plant scene.

\textbf{Mesh scale (fixed density).}
Larger meshes subtend more channels per origin, growing the BAT list and the late-pass workload (Figure~\ref{fig:scale_vs_bvh}; Table~\ref{tab:scale_variation}); at $[0.6,60]$ the average BAT count grows from 133\,K to 147\,K ($+11\%$) over the default $[0.001,30]$.
\grcacpu{} vs Embree speedup depends primarily on dyn count: more dynamic objects mean more BVH rebuilds for Embree with no additional structural cost for \grca{}; at low dyn counts Embree's per-frame rebuild is cheap enough to offset \grca{}'s filtering cost.
\grcagpu{} leads OptiX across all scale ranges, narrowing at the large-scale end as BAT growth from increased apparent area erodes the margin.

\textbf{Joint scale $\times$ density.}
BVH cost is bound by triangle count, \grca{}'s by BAT count: under $1\times$ subdivision ($4\times$ triangles) at large-scale ranges $[0.6,60]$ and $[1,100]$ (Figure~\ref{fig:scale_bat_subdiv}), \grca{}'s speedups reach $2.78\times$/$2.18\times$ GPU and $1.96\times$/$1.53\times$ CPU.
BAT count stays near-constant at 203\,K and 258\,K---even subdivided triangles remain above $\varepsilon_A$ at these scales---so \grcagpu{} rises only $57.9\!\to\!69.3$\,ms ($+20\%$) despite $4\times$ the triangles.
BVH backends are less forgiving: OptiX rises ${\sim}2.4\times$ and Embree $1.8$--$2.3\times$ because BVH construction scales super-linearly with triangle count~\cite{wald2007bvh}.

%% =============================================================================

\subsubsection{Range Variation}
\label{sec:range_variation}
%% =============================================================================

Real LiDARs reach 10--120\,m; 1000\,m is a worst case never reached in deployment.
Figure~\ref{fig:range_variation} (top; Table~\ref{tab:range_variation}) sweeps $D\qual{\max}$ from 5\,m to 1000\,m (Power~Plant, 30~dyn, $\Omega{=}8$, avg ND/OBD/SWD).
Under ND, CPU speedup declines with range as culling disengages; under SWD it \emph{increases} ($5.2\to9.3\times$ up to 100\,m) as Embree's trace through scattered geometry grows faster with range than \grcacpu{}'s work.
GPU speedup plateaus beyond ${\approx}300$\,m, once the range reaches the farthest geometry.

\subsection{Results}
\label{sec:final_analysis}
\label{sec:main_suite}

Full benchmark data across all scenes, dyn counts, $\Omega$ values, and platforms is in Tables~\ref{tab:main_suite_full}--\ref{tab:range_hybrid_suite} (Suppl.~\ref{sec:additional_data}); all runs use random per-frame motion, scale $[0.001,30]$, ND/OBD/SWD, 500 frames at 100\,Hz.
Table~\ref{tab:main_suite_full} covers standalone \grca{} at $D\qual{\max}{=}1000$\,m; Table~\ref{tab:range_suite} range-filtered \grca{} at 100\,m; Table~\ref{tab:hybrid_suite} the hybrid pipeline at 1000\,m; and Table~\ref{tab:range_hybrid_suite} Range+Hybrid at 100\,m.

\subsubsection{Cost Contributors}
\label{sec:cost_breakdown}

Frame cost has five structural contributors: \textbf{static scene} ($O(\tau)$ vs.\ $\tau$-scaling BVH rebuild); \textbf{scene extent} (BAT list and traversal depth); \textbf{triangle area} (Rungholt and Vokselia~Spawn at ${\approx}0.50\,\text{m}^2$/tri inflate the SAT count via \hyperref[obs:apparent_area]{Observation~3}); \textbf{dynamic object count} (full rebuild vs.\ dynamic re-filter); and \textbf{ray origin count} (both linear; \grca{} apparent-area-capped).

Figure~\ref{fig:cost-contributors} isolates each contributor; the two failure regimes (Rungholt CPU, San~Miguel GPU) are detailed in Suppl.~\ref{sec:scene_size} and \S\ref{sec:cross_scene_summary}.
GPU speedups are conservative: OptiX traverses on dedicated RT cores, and its time is dominated by the shader-core rebuild (${\approx}76$--$85$\,ms vs.\ $0.7$--$12.7$\,ms of trace; Power~Plant, 30~dyn, $\Omega{=}2$--$8$).

\subsubsection{Cross-Scene Performance Summary}
\label{sec:cross_scene_summary}

\textbf{Overall} (PC1, 1000\,m, Table~\ref{tab:main_suite_full}).
\grca{} leads in $84\%$ of pairs (above $1\times$: avg $\mathbf{4.34\times}$ CPU, $\mathbf{2.63\times}$ GPU); $7\%$ are at ${\pm}10\%$ parity, $9\%$ losses (avg $0.65\times$).
\textbf{Peaks:} \grcacpu{} reaches up to $\mathbf{14.55\times}$ over Embree (San~Miguel, 30~dyn, $\Omega{=}2$, SWD) and $\mathbf{10.40\times}$ (Lumberyard, 30~dyn, SWD); \grcagpu{} up to $\mathbf{7.97\times}$ over OptiX (Emerald~Square, 30~dyn, $\Omega{=}2$, ND).

\textbf{Range culling} (100\,m; Figure~\ref{fig:range_variation}, bottom; Table~\ref{tab:range_suite}).
It lifts typical speedups, not peaks: median speedup rises $2.3\to3.1\times$ (CPU) and $2.0\to2.9\times$ (GPU), while peaks barely move ($\mathbf{14.66\times}$ CPU, $\mathbf{8.33\times}$ GPU).
In the five scenes larger than the sensor range, median \grca{} frame time drops 18--52\% (CPU) and 14--45\% (GPU) vs.\ 1000\,m (up to 53\%/56\%, Rungholt).

\textbf{Origins and platforms.}
Speedup falls with $\Omega$ (median ND: $2.4\to1.2\times$ CPU, $5.1\to1.4\times$ GPU) as each origin adds \grca{} work but no rebuild; only CPU SWD stays near-constant ($6.7\to6.0\times$), as Embree's trace grows with $\Omega$ too (Figure~\ref{fig:time-vs-origins}).
On PC2--PC5 the worst case (Power~Plant, 30~dyn, $\Omega{=}8$, SWD) stays at $8.66$--$10.61\times$ CPU and $1.32$--$2.13\times$ GPU (Figure~\ref{fig:multi_platform}, Table~\ref{tab:main_suite_full}).

\textbf{Failure cases.}
Both losses are specific conditions; SWD always rescues \grcacpu{} as Embree's trace explodes (Table~\ref{tab:negative_cases}).
\emph{Rungholt} (CPU, $0.32\times$): large triangles saturate the span threshold (Suppl.~\ref{sec:scene_size}).
\emph{San~Miguel} (GPU, $0.52\times$): its compact 76\,m extent keeps the BAT count small, so scene extent, not BAT size, is the key variable. The default scale range lets the sportscar outgrow the scene (up to 137\,m); at a scene-proportionate $[0.001,5]$ (${\leq}23$\,m), \grcagpu{} wins all 36 configurations ($1.14$--$6.02\times$; \grcacpu{} $1.24$--$9.64\times$; Table~\ref{tab:sanmiguel_scale5_perf}).

\textbf{Span thresholds.}
Tuning recovers neither failure: over all 36 $(\gamma_T,\chi_T)\in\{8,16,32,64,96,128\}^2$ the default $(64,64)$ is within ${\leq}5.1\%$ GPU of the best per scene and condition (Suppl.~\ref{sec:sat_threshold_extended}, Table~\ref{tab:sat_threshold_spread}).
The spread is not negligible: for $\chi_T{\geq}32$ it is $5$--$16\%$ GPU (median $9\%$), $\chi_T{\leq}16$ can cost up to $1.9{\times}$, and $\chi_T{=}64$ is near-optimal.

\section{Accuracy}
\label{sec:accuracy}

The GACP-T intersection is conservative by design and yields $100\%$ hit-match with OptiX when all performance optimizations are disabled.
With optimizations enabled, the average Hit\% floor (OptiX hits reproduced) across all configurations is ${\geq}98\%$; the four causes of sub-threshold misses are apparent-area discard, $C_{\mathrm{span}}$ search initialization, near-grazing rounding in the GACP-T early exit, and CW/CCW seam misprediction (Suppl.~\ref{sec:accuracy_causes}).
Only San~Miguel falls persistently below $98\%$: the worst case is $95.8\%$, compounded by its compact 76\,m scene extent, widest scale range, and scene-wide deformation---a sensor-geometry mismatch, not an algorithmic defect.
Scale $[0.001,5]$ recovers it to ${\geq}97.3\%$, and at $\Omega{=}8$ the floor is $\mathbf{100.0\%}$ across all conditions (Table~\ref{tab:sanmiguel_scale5_perf}, Suppl.~\ref{sec:accuracy_results}).
Both backends give identical floors: accuracy is model-driven, not backend-driven.

%% =============================================================================

\section{Hybrid Static/Dynamic Pipeline}
\label{sec:hybrid_pipeline}

\grca{} handles dynamic geometry and a static BVH (built once, never rebuilt) the static part; hits are merged per ray by minimum distance (Figure~\ref{fig:hybrid_benchmark}; static-only costs in Suppl.~\ref{sec:scene_size}).
Across all six scenes, median speedups over ND/OBD/SWD reach $\mathbf{2.8}$--$\mathbf{11.4\times}$ CPU and $\mathbf{4.4}$--$\mathbf{9.8\times}$ GPU (Table~\ref{tab:hybrid_suite}).
The only hybrid losses are San~Miguel on the GPU under OBD/SWD at $\Omega{\geq}4$ ($0.55$--$0.97\times$, 10 of 216 configurations), as for standalone \grcagpu{} (\S\ref{sec:cross_scene_summary}).

\textbf{Range+Hybrid} (100\,m; Figure~\ref{fig:range_variation}, bottom; Table~\ref{tab:range_hybrid_suite}) gives the highest GPU speedups: up to $\mathbf{36.5\times}$ GPU and $\mathbf{17.9\times}$ CPU, median $6.8\times$ GPU and $4.7\times$ CPU (Hybrid@1000\,m: $6.0\times$, $4.5\times$).
Culling trims only the dynamic work, so the hybrid gains less than standalone; Embree's shorter SWD trace pulls its CPU peak below Hybrid@1000\,m.

\textbf{GACP-T$_g$} (Figure~\ref{fig:gacp-tg}) would lift GACP-T from a single triangle to a BVH bounding volume, pruning entire subtrees in $O(1)$ without per-frame rebuild and unifying the static and dynamic halves under one closed-form predicate; its implementation is left as future work.

\section{General Purpose Ray Caster}
\label{sec:general_purpose}

Any emitter with a known position and bounded angular extent defines a ray set that \grca{} can filter---the pipeline is not LiDAR-specific.
A spot light's emission cone maps directly to a GACP; a point light is the degenerate full-sphere case.
Applying \grca{}'s two-pass pipeline enables \emph{per-triangle} prediction of which lights each triangle can occlude, eliminating shadow rays before they are cast---with no per-frame BVH rebuild for dynamic occluders (Figure~\ref{fig:grca_shadows})~\cite{whitted1980,walter2005lightcuts}.
Area lights are approximated by distributing $N$ ray origins across the emitter surface; when multiple emitters share a location, co-locating their origins into one increases BAT overlap and cuts \grca{} frame time by $13$--$17\%$ at $\Omega{=}8$ (Suppl.~\ref{sec:combined_origin}).

The $\gamma_n{\times}\chi_n$ grid extends to non-spinning LiDARs (Suppl.~\ref{sec:nonspinning}), time-of-flight cameras~\cite{kim2023dopplertof}, sonar~\cite{jansen2026sonotrace}, and radar cross-sections.

\section{Limitations}
\label{sec:limitations}
%% =============================================================================

\textbf{Angular noise constraints.}
Distance and azimuth noise are supported provided no perturbed ray crosses an adjacent ray boundary; per-channel elevation noise is supported provided no channel offset crosses an adjacent channel boundary (Suppl.~\ref{sec:noise}). Per-ray vertical noise---distinct perturbations per ray within a channel---is not supported: GACP-T evaluates at a single nominal elevation per channel.

\textbf{Static-only scenes.}
\grca{} provides no advantage over BVH for purely static geometry (no BVH rebuild cost): speedup ${<}1$ in all static-only runs. CPU cost is dominated by SAT count, GPU cost by BAT count; large-triangle scenes (e.g.\ Rungholt, $0.17\times$ CPU) saturate the span-threshold. However, apparent-area culling is highly effective in typical scenes: Power~Plant's 12.8M triangles reduce to 837K~SAT and 125K~BAT per frame. Details: Suppl.~\ref{sec:scene_size}.

%% =============================================================================

\section{Conclusion}
%% =============================================================================

BVH construction dominates frame time in dynamic LiDAR simulation and offers no mechanism for directional or apparent-area culling.
\grca{} eliminates it entirely: each triangle predicts which rays can reach it, outperforming OptiX and Embree in dynamic practical scenes without a BVH.
The paradigm inversion, emitter-centric GACP model, CUDA/AVX2 implementations, hybrid static/dynamic pipeline, and range culling together deliver up to $36.5\times$ GPU and $22.7\times$ CPU speedup over the best-suited BVH baseline for dynamic scenes across 216 test configurations (Hybrid@1000\,m: $22.7\times$ CPU, $34.9\times$ GPU; Range+Hybrid@100\,m: $17.9\times$ CPU, $36.5\times$ GPU).

\grca{} reduces the brute-force bound (Equation~\ref{eq:brute}) to:
\begin{equation}
  \label{eq:rtic_reduced}
  \begin{aligned}
    \text{Total RTIC} \approx\;
      \sum_{n=1}^{\Omega} \gamma_n\qual{\text{predicted}} \times \chi_n\qual{\text{predicted}} \times \tau_n\qual{\text{filtered}}
  \end{aligned}
\end{equation}
where $\tau_n\qual{\text{filtered}}$ counts origin $n$'s triangles left after the early-pass filters and $\gamma_n\qual{\text{predicted}}$, $\chi_n\qual{\text{predicted}}$ are their mean predicted spans.
At 1000\,m (worst case, range culling inactive) \grcagpu{}'s per-frame RTIC ranges $4.4{\times}10^5$--$1.72{\times}10^9$ (mean $3.25{\times}10^7$)---three to seven orders of magnitude below Equation~\ref{eq:brute}.
Future work extends the hybrid pipeline (\S\ref{sec:hybrid_pipeline}) and general-purpose framing (\S\ref{sec:general_purpose}) in the OSS repository.

%% =============================================================================
\bibliographystyle{ACM-Reference-Format}
\bibliography{references}

%%% -*-BibTeX-*-
%%% Do NOT edit. File created by BibTeX with style
%%% ACM-Reference-Format-Journals [18-Jan-2012].

\begin{thebibliography}{45}

%%% ====================================================================
%%% NOTE TO THE USER: you can override these defaults by providing
%%% customized versions of any of these macros before the \bibliography
%%% command.  Each of them MUST provide its own final punctuation,
%%% except for \shownote{} and \showURL{}.  The latter two
%%% do not use final punctuation, in order to avoid confusing it with
%%% the Web address.
%%%
%%% To suppress output of a particular field, define its macro to expand
%%% to an empty string, or better, \unskip, like this:
%%%
%%% \newcommand{\showURL}[1]{\unskip}   % LaTeX syntax
%%%
%%% \def \showURL #1{\unskip}           % plain TeX syntax
%%%
%%% ====================================================================

\ifx \showCODEN    \undefined \def \showCODEN     #1{\unskip}     \fi
\ifx \showISBNx    \undefined \def \showISBNx     #1{\unskip}     \fi
\ifx \showISBNxiii \undefined \def \showISBNxiii  #1{\unskip}     \fi
\ifx \showISSN     \undefined \def \showISSN      #1{\unskip}     \fi
\ifx \showLCCN     \undefined \def \showLCCN      #1{\unskip}     \fi
\ifx \shownote     \undefined \def \shownote      #1{#1}          \fi
\ifx \showarticletitle \undefined \def \showarticletitle #1{#1}   \fi
\ifx \showURL      \undefined \def \showURL       {\relax}        \fi
% The following commands are used for tagged output and should be
% invisible to TeX
\providecommand\bibfield[2]{#2}
\providecommand\bibinfo[2]{#2}
\providecommand\natexlab[1]{#1}
\providecommand\showeprint[2][]{arXiv:#2}

\bibitem[Aila and Laine(2009)]%
        {aila2009}
\bibfield{author}{\bibinfo{person}{Timo Aila} {and} \bibinfo{person}{Samuli
  Laine}.} \bibinfo{year}{2009}\natexlab{}.
\newblock \showarticletitle{Understanding the Efficiency of Ray Traversal on
  {GPU}s}. In \bibinfo{booktitle}{\emph{Proceedings of the Conference on High
  Performance Graphics}}. \bibinfo{pages}{145--149}.
\newblock
\href{https://doi.org/10.1145/1572769.1572792}{doi:\nolinkurl{10.1145/1572769.1572792}}


\bibitem[Amanatides(1984)]%
        {amanatides1984}
\bibfield{author}{\bibinfo{person}{John Amanatides}.}
  \bibinfo{year}{1984}\natexlab{}.
\newblock \showarticletitle{Ray Tracing with Cones}.
\newblock \bibinfo{journal}{\emph{ACM SIGGRAPH Computer Graphics}}
  \bibinfo{volume}{18}, \bibinfo{number}{3} (\bibinfo{year}{1984}),
  \bibinfo{pages}{129--135}.
\newblock
\href{https://doi.org/10.1145/964965.808589}{doi:\nolinkurl{10.1145/964965.808589}}


\bibitem[Arvo(1986)]%
        {arvo1986backward}
\bibfield{author}{\bibinfo{person}{James Arvo}.}
  \bibinfo{year}{1986}\natexlab{}.
\newblock \showarticletitle{Backward Ray Tracing}. In
  \bibinfo{booktitle}{\emph{{SIGGRAPH} '86 Developments in Ray Tracing (Course
  Notes)}}. \bibinfo{pages}{259--263}.
\newblock


\bibitem[Arvo and Kirk(1987)]%
        {arvo1987ray}
\bibfield{author}{\bibinfo{person}{James Arvo} {and} \bibinfo{person}{David~B.
  Kirk}.} \bibinfo{year}{1987}\natexlab{}.
\newblock \showarticletitle{Fast Ray Tracing by Ray Classification}. In
  \bibinfo{booktitle}{\emph{Proceedings of the 14th Annual Conference on
  Computer Graphics and Interactive Techniques ({SIGGRAPH})}}.
  \bibinfo{pages}{55--64}.
\newblock
\href{https://doi.org/10.1145/37401.37409}{doi:\nolinkurl{10.1145/37401.37409}}


\bibitem[Denis et~al\mbox{.}(2023)]%
        {denis2023rasterization}
\bibfield{author}{\bibinfo{person}{Leon Denis}, \bibinfo{person}{Remco Royen},
  \bibinfo{person}{Quentin Bols\'{e}e}, \bibinfo{person}{Nicolas Vercheval},
  \bibinfo{person}{Aleksandra Pi{\v{z}}urica}, {and} \bibinfo{person}{Adrian
  Munteanu}.} \bibinfo{year}{2023}\natexlab{}.
\newblock \showarticletitle{{GPU} Rasterization-Based {3D} {LiDAR} Simulation
  for Deep Learning}.
\newblock \bibinfo{journal}{\emph{Sensors}} \bibinfo{volume}{23},
  \bibinfo{number}{19} (\bibinfo{year}{2023}), \bibinfo{pages}{8130}.
\newblock
\href{https://doi.org/10.3390/s23198130}{doi:\nolinkurl{10.3390/s23198130}}


\bibitem[Dosovitskiy et~al\mbox{.}(2017)]%
        {dosovitskiy2017carla}
\bibfield{author}{\bibinfo{person}{Alexey Dosovitskiy}, \bibinfo{person}{German
  Ros}, \bibinfo{person}{Felipe Codevilla}, \bibinfo{person}{Antonio Lopez},
  {and} \bibinfo{person}{Vladlen Koltun}.} \bibinfo{year}{2017}\natexlab{}.
\newblock \showarticletitle{{CARLA}: An Open Urban Driving Simulator}. In
  \bibinfo{booktitle}{\emph{Proceedings of the 1st Annual Conference on Robot
  Learning (CoRL)}}. \bibinfo{pages}{1--16}.
\newblock
\urldef\tempurl%
\url{https://proceedings.mlr.press/v78/dosovitskiy17a.html}
\showURL{%
\tempurl}


\bibitem[Eberly(2025)]%
        {eberly_geometrictools}
\bibfield{author}{\bibinfo{person}{David Eberly}.}
  \bibinfo{year}{2025}\natexlab{}.
\newblock \bibinfo{title}{Geometric Tools}.
\newblock
\urldef\tempurl%
\url{https://www.geometrictools.com}
\showURL{%
\tempurl}
\newblock
\shownote{Technical documentation library for geometric algorithms}.


\bibitem[Ericson(2004)]%
        {ericson2004}
\bibfield{author}{\bibinfo{person}{Christer Ericson}.}
  \bibinfo{year}{2004}\natexlab{}.
\newblock \bibinfo{booktitle}{\emph{Real-Time Collision Detection}}.
\newblock \bibinfo{publisher}{Morgan Kaufmann}.
\newblock
\showISBNx{1-55860-732-3}


\bibitem[Fang et~al\mbox{.}(2020)]%
        {fang2020alsim}
\bibfield{author}{\bibinfo{person}{Jin Fang}, \bibinfo{person}{Dingfu Zhou},
  \bibinfo{person}{Feilong Yan}, \bibinfo{person}{Tongtong Zhao},
  \bibinfo{person}{Feihu Zhang}, \bibinfo{person}{Yu Ma},
  \bibinfo{person}{Liang Wang}, {and} \bibinfo{person}{Ruigang Yang}.}
  \bibinfo{year}{2020}\natexlab{}.
\newblock \showarticletitle{Augmented {LiDAR} Simulator for Autonomous
  Driving}.
\newblock \bibinfo{journal}{\emph{IEEE Robotics and Automation Letters}}
  \bibinfo{volume}{5}, \bibinfo{number}{2} (\bibinfo{year}{2020}),
  \bibinfo{pages}{1931--1938}.
\newblock
\href{https://doi.org/10.1109/LRA.2020.2969927}{doi:\nolinkurl{10.1109/LRA.2020.2969927}}


\bibitem[Goral et~al\mbox{.}(1984)]%
        {goral1984}
\bibfield{author}{\bibinfo{person}{Cindy~M. Goral}, \bibinfo{person}{Kenneth~E.
  Torrance}, \bibinfo{person}{Donald~P. Greenberg}, {and}
  \bibinfo{person}{Bennett Battaile}.} \bibinfo{year}{1984}\natexlab{}.
\newblock \showarticletitle{Modeling the Interaction of Light Between Diffuse
  Surfaces}. In \bibinfo{booktitle}{\emph{Proceedings of the 11th Annual
  Conference on Computer Graphics and Interactive Techniques ({SIGGRAPH})}}.
  \bibinfo{pages}{213--222}.
\newblock
\href{https://doi.org/10.1145/800031.808601}{doi:\nolinkurl{10.1145/800031.808601}}


\bibitem[Guillard et~al\mbox{.}(2022)]%
        {guillard2022lidar}
\bibfield{author}{\bibinfo{person}{Benoit Guillard}, \bibinfo{person}{Sai
  Vemprala}, \bibinfo{person}{Jayesh~K. Gupta}, \bibinfo{person}{Ondrej
  Miksik}, \bibinfo{person}{Vibhav Vineet}, \bibinfo{person}{Pascal Fua}, {and}
  \bibinfo{person}{Ashish Kapoor}.} \bibinfo{year}{2022}\natexlab{}.
\newblock \showarticletitle{Learning to Simulate Realistic {LiDAR}s}. In
  \bibinfo{booktitle}{\emph{2022 IEEE/RSJ International Conference on
  Intelligent Robots and Systems ({IROS})}}. \bibinfo{pages}{8173--8180}.
\newblock
\href{https://doi.org/10.1109/IROS47612.2022.9981120}{doi:\nolinkurl{10.1109/IROS47612.2022.9981120}}


\bibitem[Guthmann(2023)]%
        {guthmann2023occupancy}
\bibfield{author}{\bibinfo{person}{Fran\c{c}ois Guthmann}.}
  \bibinfo{year}{2023}\natexlab{}.
\newblock \bibinfo{title}{Occupancy Explained}.
\newblock
\urldef\tempurl%
\url{https://gpuopen.com/learn/occupancy-explained/}
\showURL{%
\tempurl}
\newblock
\shownote{{AMD} {GPUOpen}}.


\bibitem[Heckbert and Hanrahan(1984)]%
        {heckbert1984}
\bibfield{author}{\bibinfo{person}{Paul~S. Heckbert} {and} \bibinfo{person}{Pat
  Hanrahan}.} \bibinfo{year}{1984}\natexlab{}.
\newblock \showarticletitle{Beam Tracing Polygonal Objects}. In
  \bibinfo{booktitle}{\emph{Proceedings of the 11th Annual Conference on
  Computer Graphics and Interactive Techniques (SIGGRAPH)}}.
  \bibinfo{pages}{119--127}.
\newblock
\href{https://doi.org/10.1145/800031.808588}{doi:\nolinkurl{10.1145/800031.808588}}


\bibitem[Jansen and Steckel(2026)]%
        {jansen2026sonotrace}
\bibfield{author}{\bibinfo{person}{Wouter Jansen} {and} \bibinfo{person}{Jan
  Steckel}.} \bibinfo{year}{2026}\natexlab{}.
\newblock \bibinfo{title}{Hardware-Accelerated Geometrical Simulation of
  Biological and Engineered In-Air Ultrasonic Systems}.
\newblock \bibinfo{howpublished}{arXiv:2602.19652}.
\newblock
\href{https://doi.org/10.48550/arXiv.2602.19652}{doi:\nolinkurl{10.48550/arXiv.2602.19652}}


\bibitem[Jensen(1996)]%
        {jensen1996photon}
\bibfield{author}{\bibinfo{person}{Henrik~Wann Jensen}.}
  \bibinfo{year}{1996}\natexlab{}.
\newblock \showarticletitle{Global Illumination Using Photon Maps}. In
  \bibinfo{booktitle}{\emph{Proceedings of the Eurographics Workshop on
  Rendering}}. \bibinfo{pages}{21--30}.
\newblock
\href{https://doi.org/10.1007/978-3-7091-7484-5_3}{doi:\nolinkurl{10.1007/978-3-7091-7484-5_3}}


\bibitem[Johnson and Cohen(2001)]%
        {johnson2001snch}
\bibfield{author}{\bibinfo{person}{David~E. Johnson} {and}
  \bibinfo{person}{Elaine Cohen}.} \bibinfo{year}{2001}\natexlab{}.
\newblock \showarticletitle{Spatialized Normal Cone Hierarchies}. In
  \bibinfo{booktitle}{\emph{Proceedings of the 2001 Symposium on Interactive 3D
  Graphics (I3D)}}. \bibinfo{pages}{129--134}.
\newblock
\href{https://doi.org/10.1145/364338.364380}{doi:\nolinkurl{10.1145/364338.364380}}


\bibitem[Karras(2012)]%
        {karras2012bvh}
\bibfield{author}{\bibinfo{person}{Tero Karras}.}
  \bibinfo{year}{2012}\natexlab{}.
\newblock \showarticletitle{Maximizing Parallelism in the Construction of
  {BVH}s, Octrees, and $k$-d Trees}. In \bibinfo{booktitle}{\emph{Proceedings
  of the Fourth ACM SIGGRAPH / Eurographics Conference on High-Performance
  Graphics}}. \bibinfo{pages}{33--37}.
\newblock
\href{https://doi.org/10.2312/EGGH/HPG12/033-037}{doi:\nolinkurl{10.2312/EGGH/HPG12/033-037}}


\bibitem[Keller(1997)]%
        {keller1997instant}
\bibfield{author}{\bibinfo{person}{Alexander Keller}.}
  \bibinfo{year}{1997}\natexlab{}.
\newblock \showarticletitle{Instant Radiosity}. In
  \bibinfo{booktitle}{\emph{Proceedings of the 24th Annual Conference on
  Computer Graphics and Interactive Techniques ({SIGGRAPH})}}.
  \bibinfo{pages}{49--56}.
\newblock
\href{https://doi.org/10.1145/258734.258769}{doi:\nolinkurl{10.1145/258734.258769}}


\bibitem[Kim et~al\mbox{.}(2023)]%
        {kim2023dopplertof}
\bibfield{author}{\bibinfo{person}{Juhyeon Kim}, \bibinfo{person}{Wojciech
  Jarosz}, \bibinfo{person}{Ioannis Gkioulekas}, {and} \bibinfo{person}{Adithya
  Pediredla}.} \bibinfo{year}{2023}\natexlab{}.
\newblock \showarticletitle{Doppler Time-of-Flight Rendering}.
\newblock \bibinfo{journal}{\emph{ACM Transactions on Graphics}}
  \bibinfo{volume}{42}, \bibinfo{number}{6} (\bibinfo{year}{2023}),
  \bibinfo{pages}{271:1--271:18}.
\newblock
\href{https://doi.org/10.1145/3618335}{doi:\nolinkurl{10.1145/3618335}}


\bibitem[Koenig and Howard(2004)]%
        {koenig2004gazebo}
\bibfield{author}{\bibinfo{person}{Nathan Koenig} {and} \bibinfo{person}{Andrew
  Howard}.} \bibinfo{year}{2004}\natexlab{}.
\newblock \showarticletitle{Design and Use Paradigms for {Gazebo}, an
  Open-Source Multi-Robot Simulator}. In \bibinfo{booktitle}{\emph{Proceedings
  of the IEEE/RSJ International Conference on Intelligent Robots and Systems
  (IROS)}}. \bibinfo{pages}{2149--2154}.
\newblock
\href{https://doi.org/10.1109/IROS.2004.1389727}{doi:\nolinkurl{10.1109/IROS.2004.1389727}}


\bibitem[Kopta et~al\mbox{.}(2012)]%
        {kopta2012bvhrefit}
\bibfield{author}{\bibinfo{person}{Daniel Kopta}, \bibinfo{person}{Thiago Ize},
  \bibinfo{person}{Josef Spjut}, \bibinfo{person}{Erik Brunvand},
  \bibinfo{person}{Al Davis}, {and} \bibinfo{person}{Andrew Kensler}.}
  \bibinfo{year}{2012}\natexlab{}.
\newblock \showarticletitle{Fast, Effective {BVH} Updates for Animated Scenes}.
  In \bibinfo{booktitle}{\emph{Proceedings of the ACM SIGGRAPH Symposium on
  Interactive 3D Graphics and Games ({I3D})}}. \bibinfo{pages}{197--204}.
\newblock
\href{https://doi.org/10.1145/2159616.2159649}{doi:\nolinkurl{10.1145/2159616.2159649}}


\bibitem[Lafortune and Willems(1993)]%
        {lafortune1993}
\bibfield{author}{\bibinfo{person}{Eric~P. Lafortune} {and}
  \bibinfo{person}{Yves~D. Willems}.} \bibinfo{year}{1993}\natexlab{}.
\newblock \showarticletitle{Bi-Directional Path Tracing}. In
  \bibinfo{booktitle}{\emph{Proceedings of {CompuGraphics}}}.
  \bibinfo{pages}{145--153}.
\newblock


\bibitem[Laine et~al\mbox{.}(2013)]%
        {laine2013megakernels}
\bibfield{author}{\bibinfo{person}{Samuli Laine}, \bibinfo{person}{Tero
  Karras}, {and} \bibinfo{person}{Timo Aila}.} \bibinfo{year}{2013}\natexlab{}.
\newblock \showarticletitle{Megakernels Considered Harmful: Wavefront Path
  Tracing on {GPU}s}. In \bibinfo{booktitle}{\emph{Proceedings of the Fifth ACM
  SIGGRAPH / Eurographics Conference on High-Performance Graphics}}.
  \bibinfo{pages}{137--144}.
\newblock
\href{https://doi.org/10.1145/2492045.2492060}{doi:\nolinkurl{10.1145/2492045.2492060}}


\bibitem[L\'{o}pez~Ruiz et~al\mbox{.}(2022)]%
        {lopez2022lidar}
\bibfield{author}{\bibinfo{person}{Alfonso L\'{o}pez~Ruiz},
  \bibinfo{person}{Carlos Og\'{a}yar}, \bibinfo{person}{Juan~M. Jurado}, {and}
  \bibinfo{person}{Francisco Feito}.} \bibinfo{year}{2022}\natexlab{}.
\newblock \showarticletitle{A {GPU}-accelerated framework for simulating
  {LiDAR} scanning}.
\newblock \bibinfo{journal}{\emph{IEEE Transactions on Geoscience and Remote
  Sensing}}  \bibinfo{volume}{60} (\bibinfo{year}{2022}),
  \bibinfo{pages}{1--18}.
\newblock
\href{https://doi.org/10.1109/TGRS.2022.3165746}{doi:\nolinkurl{10.1109/TGRS.2022.3165746}}


\bibitem[Manivasagam et~al\mbox{.}(2020)]%
        {manivasagam2020lidarsim}
\bibfield{author}{\bibinfo{person}{Sivabalan Manivasagam},
  \bibinfo{person}{Shenlong Wang}, \bibinfo{person}{Kelvin Wong},
  \bibinfo{person}{Wenyuan Zeng}, \bibinfo{person}{Mikita Sazanovich},
  \bibinfo{person}{Shuhan Tan}, \bibinfo{person}{Bin Yang},
  \bibinfo{person}{Wei-Chiu Ma}, {and} \bibinfo{person}{Raquel Urtasun}.}
  \bibinfo{year}{2020}\natexlab{}.
\newblock \showarticletitle{{LiDARsim}: Realistic {LiDAR} Simulation by
  Leveraging the Real World}. In \bibinfo{booktitle}{\emph{Proceedings of the
  IEEE/CVF Conference on Computer Vision and Pattern Recognition (CVPR)}}.
  \bibinfo{pages}{11167--11176}.
\newblock
\href{https://doi.org/10.1109/CVPR42600.2020.01119}{doi:\nolinkurl{10.1109/CVPR42600.2020.01119}}


\bibitem[McGuire(2017)]%
        {mcguire2017}
\bibfield{author}{\bibinfo{person}{Morgan McGuire}.}
  \bibinfo{year}{2017}\natexlab{}.
\newblock \bibinfo{title}{Computer Graphics Archive}.
\newblock
\urldef\tempurl%
\url{https://casual-effects.com/data}
\showURL{%
\tempurl}


\bibitem[Mock et~al\mbox{.}(2025)]%
        {mock2025radarays}
\bibfield{author}{\bibinfo{person}{Alexander Mock}, \bibinfo{person}{Martin
  Magnusson}, {and} \bibinfo{person}{Joachim Hertzberg}.}
  \bibinfo{year}{2025}\natexlab{}.
\newblock \showarticletitle{{RadaRays}: Real-Time Simulation of Rotating {FMCW}
  Radar for Mobile Robotics via Hardware-Accelerated Ray Tracing}.
\newblock \bibinfo{journal}{\emph{IEEE Robotics and Automation Letters}}
  \bibinfo{volume}{10}, \bibinfo{number}{3} (\bibinfo{year}{2025}),
  \bibinfo{pages}{2470--2477}.
\newblock
\href{https://doi.org/10.1109/LRA.2025.3531689}{doi:\nolinkurl{10.1109/LRA.2025.3531689}}


\bibitem[M\"{o}ller and Trumbore(1997)]%
        {moller1997}
\bibfield{author}{\bibinfo{person}{Tomas M\"{o}ller} {and} \bibinfo{person}{Ben
  Trumbore}.} \bibinfo{year}{1997}\natexlab{}.
\newblock \showarticletitle{Fast, Minimum Storage Ray-Triangle Intersection}.
\newblock \bibinfo{journal}{\emph{Journal of Graphics Tools}}
  \bibinfo{volume}{2}, \bibinfo{number}{1} (\bibinfo{year}{1997}),
  \bibinfo{pages}{21--28}.
\newblock
\href{https://doi.org/10.1080/10867651.1997.10487468}{doi:\nolinkurl{10.1080/10867651.1997.10487468}}


\bibitem[{NVIDIA}(2017)]%
        {nvidia_orca2017}
\bibfield{author}{\bibinfo{person}{{NVIDIA}}.} \bibinfo{year}{2017}\natexlab{}.
\newblock \bibinfo{title}{{ORCA}: Open Research Content Archive}.
\newblock
\urldef\tempurl%
\url{https://developer.nvidia.com/orca}
\showURL{%
\tempurl}


\bibitem[{NVIDIA}(2023)]%
        {isaacsim}
\bibfield{author}{\bibinfo{person}{{NVIDIA}}.} \bibinfo{year}{2023}\natexlab{}.
\newblock \bibinfo{title}{{Isaac Sim}: Robotics Simulation and Synthetic Data
  Generation}.
\newblock \bibinfo{howpublished}{\url{https://developer.nvidia.com/isaac/sim}}.
\newblock


\bibitem[Parker et~al\mbox{.}(2010)]%
        {parker2010optix}
\bibfield{author}{\bibinfo{person}{Steven~G. Parker}, \bibinfo{person}{James
  Bigler}, \bibinfo{person}{Andreas Dietrich}, \bibinfo{person}{Heiko
  Friedrich}, \bibinfo{person}{Jared Hoberock}, \bibinfo{person}{David Luebke},
  \bibinfo{person}{David McAllister}, \bibinfo{person}{Morgan McGuire},
  \bibinfo{person}{Keith Morley}, \bibinfo{person}{Austin Robison}, {and}
  \bibinfo{person}{Martin Stich}.} \bibinfo{year}{2010}\natexlab{}.
\newblock \showarticletitle{{OptiX}: A General Purpose Ray Tracing Engine}.
\newblock \bibinfo{journal}{\emph{ACM Transactions on Graphics}}
  \bibinfo{volume}{29}, \bibinfo{number}{4} (\bibinfo{year}{2010}),
  \bibinfo{pages}{66:1--66:13}.
\newblock
\href{https://doi.org/10.1145/1778765.1778803}{doi:\nolinkurl{10.1145/1778765.1778803}}


\bibitem[{Robotec.AI}(2023)]%
        {robotecgpulidar}
\bibfield{author}{\bibinfo{person}{{Robotec.AI}}.}
  \bibinfo{year}{2023}\natexlab{}.
\newblock \bibinfo{title}{{RobotecGPULidar}: GPU-accelerated LiDAR simulation
  using {OptiX}}.
\newblock
  \bibinfo{howpublished}{\url{https://github.com/RobotecAI/RobotecGPULidar}}.
\newblock


\bibitem[Rong et~al\mbox{.}(2020)]%
        {rong2020lgsvl}
\bibfield{author}{\bibinfo{person}{Guodong Rong}, \bibinfo{person}{Byung~Hyun
  Shin}, \bibinfo{person}{Hadi Tabatabaee}, \bibinfo{person}{Qiang Lu},
  \bibinfo{person}{Steve Lemke}, \bibinfo{person}{Ma{\v{r}}a
  Moze{\v{n}}i\v{c}}, \bibinfo{person}{Eric Li}, \bibinfo{person}{Taylor
  Sprinkle}, {and} \bibinfo{person}{Mani Ramanagopal}.}
  \bibinfo{year}{2020}\natexlab{}.
\newblock \showarticletitle{{LGSVL} Simulator: A High Fidelity Simulator for
  Autonomous Driving}. In \bibinfo{booktitle}{\emph{2020 IEEE 23rd
  International Conference on Intelligent Transportation Systems ({ITSC})}}.
  \bibinfo{pages}{1--6}.
\newblock
\href{https://doi.org/10.1109/ITSC45102.2020.9294422}{doi:\nolinkurl{10.1109/ITSC45102.2020.9294422}}


\bibitem[Sch\"{u}tz et~al\mbox{.}(2026)]%
        {schuetz2026curast}
\bibfield{author}{\bibinfo{person}{Markus Sch\"{u}tz}, \bibinfo{person}{Lukas
  Lipp}, \bibinfo{person}{Elias Kristmann}, {and} \bibinfo{person}{Michael
  Wimmer}.} \bibinfo{year}{2026}\natexlab{}.
\newblock \bibinfo{title}{{CuRast}: {CUDA}-Based Software Rasterization for
  Billions of Triangles}.
\newblock \bibinfo{howpublished}{arXiv:2604.21749}.
\newblock
\href{https://doi.org/10.48550/arXiv.2604.21749}{doi:\nolinkurl{10.48550/arXiv.2604.21749}}


\bibitem[Shah et~al\mbox{.}(2017)]%
        {shah2017airsim}
\bibfield{author}{\bibinfo{person}{Shital Shah}, \bibinfo{person}{Debadeepta
  Dey}, \bibinfo{person}{Chris Lovett}, {and} \bibinfo{person}{Ashish Kapoor}.}
  \bibinfo{year}{2017}\natexlab{}.
\newblock \showarticletitle{{AirSim}: High-Fidelity Visual and Physical
  Simulation for Autonomous Vehicles}. In \bibinfo{booktitle}{\emph{Field and
  Service Robotics ({FSR})}}. \bibinfo{pages}{621--635}.
\newblock
\href{https://doi.org/10.1007/978-3-319-67361-5_40}{doi:\nolinkurl{10.1007/978-3-319-67361-5_40}}


\bibitem[{TIER IV}(2023)]%
        {awsim}
\bibfield{author}{\bibinfo{person}{{TIER IV}}.}
  \bibinfo{year}{2023}\natexlab{}.
\newblock \bibinfo{title}{{AWSIM}: Open Source Autonomous Driving Simulator}.
\newblock \bibinfo{howpublished}{\url{https://github.com/tier4/AWSIM}}.
\newblock


\bibitem[T{\'o}th et~al\mbox{.}(2025)]%
        {toth2025hybrid}
\bibfield{author}{\bibinfo{person}{M{\'a}t{\'e} T{\'o}th},
  \bibinfo{person}{P{\'e}ter Kov{\'a}cs}, \bibinfo{person}{R{\'e}ka Bencses},
  \bibinfo{person}{Bal{\'a}zs Ter{\'e}ki}, \bibinfo{person}{Zolt{\'a}n
  Bendefy}, \bibinfo{person}{Zolt{\'a}n Hortsin}, {and}
  \bibinfo{person}{Tam{\'a}s Matuszka}.} \bibinfo{year}{2025}\natexlab{}.
\newblock \bibinfo{title}{Hybrid Rendering for Multimodal Autonomous Driving:
  Merging Neural and Physics-Based Simulation}.
\newblock \bibinfo{howpublished}{arXiv:2503.09464}.
\newblock
\href{https://doi.org/10.48550/arXiv.2503.09464}{doi:\nolinkurl{10.48550/arXiv.2503.09464}}


\bibitem[Veach(1997)]%
        {veach1997}
\bibfield{author}{\bibinfo{person}{Eric Veach}.}
  \bibinfo{year}{1997}\natexlab{}.
\newblock \emph{\bibinfo{title}{Robust Monte Carlo Methods for Light Transport
  Simulation}}.
\newblock \bibinfo{thesistype}{Ph.\,D. Dissertation}. \bibinfo{school}{Stanford
  University}.
\newblock
\href{https://doi.org/10.5555/927297}{doi:\nolinkurl{10.5555/927297}}


\bibitem[Wald(2007)]%
        {wald2007bvh}
\bibfield{author}{\bibinfo{person}{Ingo Wald}.}
  \bibinfo{year}{2007}\natexlab{}.
\newblock \showarticletitle{On fast Construction of {SAH}-based Bounding Volume
  Hierarchies}. In \bibinfo{booktitle}{\emph{IEEE Symposium on Interactive Ray
  Tracing}}. \bibinfo{pages}{33--40}.
\newblock
\href{https://doi.org/10.1109/RT.2007.4342588}{doi:\nolinkurl{10.1109/RT.2007.4342588}}


\bibitem[Wald et~al\mbox{.}(2007)]%
        {wald2007dynamic}
\bibfield{author}{\bibinfo{person}{Ingo Wald}, \bibinfo{person}{Solomon
  Boulos}, {and} \bibinfo{person}{Peter Shirley}.}
  \bibinfo{year}{2007}\natexlab{}.
\newblock \showarticletitle{Ray Tracing Deformable Scenes Using Dynamic
  Bounding Volume Hierarchies}.
\newblock \bibinfo{journal}{\emph{ACM Transactions on Graphics}}
  \bibinfo{volume}{26}, \bibinfo{number}{1} (\bibinfo{year}{2007}),
  \bibinfo{pages}{6:1--6:18}.
\newblock
\href{https://doi.org/10.1145/1189762.1206075}{doi:\nolinkurl{10.1145/1189762.1206075}}
\newblock
\shownote{Also SCI Technical Report UUSCI-2006-023, University of Utah}.


\bibitem[Wald et~al\mbox{.}(2014)]%
        {wald2014embree}
\bibfield{author}{\bibinfo{person}{Ingo Wald}, \bibinfo{person}{Sven Woop},
  \bibinfo{person}{Carsten Benthin}, \bibinfo{person}{Gregory~S. Johnson},
  {and} \bibinfo{person}{Manfred Ernst}.} \bibinfo{year}{2014}\natexlab{}.
\newblock \showarticletitle{Embree: A Kernel Framework for Efficient {CPU} Ray
  Tracing}. In \bibinfo{booktitle}{\emph{ACM SIGGRAPH 2014 Papers}}.
  \bibinfo{pages}{143:1--143:8}.
\newblock
\href{https://doi.org/10.1145/2601097.2601199}{doi:\nolinkurl{10.1145/2601097.2601199}}


\bibitem[Walter et~al\mbox{.}(2005)]%
        {walter2005lightcuts}
\bibfield{author}{\bibinfo{person}{Bruce Walter}, \bibinfo{person}{Sebastian
  Fernandez}, \bibinfo{person}{Adam Arbree}, \bibinfo{person}{Kavita Bala},
  \bibinfo{person}{Michael Donikian}, {and} \bibinfo{person}{Donald~P.
  Greenberg}.} \bibinfo{year}{2005}\natexlab{}.
\newblock \showarticletitle{Lightcuts: A Scalable Approach to Illumination}. In
  \bibinfo{booktitle}{\emph{ACM SIGGRAPH 2005 Papers}}.
  \bibinfo{pages}{1098--1107}.
\newblock
\href{https://doi.org/10.1145/1186822.1073318}{doi:\nolinkurl{10.1145/1186822.1073318}}


\bibitem[Whitted(1980)]%
        {whitted1980}
\bibfield{author}{\bibinfo{person}{Turner Whitted}.}
  \bibinfo{year}{1980}\natexlab{}.
\newblock \showarticletitle{An Improved Illumination Model for Shaded Display}.
\newblock \bibinfo{journal}{\emph{Commun. ACM}} \bibinfo{volume}{23},
  \bibinfo{number}{6} (\bibinfo{year}{1980}), \bibinfo{pages}{343--349}.
\newblock
\href{https://doi.org/10.1145/358876.358882}{doi:\nolinkurl{10.1145/358876.358882}}


\bibitem[Wu et~al\mbox{.}(2024)]%
        {wu2024dynlidar}
\bibfield{author}{\bibinfo{person}{Hanfeng Wu}, \bibinfo{person}{Xingxing Zuo},
  \bibinfo{person}{Stefan Leutenegger}, \bibinfo{person}{Or Litany},
  \bibinfo{person}{Konrad Schindler}, {and} \bibinfo{person}{Shengyu Huang}.}
  \bibinfo{year}{2024}\natexlab{}.
\newblock \showarticletitle{Dynamic {LiDAR} Re-simulation Using Compositional
  Neural Fields}. In \bibinfo{booktitle}{\emph{Proceedings of the IEEE/CVF
  Conference on Computer Vision and Pattern Recognition ({CVPR})}}.
  \bibinfo{pages}{19988--19998}.
\newblock
\href{https://doi.org/10.1109/CVPR52733.2024.01889}{doi:\nolinkurl{10.1109/CVPR52733.2024.01889}}


\bibitem[Yang et~al\mbox{.}(2023)]%
        {yang2023unisim}
\bibfield{author}{\bibinfo{person}{Ze Yang}, \bibinfo{person}{Yun Chen},
  \bibinfo{person}{Jingkang Wang}, \bibinfo{person}{Sivabalan Manivasagam},
  \bibinfo{person}{Wei-Chiu Ma}, \bibinfo{person}{Anqi~Joyce Yang}, {and}
  \bibinfo{person}{Raquel Urtasun}.} \bibinfo{year}{2023}\natexlab{}.
\newblock \showarticletitle{{UniSim}: A Neural Closed-Loop Sensor Simulator}.
  In \bibinfo{booktitle}{\emph{Proceedings of the IEEE/CVF Conference on
  Computer Vision and Pattern Recognition ({CVPR})}}.
  \bibinfo{pages}{1389--1399}.
\newblock
\href{https://doi.org/10.1109/CVPR52729.2023.00140}{doi:\nolinkurl{10.1109/CVPR52729.2023.00140}}


\end{thebibliography}

%% --- Optional figure pages (do not count against page limit) ---
%% Order matches first-citation order in the text.
\clearpage
%% Figure-page width scale: 1 by default; a wrapper may \newcommand{\figpagescale}{..} before \input{body}
\providecommand{\figpagescale}{1}
\newlength{\figpagew}\setlength{\figpagew}{\figpagescale\columnwidth}
%% Each figure is defined once; \setcounter keeps the numbers in first-citation order (7-17).
%% Target layout (as in the acmtog review build): page 1: 9, 10 | 11, 12; page 2: 16, 17, 13, 15 | 14, 7, 8.

\newcommand{\FigPageDeformation}{\setcounter{figure}{6}%
\begin{figure}[htb]
  \centering
  %% 2x2 collage on a black background (black gaps). Rendered with GRCA-CUDA by
  %% GRCA/presets/benchmark_config_vis_deformation_{preview,modes}.json (ND/OBD/SWD at frames 1270/175/269).
  {\setlength{\fboxsep}{0pt}\setlength{\lineskip}{0pt}%
   \colorbox{black}{\begin{minipage}{\figpagew}%
     \includegraphics[width=0.497\figpagew]{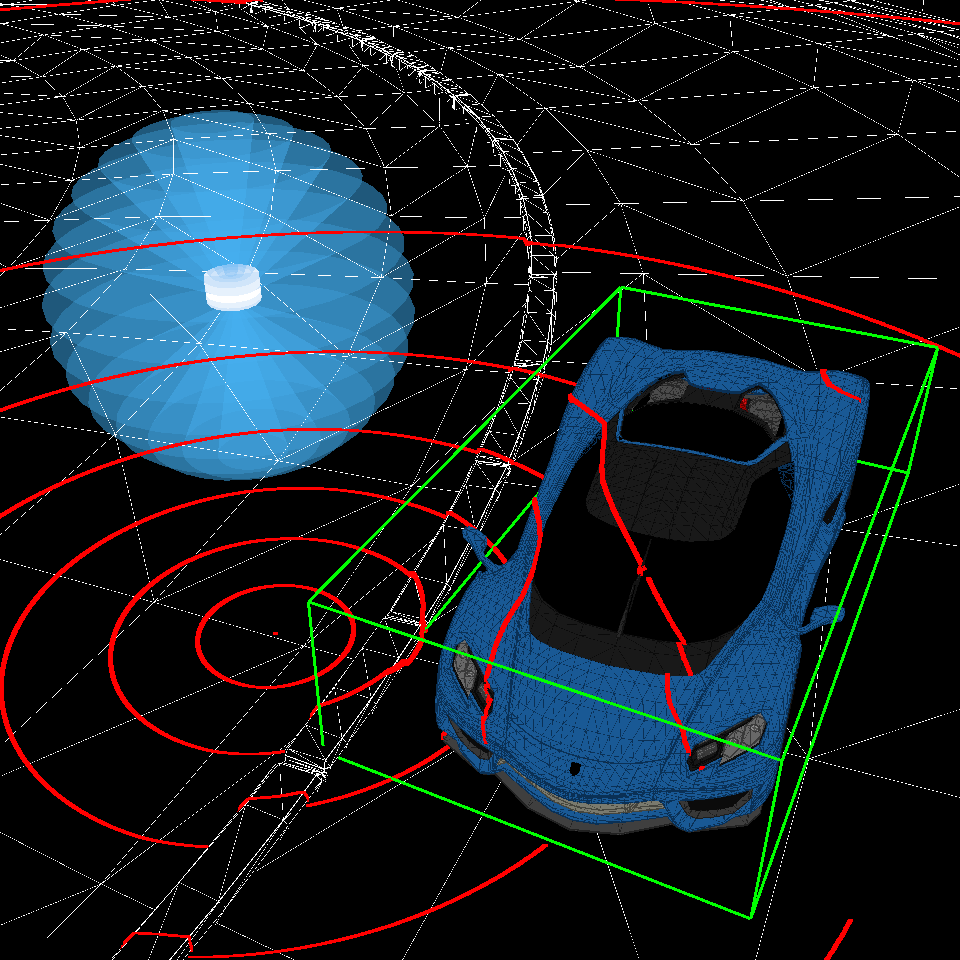}\hfill
     \includegraphics[width=0.497\figpagew]{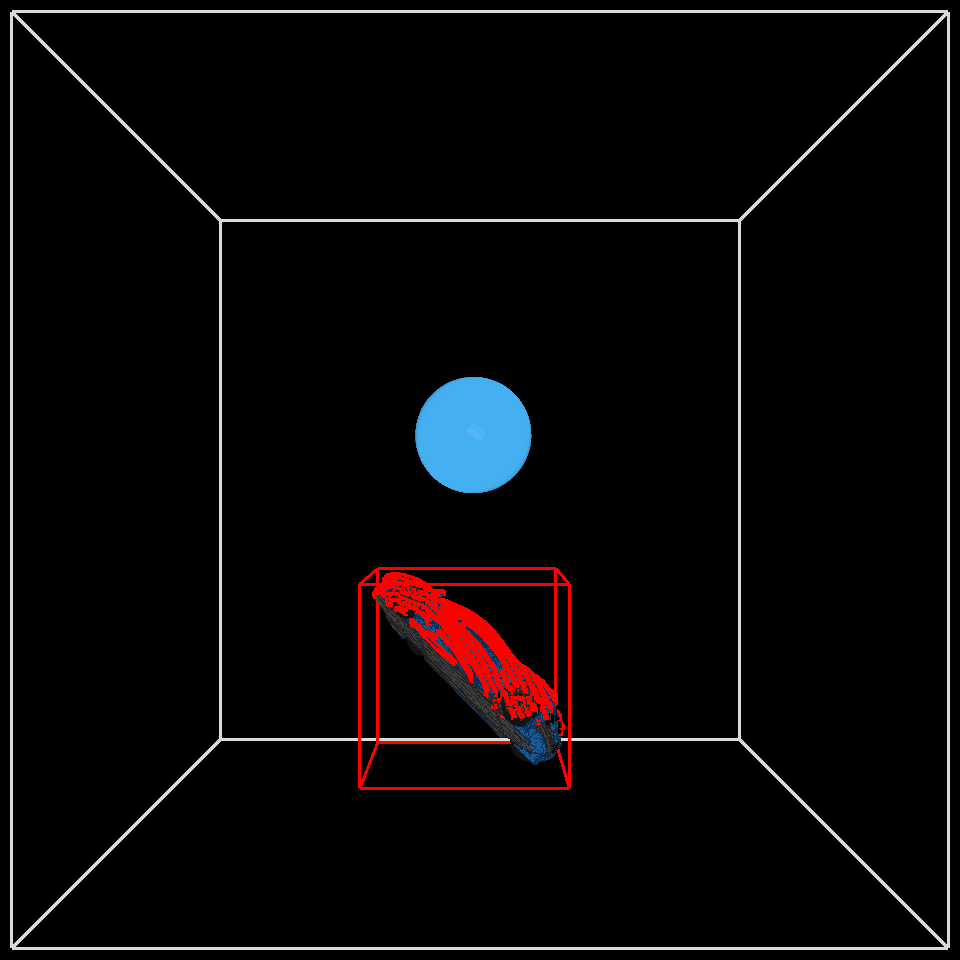}\\[0.006\figpagew]
     \includegraphics[width=0.497\figpagew]{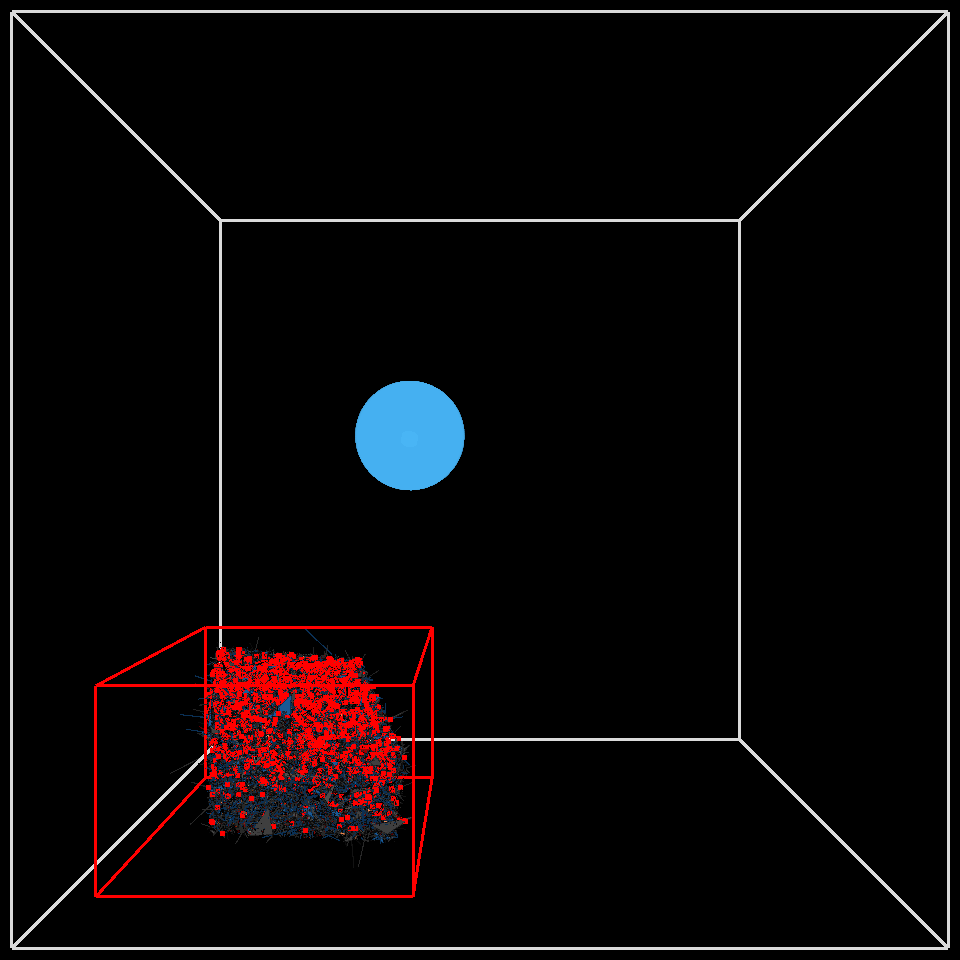}\hfill
     \includegraphics[width=0.497\figpagew]{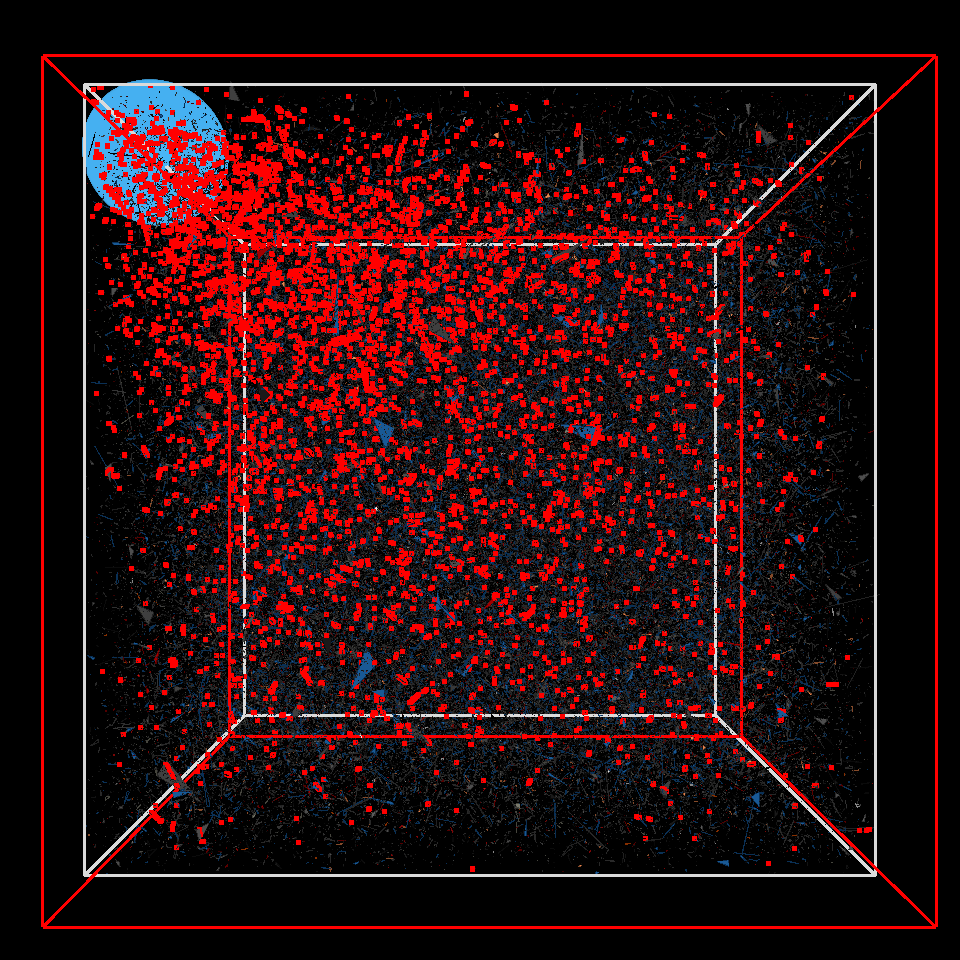}%
   \end{minipage}}}
  \caption{Top left: LiDAR emitter cones (blue) and the sportscar (green box) on a Bistro street. Top right: ND; bottom: OBD (left), SWD (right); LiDAR and car poses random per frame. Red = LiDAR hits.}
  \Description{Four-panel figure: a LiDAR sensor with blue emitter cones next to a sportscar on a street drawn as white triangle edges, then the randomly posed car under no deformation, under object-bound deformation with its triangles scattered inside its bounding box, and under scene-wide deformation with triangles scattered across the scene; LiDAR hits are shown in red.}
  \label{fig:deformation_modes}
\end{figure}}

\newcommand{\FigPageTriArea}{\setcounter{figure}{7}%
\begin{figure}[htb]
  \centering
  \includegraphics[width=\figpagew]{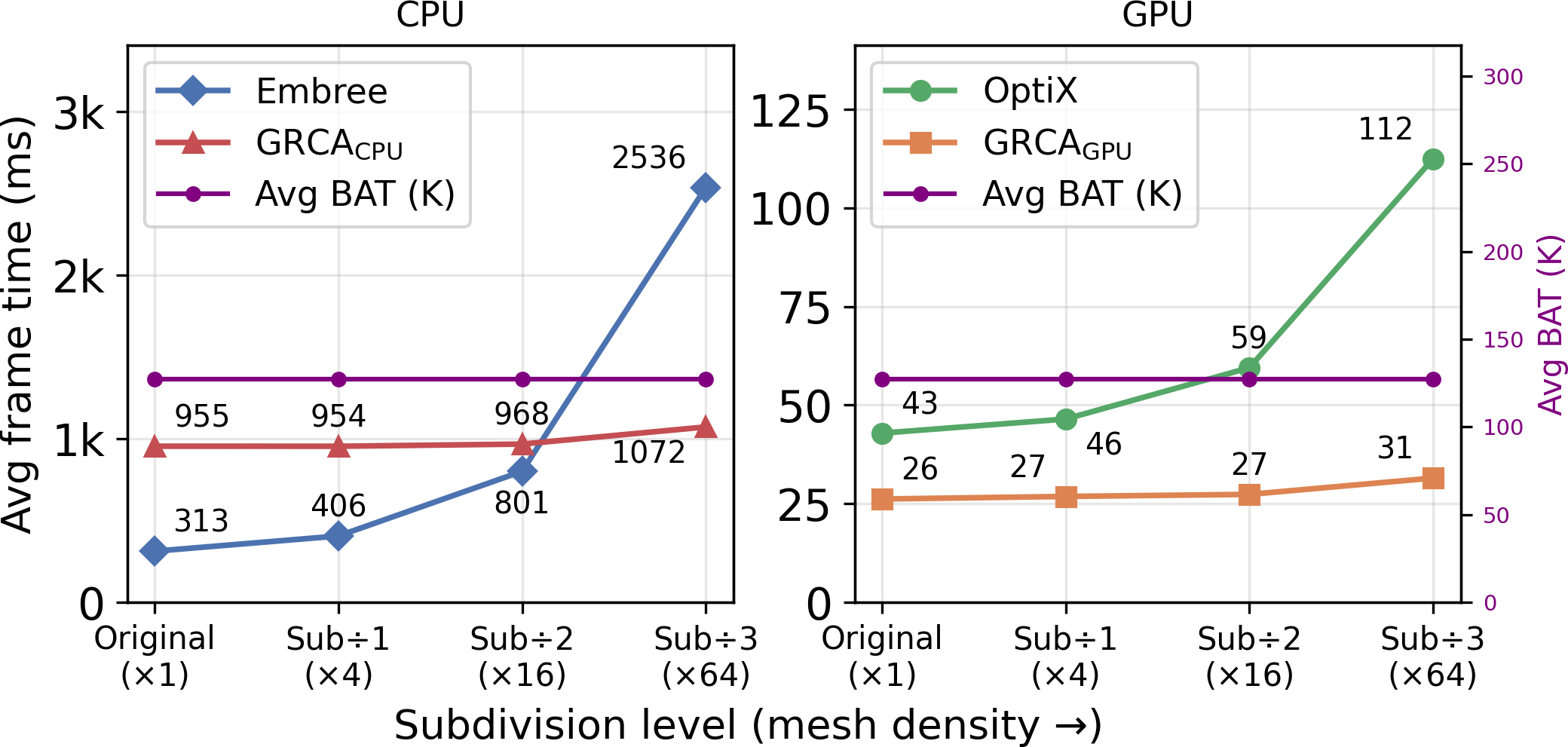}
  \caption{Time vs subdivision level (1 dyn, scale\,1.0, ND, PC1).
           Left: CPU. Right: GPU. Purple line = avg BAT count per frame (right axis).}
  \Description{Two-panel figure: left shows CPU frame time vs subdivision level with avg BAT count right-axis overlay; right shows GPU frame time vs subdivision level with avg BAT count right-axis overlay.}
  \label{fig:tri_area_culling}
\end{figure}}

\newcommand{\FigPageScaleVsBVH}{\setcounter{figure}{8}%
\begin{figure}[htb]
  \centering
  \includegraphics[width=\figpagew]{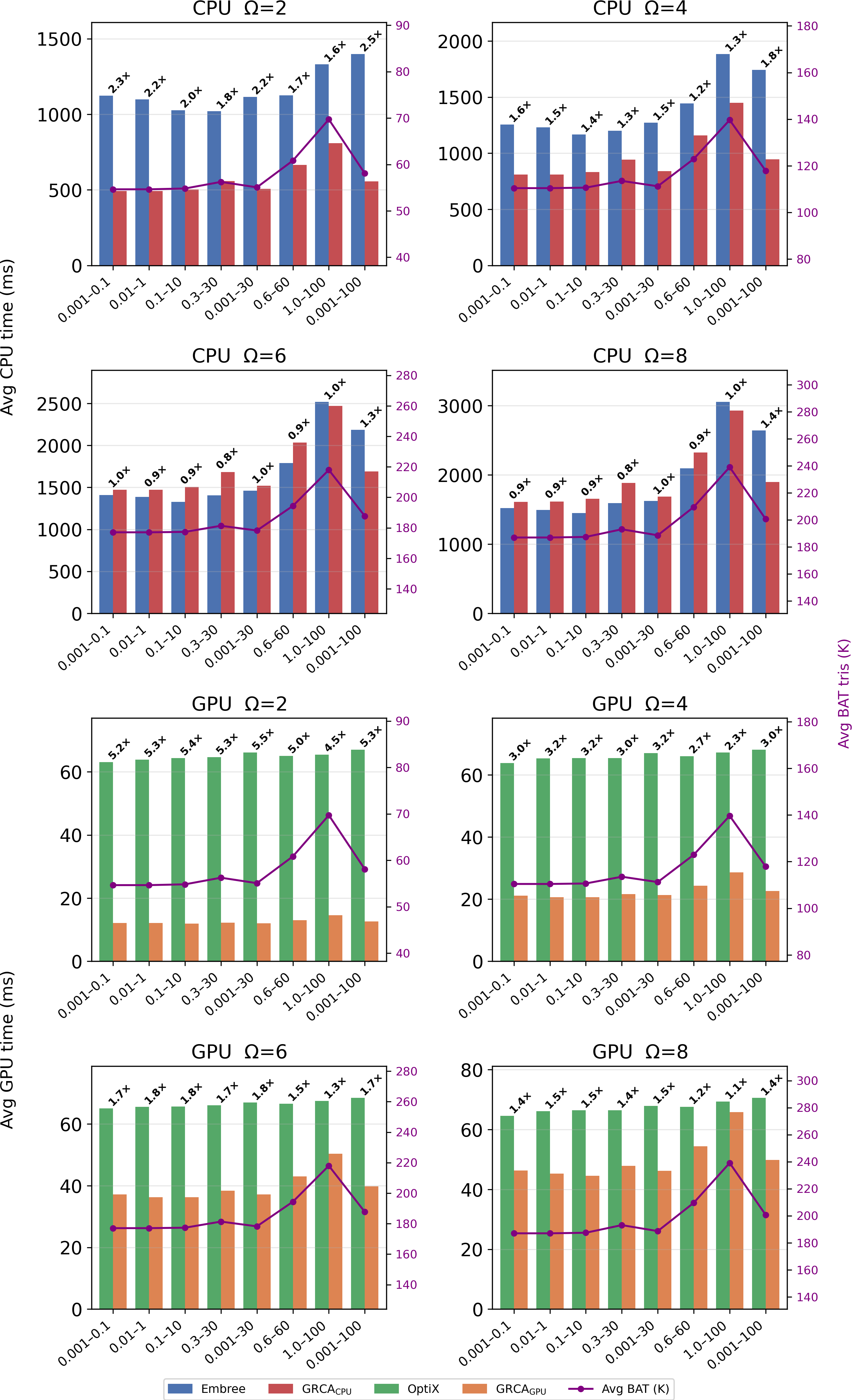}
  \caption{Time across 8 mesh scale ranges (ND, Power~Plant, avg 10/20/30~dyn, PC1).
           Top: CPU. Bottom: GPU. Purple line = avg BAT count per frame (right axis). Speedup labels above each \grca{} bar.}
  \Description{Bar charts comparing \grcacpu{} vs Embree (top) and \grcagpu{} vs OptiX (bottom) across eight mesh scale ranges with BAT count overlay.}
  \label{fig:scale_vs_bvh}
\end{figure}}

\newcommand{\FigPageScaleBAT}{\setcounter{figure}{9}%
\begin{figure}[htb]
  \centering
  \includegraphics[width=\figpagew]{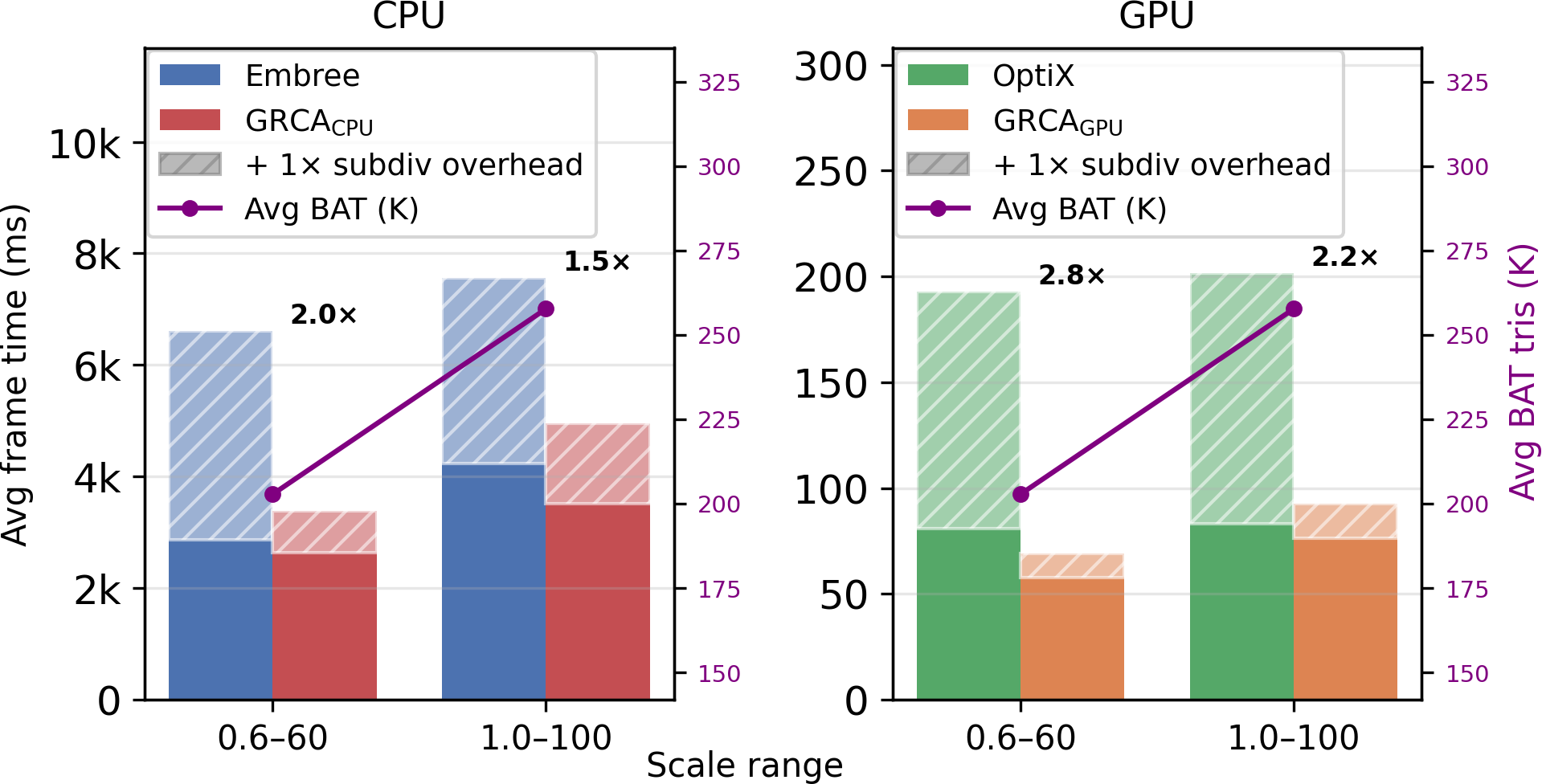}
  \caption{Joint scale $\times$ density: $1\times$ subdiv sportscar ($4\times$ tris, 48.8\,M total), 30~dyn, $\Omega{=}8$, ND, PC1.
           Left: CPU. Right: GPU. Purple line = avg BAT count per frame (right axis).}
  \Description{Side-by-side bar charts comparing \grcacpu{} vs Embree (CPU) and
               \grcagpu{} vs OptiX (GPU) at scale ranges 0.6--60 and 1--100 with
               1x subdivided mesh, with BAT count overlay.}
  \label{fig:scale_bat_subdiv}
\end{figure}}

\newcommand{\FigPageRange}{\setcounter{figure}{10}%
\begin{figure}[htb]
  \centering
  \includegraphics[width=\figpagew]{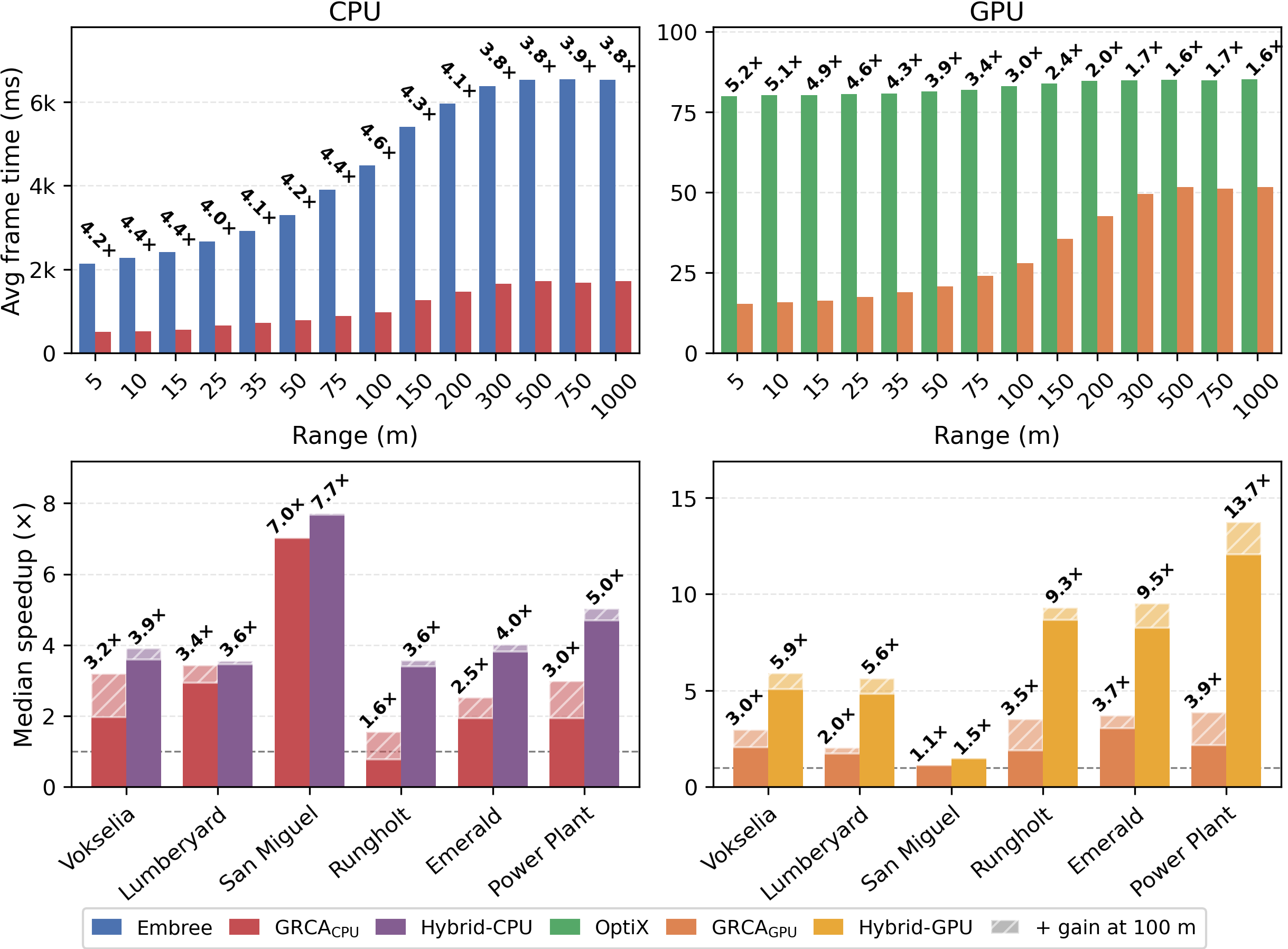}
  \caption{Range culling, PC1. Top: Power~Plant, 30~dyn, $\Omega{=}8$. Bottom: per-scene median speedup
           vs Embree/OptiX at equal range, solid = 1000\,m.}
  \Description{Four-panel figure, CPU left and GPU right. Top row: frame time vs LiDAR range for Embree and \grcacpu{}
               (CPU) and OptiX and \grcagpu{} (GPU) in Power Plant, with speedup labels. Bottom row: for each of the six
               scenes, median speedup of standalone \grca{} and of the hybrid pipeline at 1000\,m, with the extra gain at
               100\,m stacked on top; standalone gains most in the large scenes, San Miguel does not change.}
  \label{fig:range_variation}
\end{figure}}

\newcommand{\FigPageCost}{\setcounter{figure}{11}%
\begin{figure}[htb]
  \centering
  \includegraphics[width=\figpagew]{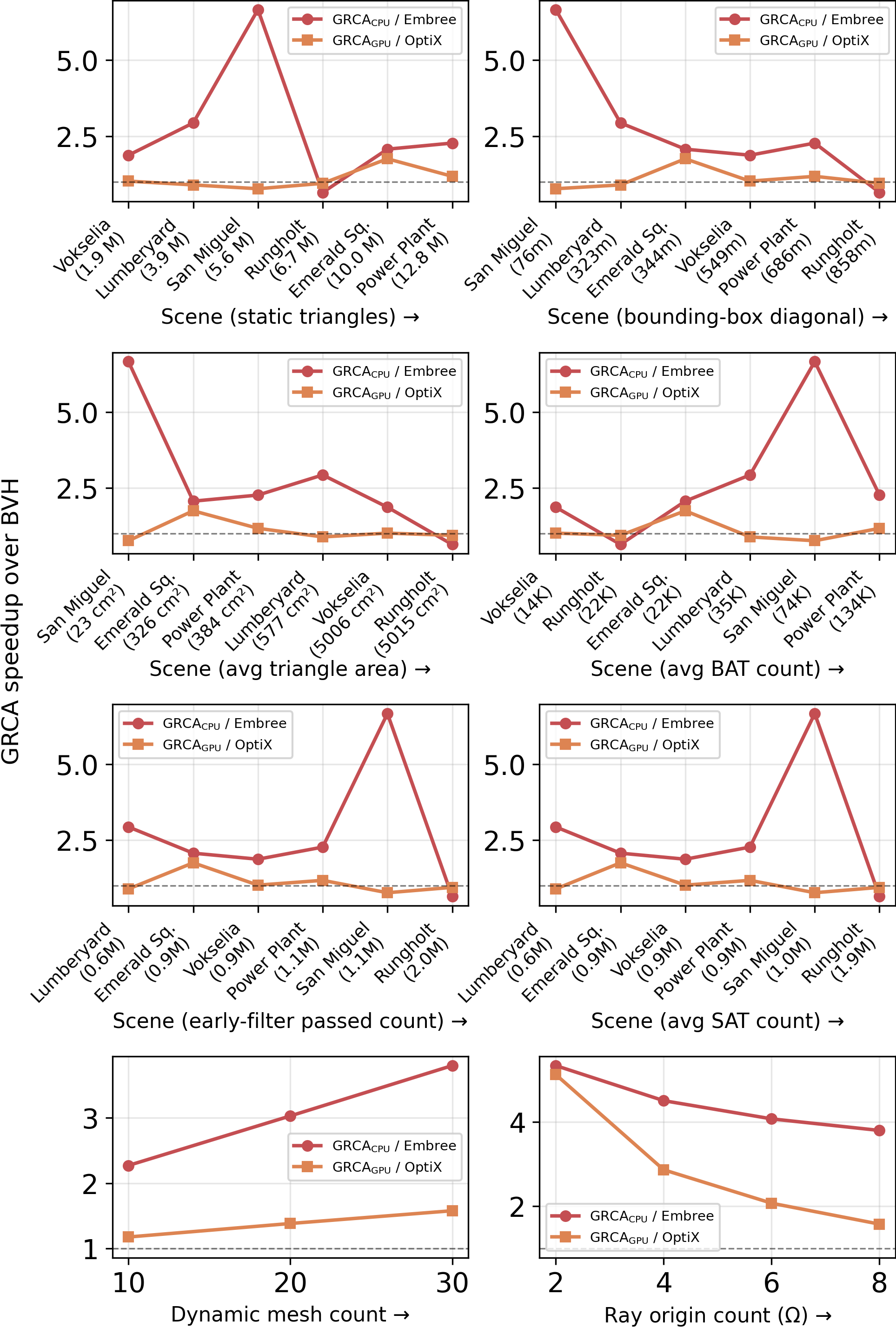}
  \caption{\grca{} speedup vs.\ independent cost variables (PC1, all avg).}
  \Description{Eight-panel 4\texttimes{}2 figure: top row shows speedup vs static triangle count and bounding-box diagonal; row 2 shows speedup vs average triangle area and per-scene avg BAT count; row 3 shows speedup vs per-scene avg early-filter count and avg SAT count; bottom row shows speedup vs dyn count and $\Omega$.}
  \label{fig:cost-contributors}
\end{figure}}

\newcommand{\FigPageOrigins}{\setcounter{figure}{12}%
\begin{figure}[htb]
  \centering
  \includegraphics[width=\figpagew]{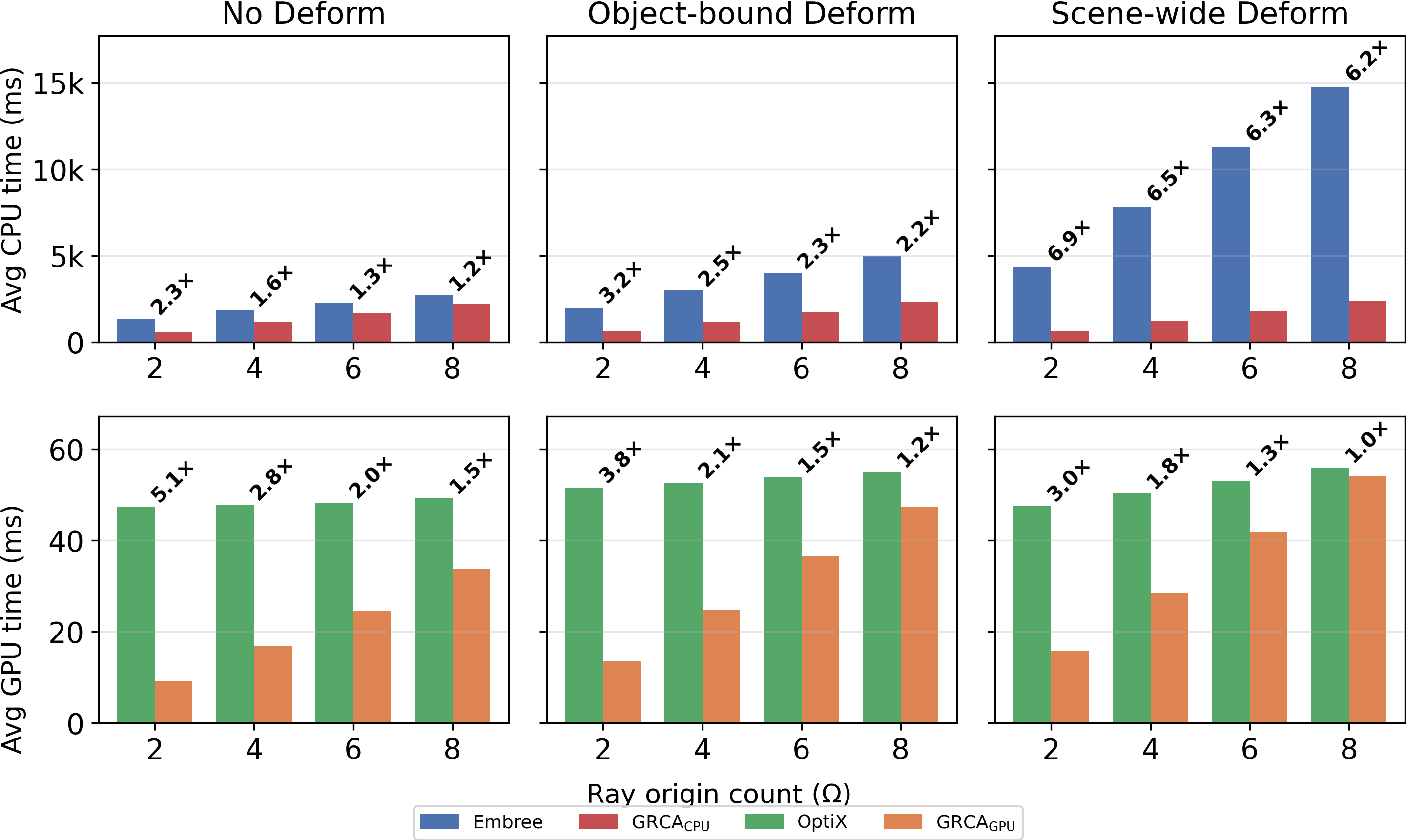}
  \caption{Time per $\Omega$, all six scenes, ND/OBD/SWD, PC1. Top: CPU. Bottom: GPU.}
  \Description{Grouped bar charts showing Embree vs \grcacpu{} CPU time and OptiX vs \grcagpu{} GPU time per $\Omega$ across three deformation conditions.}
  \label{fig:time-vs-origins}
\end{figure}}

\newcommand{\FigPageHybrid}{\setcounter{figure}{14}%
\begin{figure}[!htb]   % '!': may exceed \bottomfraction, so the taller chart keeps its slot
  \centering
  \includegraphics[width=\figpagew]{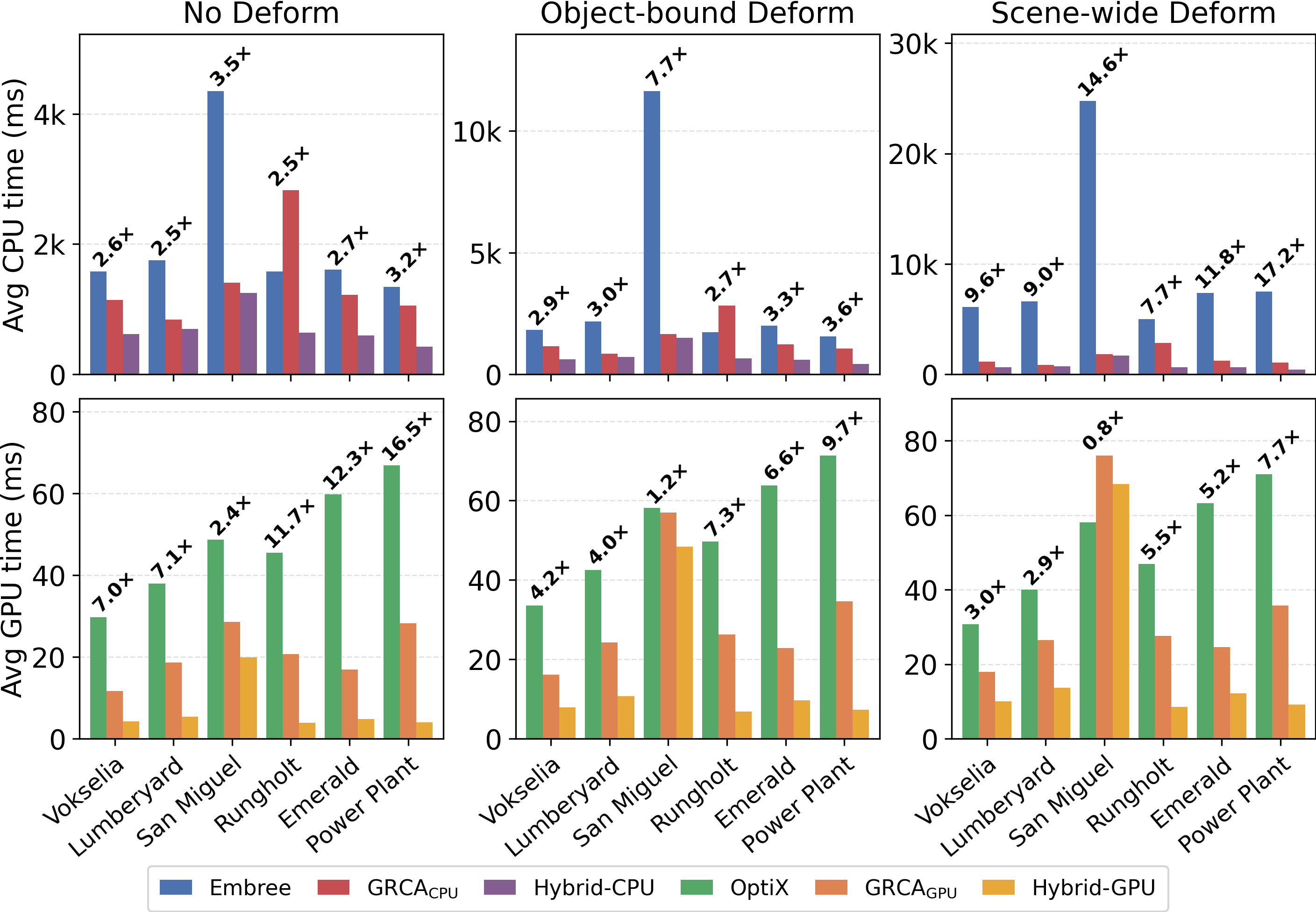}
  \caption{Hybrid vs.\ standalone, all six scenes, ND/OBD/SWD, PC1, 1000\,m; mean over dyn 10--30, $\Omega{=}2$--$8$.
           Labels: hybrid speedup vs.\ Embree/OptiX.}
  \Description{Two rows of bar charts, CPU on top and GPU below, one column per deformation condition.
               For each of the six scenes, average frame time of the rebuild-only baseline, standalone \grca{}
               and the hybrid pipeline, labeled with the hybrid's speedup; the hybrid is fastest in every scene
               except San Miguel on the GPU under deformation.}
  \label{fig:hybrid_benchmark}
\end{figure}}

\newcommand{\FigPageMultiPlatform}{\setcounter{figure}{13}%
\begin{figure}[htb]
  \centering
  \includegraphics[width=\figpagew]{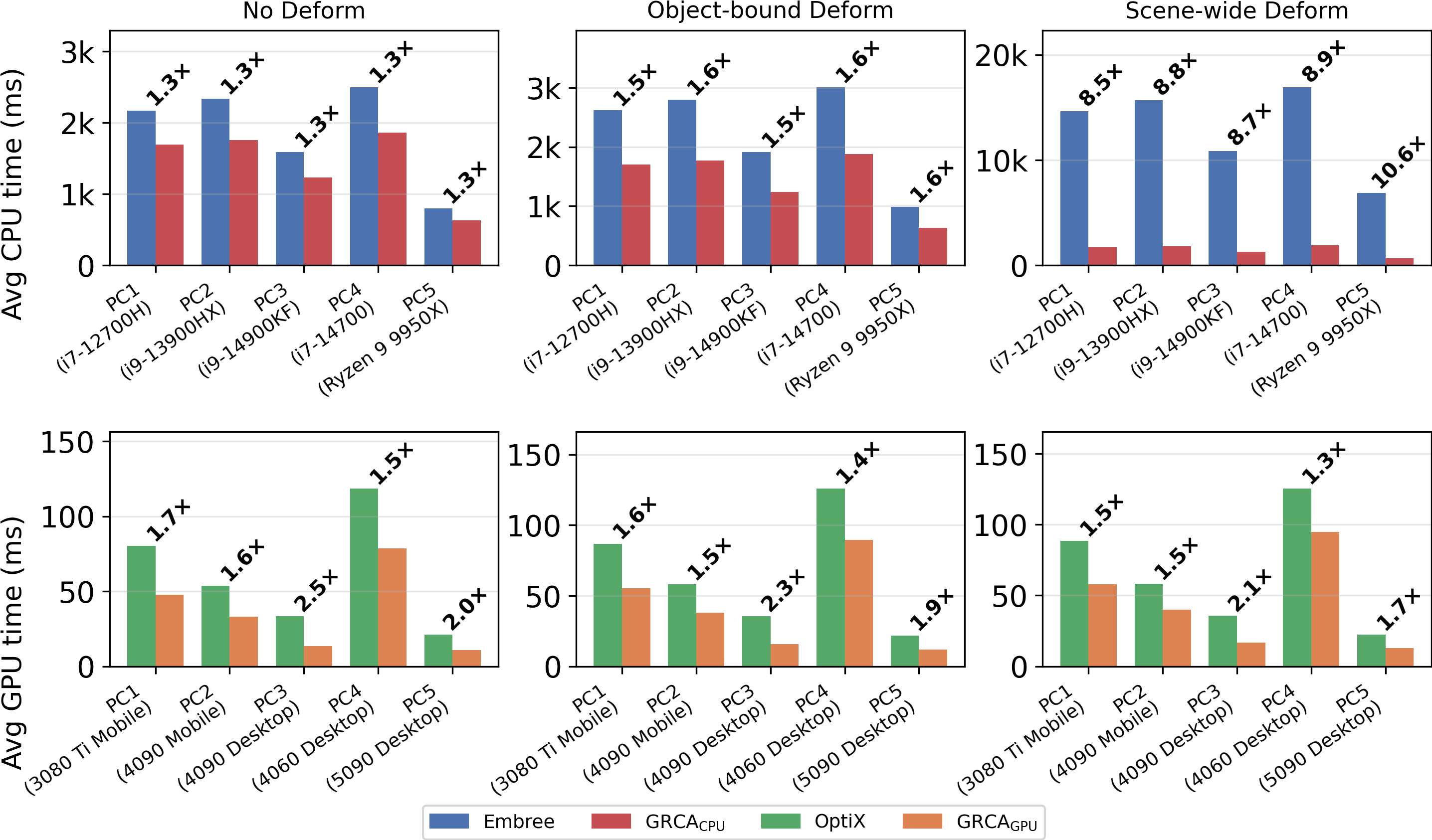}
  \caption{Multi-platform comparison (Power~Plant, 30~dyn, $\Omega{=}8$, PC1--PC5).
           Top: 5 CPU platforms. Bottom: 5 GPU platforms.
           ND/OBD/SWD.}
  \Description{Two bar charts stacked vertically: top chart compares \grcacpu{} and Embree CPU frame time across five hardware platforms; bottom chart compares \grcagpu{} and OptiX GPU frame time across five GPU platforms, both with three deformation configurations.}
  \label{fig:multi_platform}
\end{figure}}

\newcommand{\FigPageGACPTg}{\setcounter{figure}{15}%
\begin{figure}[htb]
  \centering
  \includegraphics[width=0.77\figpagew]{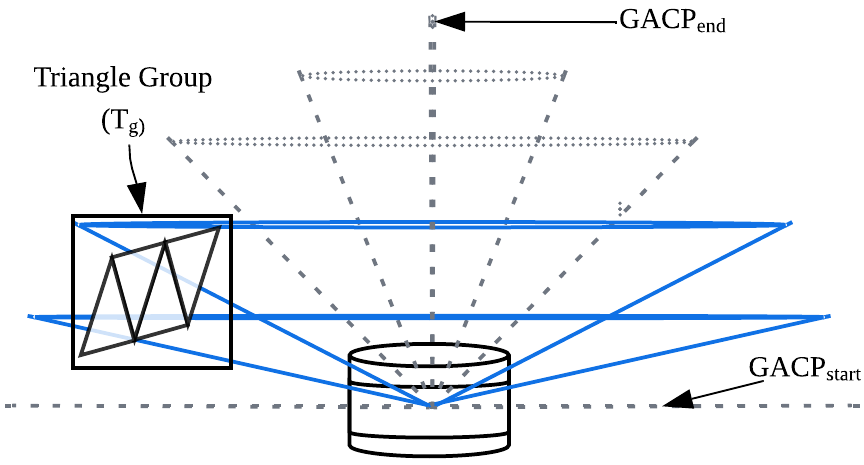}
  \caption{Multiple GACP tested against a triangle-group (T$_g$) bounding box.}
  \Description{Diagram showing GACP-Tg group-level intersection culling with multiple channels tested against a bounding box.}
  \label{fig:gacp-tg}
\end{figure}}

\newcommand{\FigPageShadows}{\setcounter{figure}{16}%
\begin{figure}[htb]
  \centering
  \includegraphics[width=\figpagew]{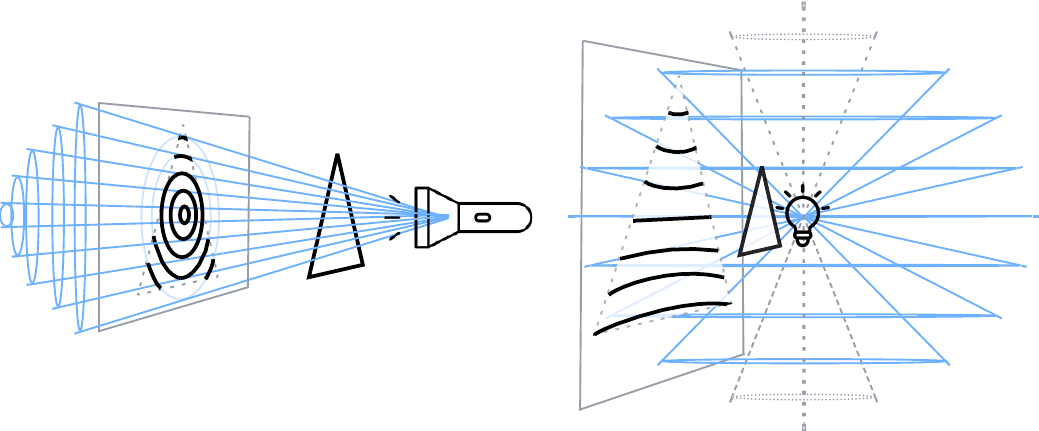}
  \caption{Light-reachability prediction for point and spot light shadows.}
  \Description{Diagram showing per-triangle light-reachability prediction for point, spot and area light shadow ray generation.}
  \label{fig:grca_shadows}
\end{figure}}

%% The acmtog builds reach the target layout from the default order; the preprint (different page
%% geometry) reaches it only when the figures come in page order, so GRCA_pre.tex defines \figpagesinlayoutorder.
\ifdefined\figpagesinlayoutorder
  \FigPageScaleVsBVH
  \FigPageScaleBAT
  \FigPageRange
  \FigPageCost
  \FigPageGACPTg
  \FigPageShadows
  \FigPageOrigins
  \FigPageHybrid
  \FigPageMultiPlatform
  \FigPageDeformation
  \FigPageTriArea
\else
  \FigPageDeformation
  \FigPageTriArea
  \FigPageScaleVsBVH
  \FigPageScaleBAT
  \FigPageRange
  \FigPageCost
  \FigPageOrigins
  \FigPageHybrid
  \FigPageMultiPlatform
  \FigPageGACPTg
  \FigPageShadows
\fi
\setcounter{figure}{17}   % back to the last figure-page number

%% Supplementary appendix: new page with a full-width title in the class title font
\makeatletter
\twocolumn[{\centering\@titlefont Supplementary Material for: GRCA: Fast Geometrically Modeled Emitter-Centric Ray-Casting Acceleration in Highly Dynamic LiDAR Simulation\par\vspace{2em}}]
\makeatother
\appendix
%% Supplementary floats and equations are numbered S1, S2, ... so they never share a number with the main paper
\setcounter{table}{0}\setcounter{figure}{0}\setcounter{equation}{0}
\renewcommand{\thetable}{S\arabic{table}}\renewcommand{\thefigure}{S\arabic{figure}}\renewcommand{\theequation}{S\arabic{equation}}
\renewcommand{\theHtable}{S\arabic{table}}\renewcommand{\theHfigure}{S\arabic{figure}}\renewcommand{\theHequation}{S\arabic{equation}}
%% Unbreakable math boxes would otherwise leave stretched gaps mid-column (flushbottom); let columns end short instead
\raggedbottom
\noindent This document provides extended derivations, implementation details, and additional benchmark data for the \grcaname{} (\grca{}) algorithm presented in the main paper. Numbered sections (\S3) refer to the main paper, lettered ones (\S C) to this supplement.

\section{Related Work (Extended)}
\label{sec:related_full}

\noindent The main paper (\S\ref{sec:related}) positions \grca{} against the most directly comparable methods.
This section covers additional related work that provides broader context.

\subsection{Spatialized normal cone hierarchies (SNCH)}
\label{sec:related_snch}
Johnson and Cohen~\cite{johnson2001snch} augment a BVH with per-node normal cones for closest-geometry queries, addressing the directional-culling gap (\hyperref[sc:360]{S2}, main paper). However, SNCH targets closest-geometry queries, as used by IR/sonar-style range sensors where per-ray directional identity is irrelevant. LiDAR needs per-ray directional identity, like a ray-traced depth image with one ray per pixel, so the benchmarks compare against hardware BVH ray tracing. Under dynamic geometry, a normal-cone hierarchy would also need rebuilding every frame, like the BVH baselines.

\subsection{Rasterization-based approaches}
Denis et al.~\cite{denis2023rasterization} back-project depth
images to generate LiDAR point clouds via GPU rasterization.  Beyond the $\ge$8 cameras per 360$^\circ$ origin noted in the main paper, multi-origin spinning LiDAR adds three costs:
the camera count scales linearly with ray origin count with no sharing;
projection distortion near frame edges affects the back-projected point cloud;
and the depth-camera pipeline shares the GPU with the simulator's rendering pipeline and onboard image sensors.

\subsection{Physics-engine-based LiDAR simulation}
CARLA~\cite{dosovitskiy2017carla}, LGSVL~\cite{rong2020lgsvl},
AirSim~\cite{shah2017airsim}, and Gazebo~\cite{koenig2004gazebo} use general-purpose collision backends (ODE, Bullet, DART, PhysX).
These expose a single-ray query interface with no GPU parallelism---ray grouping is left to the caller, and per-call overhead compounds
linearly with ray count.

\subsection{Offline and neural LiDAR simulation}
A complementary body of work targets data generation rather than real-time throughput.
LiDARsim~\cite{manivasagam2020lidarsim} re-simulates scans from real-world
reconstructions; Fang et al.~\cite{fang2020alsim}
augment real scans by ray-casting against reconstructed meshes; Guillard et
al.~\cite{guillard2022lidar} add a learned residual for physical realism;
T\'{o}th et al.~\cite{toth2025hybrid} combine neural
scene reconstruction with a ray-traced LiDAR backend.
On the neural side, UniSim~\cite{yang2023unisim} enables photorealistic closed-loop replay but is bound
to pre-captured reconstructions, and
Wu et al.~\cite{wu2024dynlidar} extend neural simulation to dynamic scenes via
per-object fields with offline training and inference.
All assume a pre-built static scene representation; arbitrary
dynamic geometry would require per-frame reconstruction.

\section{Notation}
\label{sec:notation}

Table~\ref{tab:notation} summarizes the symbols and abbreviations used throughout the main paper and this supplement.

\begin{table}[htb]
  \centering
  \caption{Notation.}
  \label{tab:notation}
  \footnotesize
  \begin{tabularx}{\columnwidth}{@{}l>{\raggedright\arraybackslash}X@{}}
    \toprule
    Symbol & Meaning \\
    \midrule
    \multicolumn{2}{@{}l}{\textit{Scene and origin}} \\
    $\Omega$                         & Ray origin count (independent LiDAR instances) \\
    $\gamma_n,\chi_n$                 & Vertical channels / rays per channel for origin $n$ \\
    $\tau$                            & Triangle count in the scene \\
    $T$                               & A scene triangle \\
    $s$                               & Per-axis mesh scale (main paper Table~\ref{tab:meshes}) \\
    $\mathbf{o}$                      & Ray origin position \\
    $\mathbf{p}$                      & World-space point \\
    $\hat{f},\hat{r},\hat{u}$         & Forward, right, up sensor axes (left-handed); $\hat{u}$ also cone spin axis \\
    \midrule
    \multicolumn{2}{@{}l}{\textit{Ray and channel geometry}} \\
    $\phi,\theta$                     & Elevation, azimuth angle \\
    $C_i,R_i$                         & Channel index (bottom${=}0$), ray index (clockwise from above) \\
    $\Delta\phi,\Delta\theta$         & Angular step sizes \\
    $\phi_0,\theta_0$                 & Center-aligned angular offsets \\
    $\hat{d}(\theta,\phi)$            & Unit ray direction (main paper Eq.~\ref{eq:ray_dir}) \\
    $\kappa$                          & Non-negative ray parameter \\
    $f(\phi,\theta)$                  & Ray as half-line $\{\mathbf{o}+\kappa\hat{d}\}$ \\
    $\Theta_{\Delta\theta}$           & Discrete azimuth sample set at spacing $\Delta\theta$ \\
    $\alpha$                          & Cone half-angle ($\tfrac{\pi}{2}-|\phi|$) \\
    $\mathbf{q}$                      & Horizontal projection of $\mathbf{p}-\mathbf{o}$ (main paper \S\ref{sec:ray_indexing}) \\
    $\lambda$                         & Position along a triangle edge, $\lambda\in[0,1]$ \\
    $c_2,c_1,c_0$                     & Edge-quadratic coefficients (main paper Eq.~\ref{eq:edge_quadratic}) \\
    \midrule
    \multicolumn{2}{@{}l}{\textit{Index spans}} \\
    $C_{\mathrm{from}},C_{\mathrm{to}}$ & Channel span endpoints \\
    $R_{\mathrm{from}},R_{\mathrm{to}}$ & Ray span endpoints (per channel) \\
    $C_{\mathrm{span}},R_{\mathrm{span}}$ & Channel / ray index spans $[C_{\mathrm{from}},C_{\mathrm{to}}]$, $[R_{\mathrm{from}},R_{\mathrm{to}}]$ \\
    $\gamma_{\mathrm{span}},\chi_{\mathrm{span}}$ & Span height / width (channel and ray counts) \\
    $\chi_{\mathrm{span}}\qual{\mathrm{widest}}$ & Largest $\chi_{\mathrm{span}}$ over a triangle's channels \\
    \midrule
    \multicolumn{2}{@{}l}{\textit{Apparent area and classification}} \\
    $A_T$                             & Triangle area \\
    $\phi_{\mathrm{view}}$            & Angle between triangle normal and line of sight \\
    $\delta$                          & Sensor-to-triangle distance \\
    $\delta\qual{\min}$               & Closest-point distance from $\mathbf{o}$ to a triangle \\
    $D\qual{\min},D\qual{\max}$       & Sensor minimum / maximum range \\
    $\hat{n},\mathbf{c}$              & Triangle unit normal (clockwise winding: points away from the sensor for front faces) and centroid \\
    $A_{\mathrm{origin}}$             & Origin-centric apparent area (main paper Eq.~\ref{eq:aorigin}) \\
    $A_{\mathrm{emitter}}$            & Emitter-centric apparent area (main paper Eq.~\ref{eq:aemitter}) \\
    $\gamma_T,\chi_T$                 & SAT/BAT classification thresholds \\
    $\varepsilon_A$                   & Minimum apparent-area threshold \\
    SAT / BAT                         & Smaller-/Bigger-Appearing Triangle \\
    \midrule
    \multicolumn{2}{@{}l}{\textit{Costs}} \\
    RTIC                              & Ray--Triangle Intersection Check \\
    $\tau_n\qual{\text{filtered}}$     & Origin $n$'s triangles left after the early-pass filters \\
    $\gamma_n\qual{\text{predicted}},\chi_n\qual{\text{predicted}}$ & Their mean predicted channel / ray spans (main paper Eq.~\ref{eq:rtic_reduced}) \\
    $\nu$                             & Angular cell count in ray classification (main paper \S\ref{sec:related}) \\
    \midrule
    \multicolumn{2}{@{}l}{\textit{Standard abbreviations}} \\
    AVX2                              & Advanced Vector Extensions 2 (256-bit x86 SIMD) \\
    BVH                               & Bounding volume hierarchy \\
    CW / CCW                          & Clockwise / counter-clockwise \\
    DOF                               & Degree of freedom \\
    FMCW                              & Frequency-modulated continuous-wave (radar) \\
    FOV                               & Field of view \\
    IAS / BLAS                        & Instance / bottom-level acceleration structure: the two levels of an instanced BVH \\
    ML                                & Machine learning \\
    RT                                & Ray tracing; RT cores are the GPU's ray-tracing hardware \\
    SIMD                              & Single instruction, multiple data \\
    \bottomrule
  \end{tabularx}
\end{table}

\section{GACP-T Intersection}
\label{sec:gacp_t_math}

\noindent\textit{Derives the cone/plane--triangle intersection test shared by both pipeline passes.
Throughout, $\mathbf{v}_0, \mathbf{v}_1, \mathbf{v}_2$ are the triangle vertices, $(i_0, i_1)$ an edge index pair, and $\mathbf{o}$ the ray origin.}

\subsection{Plane Case (\texorpdfstring{$\phi=0$}{φ = 0})}

When the channel elevation is zero, the GACP degenerates to the horizontal plane
through $\mathbf{o}$.  Each vertex has signed axial height
$h_k=(\mathbf{v}_k-\mathbf{o})\cdot\hat{u}$.  An edge crosses the plane when
$h_{i_0}\cdot h_{i_1}<0$.

\begin{tcolorbox}[mathbox, title={Plane crossing point}]
\vspace{-8pt}
\begin{equation}
  \mathbf{x} = \mathbf{v}_{i_0}
             + \frac{h_{i_0}}{h_{i_0}-h_{i_1}}
               (\mathbf{v}_{i_1}-\mathbf{v}_{i_0})
\end{equation}
\vspace{-8pt}
\end{tcolorbox}
No quadratic is needed; at most two edges cross.

\subsection{Cone Case (\texorpdfstring{$\phi \neq 0$}{φ ≠ 0})}
\label{sec:precise_cone}

For non-zero elevation, the GACP is a cone. Each triangle edge is tested for intersection by solving a quadratic in the edge parameter $\lambda\in[0,1]$.

For edge $\mathbf{v}_{i_0} \to \mathbf{v}_{i_1}$~\cite{eberly_geometrictools}, define:
\begin{align}
  \mathbf{e} &= \mathbf{v}_{i_1} - \mathbf{v}_{i_0},
  \quad e_A = \mathbf{v}_{i_1}\cdot\hat{u} - \mathbf{v}_{i_0}\cdot\hat{u}
\end{align}
Let $\mathbf{p}(\lambda) = \mathbf{v}_{i_0}+\lambda\mathbf{e}$.
Substituting into the cone equation $\bigl((\mathbf{p}-\mathbf{o})\cdot\hat{u}\bigr)^2 = \sin^2\!\phi\;|\mathbf{p}-\mathbf{o}|^2$ yields (main paper Eq.~\ref{eq:edge_quadratic}):

\begin{tcolorbox}[mathbox, title={Edge--cone quadratic}]
\vspace{-8pt}
\begin{equation}
  c_2 \lambda^2 + c_1 \lambda + c_0 = 0
\end{equation}
\begin{align}
  c_2 &= e_A^2 - \sin^2\!\phi\;(\mathbf{e}\cdot\mathbf{e}) \\
  c_1 &= 2\bigl(e_A\,((\mathbf{v}_{i_0}-\mathbf{o})\cdot\hat{u}) - \sin^2\!\phi\;((\mathbf{v}_{i_0}-\mathbf{o})\cdot\mathbf{e})\bigr) \\
  c_0 &= ((\mathbf{v}_{i_0}-\mathbf{o})\cdot\hat{u})^2 - \sin^2\!\phi\;\|\mathbf{v}_{i_0}-\mathbf{o}\|^2
\end{align}
\begin{equation}
  \lambda = \frac{-c_1 \pm \sqrt{c_1^2 - 4c_2 c_0}}{2c_2}
\end{equation}
\vspace{-8pt}
\end{tcolorbox}

A root is valid when $\lambda\in[0,1]$ \emph{and} the point lies on the forward half of the
cone (rejecting the mirror cone behind the apex):
$(\mathbf{v}_{i_0}-\mathbf{o})\cdot\hat{u}+\lambda e_A\ge 0$ for $\phi>0$,
or $\le 0$ for $\phi<0$.

\subsection{Vertex Classification and Early Exit}

Before running the per-edge quadratic, vertex positions relative to the cone classify the triangle into one of four cases.
Let $\sin\phi_k = \widehat{\mathbf{v}_k-\mathbf{o}}\cdot\hat{u}$ be the normalized elevation of vertex $k$:

\begin{tcolorbox}[mathbox, title={Classification cases (checked in order)}]
\begin{enumerate}[nosep,leftmargin=*,labelsep=0.4em]
  \item \textbf{Wrong half-space} --- $(\mathbf{v}_k-\mathbf{o})\cdot\hat{u}<0$ for all $k$
        (when $\phi>0$), or $>0$ for all $k$ (when $\phi<0$).
        \emph{Result: no intersection.}
  \item \textbf{All inside} --- $\sin\phi_k\ge\sin\phi$ for all $k$ (when $\phi>0$),
        or $-\sin\phi_k\ge\sin|\phi|$ for all $k$ (when $\phi<0$).
        \emph{Result: no crossing.}
  \item \textbf{Mixed} --- at least one vertex inside, at least one outside.
        \emph{Result: proceed to quadratic.}
  \item \textbf{All outside}:
        \begin{itemize}[nosep,leftmargin=*,labelsep=0.4em]
          \item The apex ray $\mathbf{o}+\kappa\,\mathrm{sgn}(\phi)\,\hat{u}$ hits the triangle and $\delta\qual{\min}\le D\qual{\max}\sin|\phi|$
                $\to$ the triangle spans the cone axis, so every ray of the channel crosses it. \emph{Result: whole channel (boolean: true); no quadratic.}
          \item Otherwise $\to$ run per-edge quadratic directly.
        \end{itemize}
\end{enumerate}
\end{tcolorbox}

\subsection{Boolean Predicate (Early Pass)}
\label{sec:supp_bool}

The early pass needs only yes/no, enabling short-circuits that skip the quadratic:

\begin{tcolorbox}[mathbox, title={Short-circuit rules}]
\begin{itemize}[nosep,leftmargin=*,labelsep=0.4em]
  \item All-inside $\to$ \textbf{false} (no crossing possible).
  \item Mixed or apex-ray hit $\to$ \textbf{true} (crossing guaranteed).
  \item All-outside fallthrough $\to$ run quadratic, return \textbf{true} on first valid root.
\end{itemize}
\end{tcolorbox}

\subsection{Channel-Span Prediction}
\label{sec:supp_cspan}

The midpoint channel is estimated from the three vertex elevation indices (using
the channel indexing formula from \S\ref{sec:channel_indexing}):

\begin{tcolorbox}[mathbox, title={Per-vertex channel index and midpoint}]
\vspace{-8pt}
\begin{equation}
  \underset{k\,\in\,\{0,1,2\}}{C_i^{(k)}} = \operatorname{clamp}\!\left(
    \operatorname{round}\!\left(
      \frac{\arcsin\!\bigl(\widehat{\mathbf{v}_k-\mathbf{o}}\cdot\hat{u}\bigr) - \phi_0}{\Delta\phi}
    \right),\;0,\;\gamma_n-1\right)
\end{equation}
\begin{equation}
  C_{\text{mid}} = \left\lceil
    \frac{\max_k C_i^{(k)} + \min_k C_i^{(k)}}{2}
  \right\rceil
\end{equation}
\vspace{-8pt}
\end{tcolorbox}

\subsection{Ray-Span Prediction}
\label{sec:supp_rspan}

For FOV $> 180^\circ$, let $\hat{d}_{\text{ctr}}$ be the center ray.
A right vector is built from $\hat{d}_{\text{ctr}}$ via Gram--Schmidt:
\begin{equation}
  \hat{u}_\perp = \operatorname{normalize}\!\bigl(
    \hat{w} - (\hat{w}\cdot\hat{d}_{\text{ctr}})\hat{d}_{\text{ctr}}\bigr),
  \qquad
  \hat{r}_\perp = \operatorname{normalize}\!\bigl(
    \hat{d}_{\text{ctr}} \times \hat{u}_\perp\bigr)
\end{equation}
where $\hat{w} = (0,1,0)$ if $|\hat{d}_{\text{ctr},y}| < 0.9$, else $(1,0,0)$.
Three dot-product tests check whether the triangle lies entirely on one side of the ray origin:

\begin{tcolorbox}[mathbox, title={CW half-space tests}]
\vspace{-8pt}
\begin{align}
  \text{all-front} &= \forall k:\;(\mathbf{v}_k-\mathbf{o})\cdot\hat{d}_{\text{ctr}} > 0 \\
  \text{all-right} &= \forall k:\;(\mathbf{v}_k-\mathbf{o})\cdot\hat{r}_\perp > 0 \\
  \text{all-left}  &= \forall k:\;(\mathbf{v}_k-\mathbf{o})\cdot\hat{r}_\perp < 0
\end{align}
\begin{equation}
  \text{all-CW} =
    \text{all-front} \;\vee\; \text{all-right} \;\vee\; \text{all-left}
\end{equation}
\vspace{-8pt}
\end{tcolorbox}

If none hold, the triangle straddles the seam and its ray span cannot be resolved by projection alone.
Otherwise, project each vertex to a ray index using the ray indexing formula
from \S\ref{sec:ray_indexing} with
$\theta_k = \operatorname{atan2}((\mathbf{v}_k-\mathbf{o})\cdot\hat{r},\;(\mathbf{v}_k-\mathbf{o})\cdot\hat{f})$:

\begin{tcolorbox}[mathbox, title={Per-vertex ray index and span}]
\vspace{-8pt}
\begin{equation}
  R_i^{(k)} = \operatorname{clamp}\!\left(
    \operatorname{round}\!\left(
      \frac{\theta_k - \theta_0}{\Delta\theta}
    \right),\;0,\;\chi_n-1\right),\quad k \in \{0,1,2\}
\end{equation}
\begin{equation}
  R_{\mathrm{from}} = \min\!\bigl(R_i^{(0)},R_i^{(1)},R_i^{(2)}\bigr),
  \qquad
  R_{\mathrm{to}}   = \max\!\bigl(R_i^{(0)},R_i^{(1)},R_i^{(2)}\bigr)
\end{equation}
\vspace{-8pt}
\end{tcolorbox}

\subsection{CW/CCW Arc Disambiguation}
\label{sec:supp_cw_ccw}

Let $\hat{n}$ and $\mathbf{c}$ be the triangle's unit normal and centroid (Table~\ref{tab:notation}), and $\hat{d}[R]$ the direction of ray index $R$ in the current channel.
When all-CW is false, the GACP-T intersection produces two span endpoints but the CW/CCW arc direction is ambiguous.
An arc-midpoint plane test resolves this:

\begin{tcolorbox}[mathbox, title={Arc disambiguation procedure}]
\begin{enumerate}[nosep,leftmargin=*,labelsep=0.4em]
  \item \textbf{CW midpoint}: $\chi_{\mathrm{span}}\qual{\mathrm{CW}} = R_{\mathrm{to}} - R_{\mathrm{from}} + 1$:
        \[
          R_{\mathrm{mid}}\qual{\mathrm{CW}} = \bigl(R_{\mathrm{from}} + \lfloor \chi_{\mathrm{span}}\qual{\mathrm{CW}}/2\rfloor\bigr)
            \bmod \chi_n
        \]

  \item \textbf{CCW midpoint} by reflection across $\hat{u}$ (a direction, not a buffer ray):
        \[
          \hat{d}_m\qual{\mathrm{CCW}}
            = 2\,(\hat{d}_m\qual{\mathrm{CW}}\cdot\hat{u})\,\hat{u}
              - \hat{d}_m\qual{\mathrm{CW}},\qquad \hat{d}_m\qual{\mathrm{CW}} = \hat{d}[R_{\mathrm{mid}}\qual{\mathrm{CW}}]
        \]

  \item \textbf{Plane test}: for each candidate $\hat{d}_m\in\{\hat{d}_m\qual{\mathrm{CW}},\hat{d}_m\qual{\mathrm{CCW}}\}$, with $\rho$ the ray parameter (cf.\ $\kappa$) of its hit on the triangle plane:
        \[
          \rho = \frac{\hat{n}\cdot(\mathbf{c}-\mathbf{o})}
                   {\hat{n}\cdot\hat{d}_m}
        \]
        $\operatorname{hit}(\hat{d}_m) = (\hat{n}\cdot\hat{d}_m\ge\varepsilon_{\mathrm{face}})
        \;\wedge\; (\rho\ge 0)$, \quad $\varepsilon_{\mathrm{face}}=10^{-4}$.

  \item \textbf{Decision}:
        \[
          \text{arc-CW} = \begin{cases}
            \text{true}  & \text{if } \operatorname{hit}(\hat{d}_m\qual{\mathrm{CW}})
                            \text{ and } \neg\operatorname{hit}(\hat{d}_m\qual{\mathrm{CCW}}) \\
            \text{false} & \text{if } \operatorname{hit}(\hat{d}_m\qual{\mathrm{CCW}})
                            \text{ and } \neg\operatorname{hit}(\hat{d}_m\qual{\mathrm{CW}}) \\
            \multicolumn{2}{l}{\text{full Möller--Trumbore on }
              \hat{d}_m\qual{\mathrm{CW}}\text{ (tiebreaker)}}
          \end{cases}
        \]
\end{enumerate}
\end{tcolorbox}

\section{Early Pass}
\label{sec:early_pass}
\label{sec:pipeline_detail}

\noindent\textit{Processes every triangle cheaply; most are rejected. Triangles too large for conservative span prediction are deferred to the late pass.}

\subsection{Origin-Centric Filters}
\label{sec:supp_filters}

Three cheap filters reject triangles before any span computation:

\begin{tcolorbox}[mathbox, title={Rejection filters (applied per triangle, per origin)}]
\begin{itemize}[nosep,leftmargin=*,labelsep=0.4em]
  \item \textbf{Back-face}: skip if $(\mathbf{c}-\mathbf{o})\cdot\hat{n} \le 0$ (or ${\approx}0$: origin on the triangle plane).
  \item \textbf{Apparent area}: squaring eliminates the square root (both sides positive after back-face):
        \[
          \bigl(A_T\cdot(\mathbf{c}-\mathbf{o})\cdot\hat{n}\bigr)^2 < \varepsilon_A^2\;\|\mathbf{c}-\mathbf{o}\|^6
        \]
        Skip if below threshold $\varepsilon_A$ (main paper \S\ref{sec:tri_area}).
  \item \textbf{Range}: skip if $\delta\qual{\min} \notin [D\qual{\min}, D\qual{\max}]$, where $\delta\qual{\min}$~\cite{ericson2004} is the closest-point distance. Stored for reuse.
\end{itemize}
\end{tcolorbox}

\subsection{Span Prediction}
\label{sec:step3_rspan}

\begin{itemize}[nosep,leftmargin=*,labelsep=0.4em]
  \item \textbf{Channel span}: computes $[C_{\mathrm{from}}, C_{\mathrm{to}}]$ (main paper \S\ref{sec:cspan_predict}). Discard if no channel intersects.
  \item \textbf{Ray span}: computes $[R_{\mathrm{from}}, R_{\mathrm{to}}]$ (main paper \S\ref{sec:rspan_predict}). Seam-straddling triangles (all-CW $=$ false) are classified BAT.
\end{itemize}

\subsection{SAT/BAT Classification}
\label{sec:classification}

Applied independently per (triangle, origin) pair.

\begin{tcolorbox}[mathbox, title={SAT condition}]
A triangle is \textbf{SAT} for origin $n$ if:
\vspace{-4pt}
\begin{equation}
  \text{all-CW}
  \;\wedge\; \gamma_{\mathrm{span}} \le \gamma_T
  \;\wedge\; \chi_{\mathrm{span}} \le \chi_T
  \label{eq:sat_cond}
\end{equation}
\vspace{-8pt}
\end{tcolorbox}

Otherwise it is \textbf{BAT}.
A triangle enters the deferred list only if at least one origin classifies it as BAT; SAT origins have their RTIC handled inline.
Thresholds $(\gamma_T, \chi_T)$ are tunable (\S\ref{sec:sat_threshold_detail}).

\subsection{Inline SAT RTIC}

For SAT triangles, M\"{o}ller--Trumbore~\cite{moller1997} runs over every (channel, ray)
pair in the predicted spans.

\begin{tcolorbox}[mathbox, title={Global ray index and atomic hit recording}]
\vspace{-8pt}
\begin{equation}
  g(j,i) = j \cdot \chi_n
           + \bigl(R_{\mathrm{from}} + i\bigr) \bmod \chi_n + O_n
\end{equation}
for channel $j\in[C_{\mathrm{from}},C_{\mathrm{to}}]$ and span offset $i\in[0,\chi_{\mathrm{span}})$, where $O_n$ is the ray-index offset of origin $n$.
\begin{equation}
  D_{g(j,i)} \leftarrow \min\!\bigl(D_{g(j,i)},\;f_{\mathrm{sort}}(\kappa)\bigr)
\end{equation}
where $D_g$ is the stored hit of ray $g$, $\kappa$ the new hit's ray parameter and $f_{\mathrm{sort}}$ maps a non-negative float to an order-preserving unsigned integer, so one atomic min keeps the closest hit.
\vspace{-8pt}
\end{tcolorbox}

%% =============================================================================

\section{Late Pass}
\label{sec:late_pass}

\noindent\textit{Resolves BAT triangles---those too large or seam-straddling for conservative spans. Both passes record hits in the same per-ray buffer through atomic min (\S\ref{sec:early_pass}), so their order does not matter.}

\subsection{Deferred-List Structure}

Each deferred entry carries the data pre-computed by the early pass (filters, spans and flags are not recomputed; only the vertices are reloaded):

\begin{tcolorbox}[mathbox, title={BATEntry struct}]
{\footnotesize
\begin{algorithmic}[0]
\State \textbf{struct} BATEntry \{
\State \quad triIndex \Comment{triangle ID}
\State \quad originActive[$\Omega$] \Comment{1 bit: BAT for this origin}
\State \quad apexHitP[$\Omega$], apexHitN[$\Omega$] \Comment{1 bit each: apex-ray hit along $+\hat{u}$ / $-\hat{u}$ (for $\phi{\ge}0$ / $\phi{<}0$ channels)}
\State \quad allCW[$\Omega$] \Comment{1 bit: skip arc disambiguation if true}
\State \quad $\delta\qual{\min}$[$\Omega$] \Comment{fp16: closest-point distance}
\State \quad $C_{\mathrm{from}}$[$\Omega$], $C_{\mathrm{to}}$[$\Omega$] \Comment{uint16: channel span}
\State \}
\end{algorithmic}
}
\end{tcolorbox}

\subsection{Precise GACP-T Per Channel}

For each channel in $[C_{\mathrm{from}}, C_{\mathrm{to}}]$, the full GACP-T intersection (main paper \S\ref{sec:gacp_t}) runs.

\begin{tcolorbox}[mathbox, title={Crossing-count dispatch}]
{\footnotesize
\begin{algorithmic}[0]
\State $k \gets$ GACP-T crossing count for this channel
\If{apex-ray case \textbf{or} $k > 2$} $[R_{\mathrm{from}},R_{\mathrm{to}}] \gets [0,\chi_n{-}1]$ \Comment{whole channel}
\ElsIf{$k = 0$} skip channel \Comment{no intersection}
\ElsIf{$k = 1$} $R_{\mathrm{from}} \gets R_{\mathrm{to}} \gets$ projected ray \Comment{tangent touch}
\Else{} $R_{\mathrm{from}}, R_{\mathrm{to}} \gets$ crossing points \Comment{exact span}
\EndIf
\end{algorithmic}
}
\end{tcolorbox}

\subsection{Arc Disambiguation and RTIC}

For seam-straddling triangles (all-CW $=$ false), the arc disambiguation (main paper \S\ref{sec:cw_ccw}) selects between the CW and CCW arc via a mid-ray plane test.
RTIC is then identical to Early Pass inline RTIC, over the exact spans.

%% =============================================================================

%% =============================================================================

\section{Benchmarks}

\subsection{Preliminary Study}

\subsubsection{Pose Sensitivity and Primary Configuration Selection}
\label{sec:pose_sensitivity_detail}

All backends were evaluated over 50 recorded frames across eight motion configurations covering the full combination of per-DOF interpolation modes (Table~\ref{tab:motion_cfg}: rot, r+sc, p+sc, pos, f.sm, p+r, f.i, sc) and three deformation conditions (scale $[0.001,30]$, full-rebuild, PC1). Full instant (\texttt{f.i}: position, rotation and scale all random each frame) is the main paper's random per-frame motion and is used in every other experiment.

\begin{table}[htb]
  \centering
  \caption{Motion configurations (Smooth: 10-frame interp; Instant: random each frame).}
  \label{tab:motion_cfg}
  \resizebox{0.9\columnwidth}{!}{%
  \scriptsize
  \setlength{\tabcolsep}{4pt}
  \begin{tabular}{lllp{2cm}}
    \toprule
    Position & Rotation & Non-uniform scale & Label \\
    \midrule
    smooth (10f) & instant      & instant      & pos (position only) \\
    instant      & smooth (10f) & instant      & rot (rotation only) \\
    instant      & instant      & smooth (10f) & sc (scale only) \\
    smooth (10f) & smooth (10f) & instant      & p+r (pos+rot) \\
    instant      & smooth (10f) & smooth (10f) & r+sc (rot+scale) \\
    smooth (10f) & instant      & smooth (10f) & p+sc (pos+scale) \\
    smooth (10f) & smooth (10f) & smooth (10f) & f.sm (full smooth) \\
    instant      & instant      & instant      & f.i (full instant) \\
    \bottomrule
  \end{tabular}}
\end{table}

\begin{table}[htb]
  \centering
  \caption{CPU pose sensitivity (avg ms/frame): Power~Plant, 30~dyn, $\Omega{=}8$, PC1, 50~frames.
           Columns ordered by closeness to avg ND/OBD/SWD mean (most $\to$ least representative).
           Vertical rule = \textbf{f.i} (full instant, primary). \textbf{Bold} = row min.}
  \label{tab:pose_sensitivity_cpu}
  \resizebox{\columnwidth}{!}{%
  \small
  \begin{tabular}{l rrrrrr !{\vrule width 1.2pt} r !{\vrule width 1.2pt} r r}
    \toprule
    & \multicolumn{8}{c}{most $\to$ least representative (CPU)} & \\
    \cmidrule(lr){2-9}
    Backend & rot & r+sc & p+sc & pos & f.sm & p+r & \textbf{f.i} & sc & Mean \\
    \midrule
    \multicolumn{10}{l}{\textit{No Deformation (ND)}} \\
    \quad Embree      & 2{,}147 & \textbf{2{,}015} & 2{,}078 & 2{,}225 & 2{,}073 & 2{,}249 & 2{,}211 & 2{,}026 & 2{,}128 \\
    \quad \grcacpu{}   & 1{,}791 & 1{,}797 & 2{,}000 & 2{,}028 & 2{,}098 & 2{,}087 & \textbf{1{,}665} & 1{,}671 & 1{,}892 \\
    \midrule
    \multicolumn{10}{l}{\textit{Object-bound Deform (per-tri)}} \\
    \quad Embree      & 2{,}695 & 2{,}660 & 2{,}813 & 2{,}760 & 2{,}820 & 2{,}771 & 2{,}647 & \textbf{2{,}600} & 2{,}721 \\
    \quad \grcacpu{}   & 1{,}800 & 1{,}798 & 1{,}987 & 1{,}994 & 2{,}079 & 2{,}079 & \textbf{1{,}666} & 1{,}686 & 1{,}886 \\
    \midrule
    \multicolumn{10}{l}{\textit{Scene-wide Deform (per-tri)}} \\
    \quad Embree      & 14{,}927 & 14{,}945 & 16{,}113 & 16{,}113 & 16{,}059 & 16{,}036 & \textbf{14{,}741} & 14{,}762 & 15{,}462 \\
    \quad \grcacpu{}   & 1{,}809 & 1{,}809 & 1{,}986 & 1{,}985 & 2{,}085 & 2{,}083 & \textbf{1{,}684} & 1{,}700 & 1{,}893 \\
    \bottomrule
  \end{tabular}}
\end{table}
\begin{table}[htb]
  \centering
  \caption{GPU pose sensitivity (avg ms/frame): same configuration as Table~\ref{tab:pose_sensitivity_cpu}.
           Columns ordered by closeness to avg ND/OBD/SWD mean (most $\to$ least representative); GPU ordering is scale-dominated and differs from CPU.
           Vertical rule = \textbf{f.i} (full instant, primary). \textbf{Bold} = row min.}
  \label{tab:pose_sensitivity_gpu}
  \resizebox{\columnwidth}{!}{%
  \small
  \begin{tabular}{l r !{\vrule width 1.2pt} r !{\vrule width 1.2pt} rrrrrr r}
    \toprule
    & \multicolumn{8}{c}{most $\to$ least representative (GPU)} & \\
    \cmidrule(lr){2-9}
    Backend & sc & \textbf{f.i} & p+sc & f.sm & rot & pos & p+r & r+sc & Mean \\
    \midrule
    \multicolumn{10}{l}{\textit{No Deformation (ND)}} \\
    \quad OptiX       & 80.5 & 81.2 & \textbf{79.9} & 80.4 & 81.1 & 81.2 & 80.1 & \textbf{79.9} & 80.5 \\
    \quad \grcagpu{}  & 44.6 & 43.9 & 48.9 & 49.3 & 43.0 & 50.0 & 49.9 & \textbf{42.0} & 46.4 \\
    \midrule
    \multicolumn{10}{l}{\textit{Object-bound Deform (per-tri)}} \\
    \quad OptiX       & 85.8 & 86.5 & 85.9 & 85.4 & 87.0 & 87.9 & 87.9 & \textbf{85.0} & 86.4 \\
    \quad \grcagpu{}  & 51.1 & 50.1 & 56.6 & 56.8 & \textbf{49.2} & 56.5 & 56.8 & 49.6 & 53.3 \\
    \midrule
    \multicolumn{10}{l}{\textit{Scene-wide Deform (per-tri)}} \\
    \quad OptiX       & 87.7 & 88.0 & 89.1 & 89.7 & \textbf{87.6} & 89.1 & 89.6 & 87.9 & 88.6 \\
    \quad \grcagpu{}  & 54.4 & 53.2 & 59.6 & 59.7 & \textbf{52.6} & 59.6 & 59.7 & \textbf{52.6} & 56.4 \\
    \bottomrule
  \end{tabular}}
\end{table}

Under SWD, Embree's frame time grows ${\sim}7\times$ with scene-wide scatter while \grcacpu{}'s changes by $<$1\% (Table~\ref{tab:pose_sensitivity_cpu}); on the GPU, OptiX rises ${\sim}10\%$ from ND to SWD (80.5 to 88.6\,ms) and \grcagpu{} ${\sim}22\%$ (46.4 to 56.4\,ms; Table~\ref{tab:pose_sensitivity_gpu}).

\subsubsection{BVH Mode Selection}
\label{sec:bvh_mode_detail}

Seventeen BVH update policies are evaluated under ND, OBD, and SWD (Table~\ref{tab:bvhmode}): refit (\texttt{rf}), full rebuild (\texttt{rb}), and periodic rebuild (\texttt{pr1}--\texttt{pr15}, alternating rebuild and refit at varying frequencies); per-mode build/update flags are listed in Table~\ref{tab:backend_config}.
Each cell: avg~ms / \textpm20\% / \%\,below~$\mu$.

\begin{table}[htb]
  \centering
  \caption{Backend BVH configuration per update mode.}
  \label{tab:backend_config}
  \resizebox{\columnwidth}{!}{%
  \footnotesize
  \setlength{\tabcolsep}{3pt}
  \begin{tabular}{llll}
    \toprule
    Backend & Mode & Build flag & Update flag \\
    \midrule
    OptiX  & refit   & \texttt{PREFER\_FAST\_TRACE} & \texttt{OPERATION\_UPDATE} \\
    OptiX  & rebuild & \texttt{PREFER\_FAST\_BUILD} & \texttt{OPERATION\_BUILD} (per frame) \\
    OptiX  & periodic rebuild & \texttt{PREFER\_FAST\_BUILD} & \texttt{BUILD} (rebuild frames), \texttt{UPDATE} (between) \\
    \midrule
    Embree & all     & \multicolumn{2}{l}{\texttt{RTC\_SCENE\_FLAG\_DYNAMIC}} \\
    Embree & refit   & \texttt{BUILD\_QUALITY\_HIGH} (init) & \texttt{BUILD\_QUALITY\_REFIT} (per frame) \\
    Embree & rebuild & \texttt{BUILD\_QUALITY\_LOW}  & \texttt{BUILD\_QUALITY\_LOW} \\
    Embree & periodic rebuild  & \texttt{LOW} (rebuild frames) & \texttt{REFIT} (between) \\
    \bottomrule
  \end{tabular}}
\end{table}

\subsubsection{Additional backend details}
\label{sec:backend_details}
OptiX uses triangle primitives (\texttt{OPTIX\_BUILD\_INPUT\_TYPE\_TRIANGLES}), hardware RT-core traversal, and \texttt{\_\_closesthit\_\_} intersection. Embree uses 16-ray packets (\texttt{rtcIntersect16}) on all platforms, matching \grcacpu{}. The static half of the hybrid pipeline (\S\ref{sec:hybrid_pipeline}) is built once at init in refit mode (OptiX \texttt{PREFER\_FAST\_TRACE}; Embree \texttt{BUILD\_QUALITY\_HIGH}) and never updated, since it holds no dynamic triangles.

\begin{table}[htb]
  \centering
  \caption{BVH update policies, avg ms/frame, PC1, $\Omega{=}8$, Power~Plant~$+$~30~dyn, 50~frames.
           \textbf{Bold} = best avg per column.
           Top: SWD (rf/rb/pr1/pr2 only; rf and pr2 Embree terminated $>$5h; pr3--pr15 not run under SWD).
           Bottom: ND and OBD (all 17 policies).}
  \label{tab:bvhmode}
  %% --- SWD sub-table ---
  \setlength{\tabcolsep}{3pt}%
  \renewcommand{\arraystretch}{0.82}%
  \resizebox{0.72\columnwidth}{!}{%
  \tiny
  \begin{tabular}{l rrr rrr}
    \toprule
    & \multicolumn{3}{c}{\textit{Embree SWD}} & \multicolumn{3}{c}{\textit{OptiX SWD}} \\
    \cmidrule(lr){2-4}\cmidrule(lr){5-7}
    Mode & ms & \textpm20 & \textless{}$\mu$ & ms & \textpm20 & \textless{}$\mu$ \\
    \midrule
    rf  & {$>$}5h & --- & ---           & 159k  & 90\% & 40\% \\
    rb  & \textbf{14.9k} & 100\% & 50\% & \textbf{88.6}  & 100\% & 40\% \\
    pr1 & 16.1k & 100\% & 60\%          & 101k  &  0\% & 50\% \\
    pr2 & {$>$}5h & --- & ---           & 124k  & 10\% & 40\% \\
    \bottomrule
  \end{tabular}}

  \smallskip

  %% --- ND + OBD main table ---
  \setlength{\tabcolsep}{2pt}%
  \renewcommand{\arraystretch}{0.80}%
  \resizebox{\columnwidth}{!}{%
  \tiny
  \begin{tabular}{l rrr rrr rrr rrr}
    \toprule
    & \multicolumn{6}{c}{\textit{Embree}} & \multicolumn{6}{c}{\textit{OptiX}} \\
    \cmidrule(lr){2-7}\cmidrule(lr){8-13}
    & \multicolumn{3}{c}{ND} & \multicolumn{3}{c}{OBD}
    & \multicolumn{3}{c}{ND} & \multicolumn{3}{c}{OBD} \\
    \cmidrule(lr){2-4}\cmidrule(lr){5-7}\cmidrule(lr){8-10}\cmidrule(lr){11-13}
    Mode
      & ms & \textpm20 & \textless{}$\mu$
      & ms & \textpm20 & \textless{}$\mu$
      & ms & \textpm20 & \textless{}$\mu$
      & ms & \textpm20 & \textless{}$\mu$ \\
    \midrule
    rf   & 9.9k  & 14\% & 88\% & 292k  & 26\% & 64\% & 18.8 & 64\% & 80\% & 328.2 & 30\% & 64\% \\
    rb   & \textbf{2.2k}  & 96\% & 62\% & \textbf{2.6k}  & 96\% & 54\% & 80.0  &100\% & 52\% & \textbf{85.6}  &100\% & 56\% \\
    pr1  & 4.4k  &  0\% & 50\% & 5.6k  &  0\% & 50\% & 47.2  &  2\% & 50\% & 195.2 & 10\% & 66\% \\
    pr2  & 4.5k  &  6\% & 54\% & 90.2k &  6\% & 76\% & 37.3  &  6\% & 64\% & 259.5 & 18\% & 64\% \\
    pr3  & 4.7k  &  4\% & 62\% & 153k  &  4\% & 66\% & 31.6  &  4\% & 70\% & 273.2 & 22\% & 62\% \\
    pr4  & 3.6k  &  8\% & 62\% & 185k  & 10\% & 58\% & 30.3  &  8\% & 72\% & 303.8 & 28\% & 60\% \\
    pr5  & 10.2k &  0\% & 94\% & 180k  &  8\% & 58\% & 26.5  &  6\% & 76\% & 287.5 & 22\% & 64\% \\
    pr6  & 4.7k  & 12\% & 70\% & 185k  & 14\% & 58\% & 24.7  &  8\% & 78\% & 291.8 & 24\% & 62\% \\
    pr7  & 3.9k  &  8\% & 72\% & 217k  & 18\% & 56\% & 23.5  &  2\% & 80\% & 298.0 & 26\% & 60\% \\
    pr8  & 11.6k &  2\% & 92\% & 230k  & 16\% & 62\% & 25.4  & 10\% & 76\% & 308.6 & 26\% & 62\% \\
    pr9  & 3.0k  & 10\% & 74\% & 235k  & 20\% & 58\% & 22.0  &  4\% & 80\% & 317.1 & 30\% & 64\% \\
    pr10 & 6.4k  & 18\% & 70\% & 203k  & 22\% & 52\% & 20.2  &  8\% & 78\% & 274.6 & 40\% & 62\% \\
    pr11 & 6.4k  & 16\% & 76\% & 223k  & 18\% & 60\% & 20.9  & 10\% & 82\% & 302.9 & 28\% & 62\% \\
    pr12 & 6.4k  & 12\% & 78\% & 256k  & 28\% & 64\% & 19.0  &  8\% & 84\% & 304.7 & 30\% & 62\% \\
    pr13 & 4.3k  & 12\% & 72\% & 242k  & 20\% & 60\% & 21.7  & 14\% & 78\% & 314.0 & 32\% & 60\% \\
    pr14 & 3.6k  & 16\% & 68\% & 252k  & 24\% & 62\% & 17.5  & 12\% & 84\% & 315.5 & 30\% & 62\% \\
    pr15 & 4.1k  & 16\% & 66\% & 239k  & 24\% & 60\% & \textbf{17.2}  & 10\% & 88\% & 301.4 & 34\% & 62\% \\
    \bottomrule
  \end{tabular}}
\end{table}

Under SWD, Embree \texttt{rf} and \texttt{pr2} both exceeded 5\,h.
Only \texttt{rb} achieves ${\geq}96\%$ frame pacing within ${\pm}20\%$ across all conditions and both backends; it is the reference baseline throughout.

\subsection{Performance Analysis}
\label{sec:additional_benchmarks}

\subsubsection{Default tunable parameters}
\label{sec:default_params}
Unless stated otherwise, all main-paper and supplementary experiments use the defaults in Table~\ref{tab:default_params}.

\begin{table}[htb]
  \centering
  \caption{Default \grca{} tunable parameters (all experiments unless stated otherwise).}
  \label{tab:default_params}
  \scriptsize
  \setlength{\tabcolsep}{4pt}
  \begin{tabularx}{\columnwidth}{@{}l>{\raggedright\arraybackslash}Xll@{}}
    \toprule
    Symbol & Meaning & Default & Selection \\
    \midrule
    $\gamma_T$        & SAT/BAT channel-span threshold           & $64$        & \S\ref{sec:sat_threshold_detail} \\
    $\chi_T$          & SAT/BAT ray-span threshold               & $64$        & \S\ref{sec:sat_threshold_detail} \\
    $\varepsilon_A$   & Minimum apparent-area threshold          & $10^{-6}$   & fixed \\
    BAT capacity      & Deferred-list capacity (GPU)             & $8{\times}10^5$ & fixed; unbounded on CPU \\
    Motion config     & Per-DOF interpolation mode               & \texttt{f.i} (full instant) & \S\ref{sec:pose_sensitivity_detail} \\
    Baseline mode     & OptiX / Embree build mode                & \texttt{rb} (full rebuild) & \S\ref{sec:bvh_mode_detail} \\
    \bottomrule
  \end{tabularx}
\end{table}

\subsubsection{Span Threshold Selection}
\label{sec:sat_threshold_detail}

\begin{figure}[htb]
  \centering
  \includegraphics[width=\columnwidth]{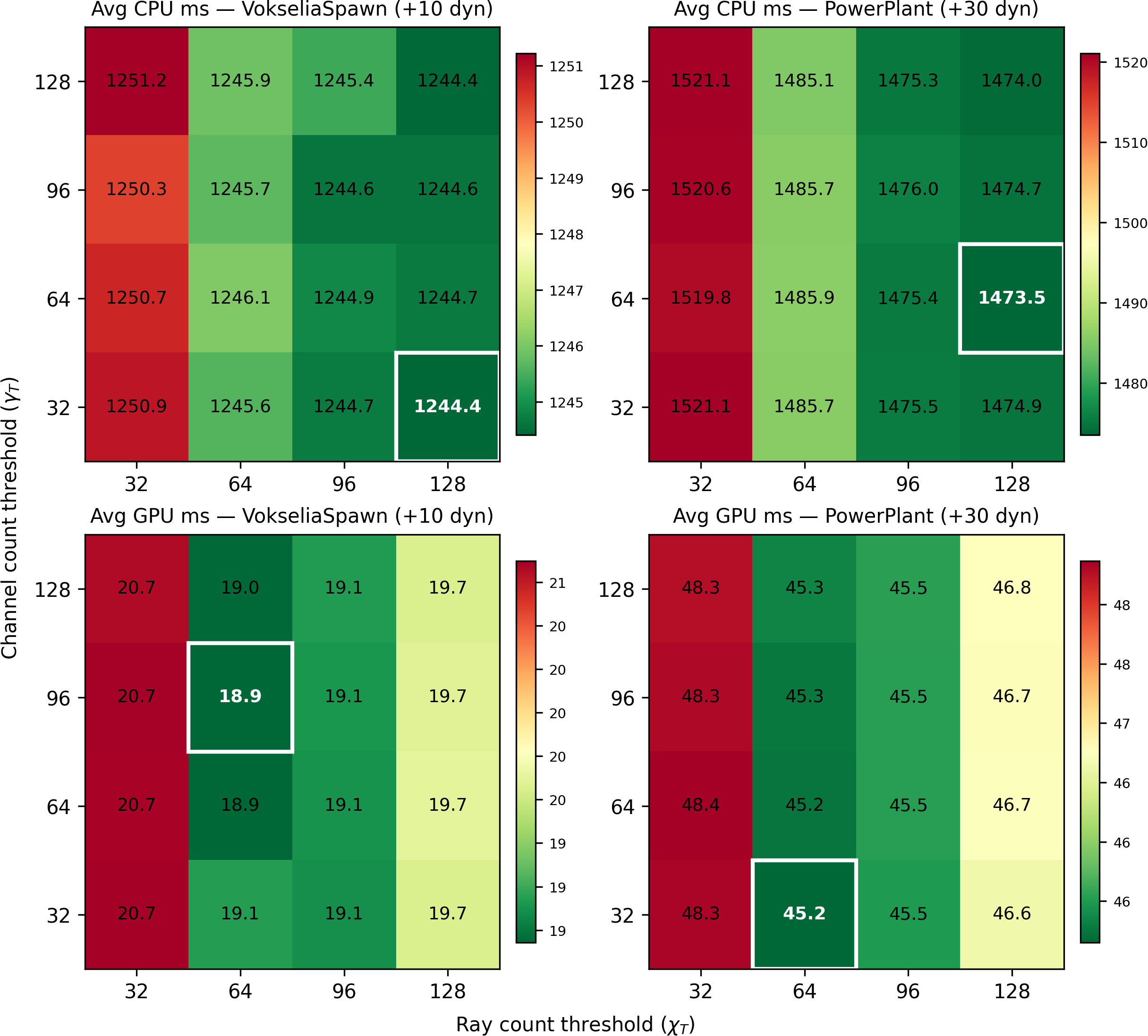}
  \caption{Avg ms/frame per $(\gamma_T, \chi_T)$, avg over 8 scale ranges, $\Omega{=}2/4/6/8$.
           Top row: \grcacpu{}. Bottom row: \grcagpu{}.
           Left: Vokselia~Spawn~$+$10~dyn. Right: Power~Plant~$+$30~dyn. PC1. White box = optimum.}
  \Description{Heatmaps of \grcacpu{} (top row) and \grcagpu{} (bottom row) frame time across SAT threshold combinations for two scenes.}
  \label{fig:sat_threshold}
\end{figure}

All 16 combinations of $(\gamma_T, \chi_T) \in \{32, 64, 96, 128\}^2$ were evaluated
across 8 mesh scale ranges, $\Omega{=}2/4/6/8$, and two bounding scenes
(Vokselia~Spawn~$+$10~dyn and Power~Plant~$+$30~dyn), under SWD~$+$~f.i, 50~frames each;
$(64,64)$ is within ${\leq}1.1\%$ of the per-range optimum for \grcacpu{} and ${\leq}2.5\%$ for \grcagpu{} across all scale ranges (Figure~\ref{fig:sat_threshold}, Table~\ref{tab:sat_threshold_per_scale}).

\begin{table}[htb]
  \centering
  \caption{Avg ms/frame at $(\gamma_T,\chi_T){=}(64,64)$ and per-range optimum.
           PC1, $\Omega{=}2/4/6/8$. \textbf{Bold} (GPU) = $(64,64)$ is the optimum.}
  \label{tab:sat_threshold_per_scale}
  \footnotesize
  \setlength{\tabcolsep}{4pt}
  \begin{tabular}{lrlrl}
    \toprule
    & \multicolumn{2}{c}{\textit{Vokselia~Spawn $+$10~dyn}} & \multicolumn{2}{c}{\textit{Power~Plant $+$30~dyn}} \\
    \cmidrule(lr){2-3}\cmidrule(lr){4-5}
    Scale & $(64,64)$ & Opt.\ (ms\ $\gamma_T{\times}\chi_T$) & $(64,64)$ & Opt.\ (ms\ $\gamma_T{\times}\chi_T$) \\
    \midrule
    \multicolumn{5}{l}{\textit{\grcacpu{}}} \\
    0.001--0.1 & $1{,}070.4$ & $1{,}068.1\ (128{\times}128)$ & $1{,}085.8$ & $1{,}074.7\ (96{\times}128)$  \\
    0.01--1    & $1{,}069.8$ & $1{,}068.5\ (64{\times}128)$  & $1{,}086.1$ & $1{,}075.1\ (64{\times}128)$  \\
    0.1--10    & $1{,}097.1$ & $1{,}094.3\ (32{\times}128)$  & $1{,}135.9$ & $1{,}124.8\ (64{\times}128)$  \\
    0.3--30    & $1{,}216.9$ & $1{,}215.6\ (96{\times}128)$  & $1{,}402.7$ & $1{,}391.2\ (96{\times}128)$  \\
    0.001--30  & $1{,}105.6$ & $1{,}104.6\ (64{\times}128)$  & $1{,}162.1$ & $1{,}150.9\ (64{\times}128)$  \\
    0.6--60    & $1{,}437.6$ & $1{,}435.3\ (128{\times}96)$  & $1{,}920.2$ & $1{,}907.7\ (96{\times}128)$  \\
    1.0--100   & $1{,}744.8$ & $1{,}741.8\ (32{\times}128)$  & $2{,}660.6$ & $2{,}641.9\ (64{\times}128)$  \\
    0.001--100 & $1{,}226.4$ & $1{,}225.2\ (32{\times}96)$   & $1{,}433.8$ & $1{,}421.1\ (64{\times}128)$  \\
    \midrule
    \multicolumn{5}{l}{\textit{\grcagpu{}}} \\
    0.001--0.1 & 12.5  & $12.2\ (32{\times}96)$   & 35.6  & $35.3\ (64{\times}96)$  \\
    0.01--1    & 10.5  & $10.3\ (32{\times}96)$   & 31.0  & $30.8\ (32{\times}96)$  \\
    0.1--10    & 11.7  & $11.6\ (64{\times}96)$   & 32.2  & $32.0\ (64{\times}96)$  \\
    0.3--30    & \textbf{16.7} & $16.7\ (64{\times}64)$  & 40.1  & $40.1\ (32{\times}64)$  \\
    0.001--30  & 13.90 & $13.89\ (128{\times}64)$ & 37.14 & $37.09\ (96{\times}64)$ \\
    0.6--60    & 25.8  & $25.8\ (128{\times}64)$  & \textbf{55.0} & $55.0\ (64{\times}64)$  \\
    1.0--100   & \textbf{38.3} & $38.3\ (64{\times}64)$  & 79.4  & $79.2\ (32{\times}64)$  \\
    0.001--100 & 22.2  & $22.2\ (96{\times}64)$   & 51.4  & $50.6\ (32{\times}64)$  \\
    \bottomrule
  \end{tabular}
\end{table}

$(\gamma_T, \chi_T) = (64, 64)$ is the empirical default, used for all benchmarks, unless explicitly specified.
Vokselia~Spawn and Power~Plant were chosen as the two bounding scenes (lowest and highest
static triangle counts in the suite); optimal thresholds depend on scene geometry and
sensor resolution (\S\ref{sec:sat_threshold_extended}).

\subsubsection{Ray Origin Co-location: Performance and Gap Filling}
\label{sec:combined_origin}
By Observation~3 (main paper), SAT/BAT classification is per ray-origin, so co-locating origins
increases their BAT overlap and shrinks the late-pass workload.
All origins are collapsed into a single co-located ray origin with independently
oriented LiDARs (Table~\ref{tab:combined_origin}).

\begin{table}[htb]
  \centering
  \caption{Distributed vs co-located ray origins (ms/frame), $\Omega{=}1$--$8$.
           Power~Plant, 30~dyn, avg ND/OBD/SWD, scale~1, PC1.}
  \label{tab:combined_origin}
  \renewcommand{\arraystretch}{0.78}%
  \setlength{\tabcolsep}{1.5pt}%
  \resizebox{\columnwidth}{!}{%
  \begin{tabular}{r rrrr c rrrr}
    \toprule
    & \multicolumn{4}{c}{CPU (ms)} & & \multicolumn{4}{c}{GPU (ms)} \\
    \cmidrule(lr){2-5}\cmidrule(lr){7-10}
    & \multicolumn{2}{c}{Distributed} & \multicolumn{2}{c}{Co-located} &
    & \multicolumn{2}{c}{Distributed} & \multicolumn{2}{c}{Co-located} \\
    \cmidrule(lr){2-3}\cmidrule(lr){4-5}\cmidrule(lr){7-8}\cmidrule(lr){9-10}
    $\Omega$ & Embree & \grcacpu{} & Embree & \grcacpu{} & & OptiX & \grcagpu{} & OptiX & \grcagpu{} \\
    \midrule
    1 & 1{,}967 & 264        & 1{,}961 & 250        & & 77.0 &  9.3 & 77.4 &  9.3 \\
    2 & 2{,}464 & 570        & 2{,}525 & 414        & & 77.5 & 14.8 & 77.6 & 13.2 \\
    3 & 3{,}011 & 701        & 3{,}087 & 570        & & 77.6 & 19.7 & 77.8 & 17.6 \\
    4 & 3{,}544 & 896        & 3{,}648 & 737        & & 78.9 & 24.5 & 79.2 & 22.4 \\
    5 & 4{,}071 & 953        & 4{,}212 & 886        & & 79.2 & 28.3 & 79.5 & 26.6 \\
    6 & 4{,}623 & 1{,}590    & 4{,}776 & 1{,}057    & & 80.8 & 41.4 & 81.0 & 30.4 \\
    7 & 5{,}160 & 1{,}331    & 5{,}336 & 1{,}213    & & 81.0 & 38.9 & 81.4 & 34.3 \\
    8 & 5{,}666 & 1{,}624    & 5{,}895 & 1{,}411    & & 80.9 & 45.7 & 81.0 & 38.2 \\
    \bottomrule
  \end{tabular}}
\end{table}

\begin{figure}[htb]
  \centering
  \includegraphics[width=\columnwidth]{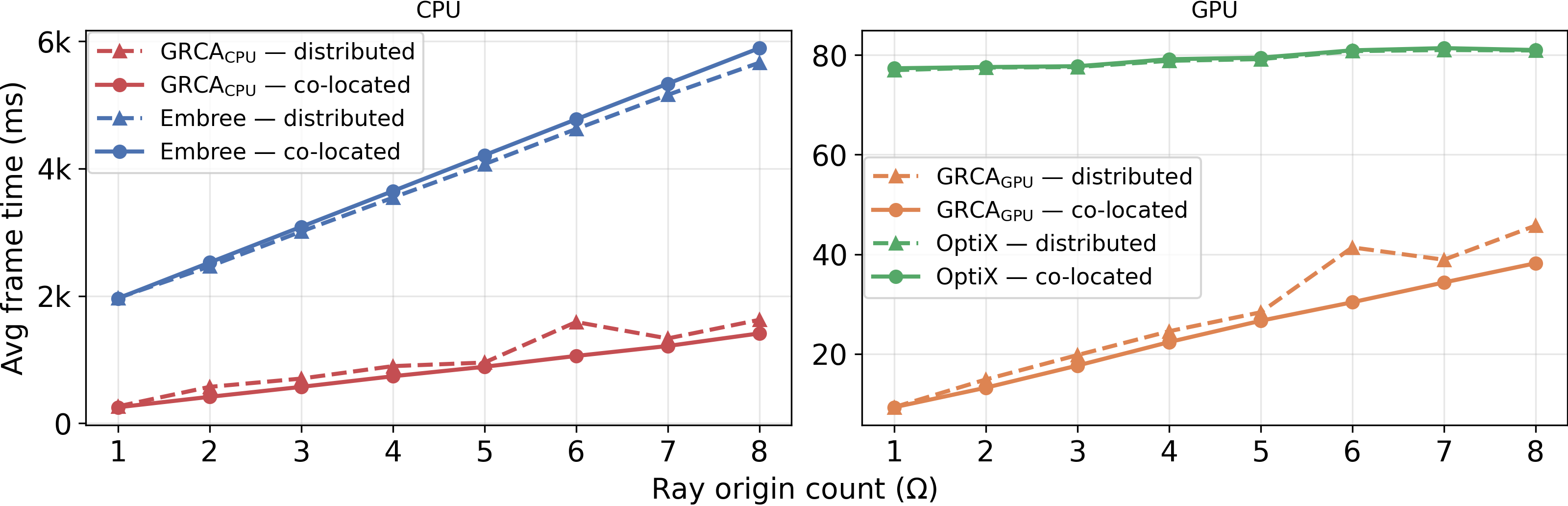}
  \caption{Time vs $\Omega$, distributed vs co-located origins.
           Power~Plant, 30~dyn, avg ND/OBD/SWD, scale~1, PC1.
           Left: CPU. Right: GPU}
  \Description{Line charts comparing CPU and GPU time for distributed versus co-located ray origins across ray origin counts.}
  \label{fig:origin-scaling}
\end{figure}

Co-location increases BAT overlap: when origins coincide, a triangle that is BAT for one is likely BAT for all, so the union of BAT lists grows sub-linearly rather than linearly with $\Omega$.
The distributed slowdown spike at $\Omega{=}6$ (CPU and GPU) is consistent across all deformation conditions and absent in co-located mode, pointing to the spatially diverse BAT lists of distributed origins (Figure~\ref{fig:origin-scaling}).

\begin{figure}[htb]
  \centering
  \includegraphics[width=0.82\columnwidth]{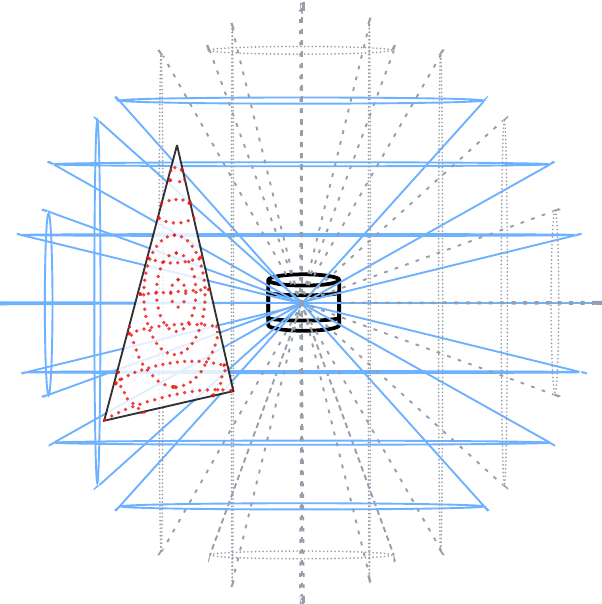}
  \caption{Non-monotonic scan patterns from two co-located ray origins with 90\textdegree{} orientation difference.}
  \Description{Two-panel diagram showing non-monotonic LiDAR scan patterns produced by two co-located ray origins rotated 90 degrees relative to each other.}
  \label{fig:pattern_diversity}
\end{figure}

Beyond the performance gain, co-location doubles as a scan-pattern tool: independently
orienting the co-located LiDARs breaks the regular scan grid and reduces coverage
gaps---particularly at distance, where channel spacing widens
(Figure~\ref{fig:pattern_diversity}). Adding more co-located origins with varied
orientations extends this further, providing a simpler alternative to the per-ray
noise model of \S\ref{sec:noise} for non-monotonic patterns.

%% =============================================================================

\subsubsection{Scene Size Robustness}
\label{sec:scene_size}

\begin{figure}[htb]
  \centering
  \includegraphics[width=\columnwidth]{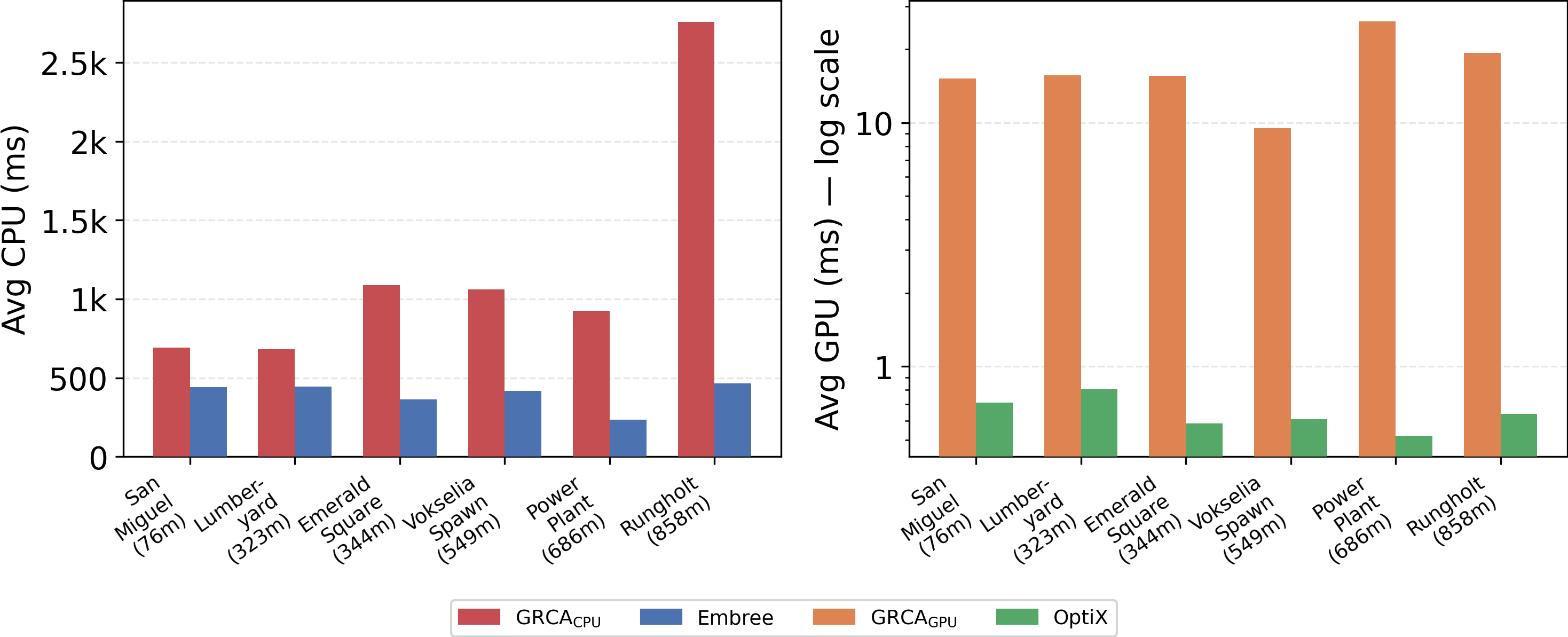}
  \caption{Time vs scene diagonal, static-only, avg $\Omega{=}2/4/6/8$, PC1.
           Left: CPU. Right: GPU (log scale).}
  \Description{Two-panel bar chart: CPU and GPU frame times across six scenes sorted by bounding-box diagonal.}
  \label{fig:scene_size}
\end{figure}

This test isolates \grca{}'s worst case: static-only geometry with no dynamic objects, so the BVH baselines pay no rebuild penalty and all speedup values are ${<}1$.
All six scenes are tested on PC1 (avg across $\Omega{=}2/4/6/8$).
As scene dimensions grow, triangles recede and their apparent area shrinks, so more are culled before any RTIC is attempted---Power~Plant's relatively small triangles (${{\approx}}384\,\text{cm}^2$/tri avg) at a 686\,m diagonal reduce its 12.8M triangles to just 837K~SAT and 125K~BAT (Table~\ref{tab:scene_size_perf}).
Figure~\ref{fig:scene_size} shows frame time vs.\ scene diagonal; the failure-case root causes are discussed in main paper \S\ref{sec:cross_scene_summary}.

\begin{table}[htb]
  \centering
  \caption{Avg ms/frame across $\Omega{=}2/4/6/8$, PC1. Sp${<}1$ = baseline wins.
           SAT/BAT avg counts per frame; Hit\% floor = worst-case match rate vs OptiX.}
  \label{tab:scene_size_perf}
  \resizebox{\columnwidth}{!}{%
  \setlength{\tabcolsep}{1.5pt}%
  \begin{tabular}{l ccc ccc ccc}
    \toprule
    & \multicolumn{3}{c}{CPU (ms)} & \multicolumn{3}{c}{GPU (ms)} & \multicolumn{3}{c}{Filter (avg/frame)} \\
    \cmidrule(lr){2-4}\cmidrule(lr){5-7}\cmidrule(lr){8-10}
    Scene & \grcacpu{} & Embree & Sp($\times$) & \grcagpu{} & OptiX & Sp($\times$) & SAT & BAT & \shortstack[c]{Hit\,(\%)\\floor} \\
    \midrule
    San~Miguel     & 693     & 441 & 0.64 & 15.2 & 0.7 & 0.05 & 414{,}431  & 15{,}468  & 97.1 \\
    Lumberyard     & 682     & 444 & 0.65 & 15.7 & 0.8 & 0.05 & 490{,}685  & 28{,}965  & 98.4 \\
    Emerald Square & 1{,}087 & 364 & 0.34 & 15.6 & 0.6 & 0.04 & 790{,}028  & 18{,}621  & 97.8 \\
    Vokselia Spawn & 1{,}060 & 417 & 0.39 &  9.5 & 0.6 & 0.06 & 861{,}863  & 11{,}972  & 98.7 \\
    Power~Plant    &   924   & 236 & 0.26 & 26.1 & 0.5 & 0.02 & 836{,}879  & 125{,}051 & 99.6 \\
    Rungholt       & 2{,}756 & 464 & 0.17 & 19.3 & 0.6 & 0.03 & 1{,}889{,}647 & 19{,}395  & 98.5 \\
    \bottomrule
  \end{tabular}}
\end{table}

%% =============================================================================

\subsection{Results}

\subsubsection{50-Frame Analysis}
\label{sec:frame_pacing}

Low baseline ${\pm}20\%$ in the Mesh Density (main paper \S\ref{sec:tri_area}) and Scene Size (\S\ref{sec:scene_size}) sections is expected: both aggregate sessions with widely differing per-frame costs.
Scene Size (static geometry, no BVH rebuild penalty) is the only section where \grca{}'s $\mu$ exceeds the rebuild-only baselines.
For \grca{}, below-$\mu$ is more informative than ${\pm}20\%$: more than half of all frames finish faster than the mean in every section (Table~\ref{tab:frame_pacing}).

\begin{table}[htb]
  \centering
  \caption{Frame pacing per section (excl.\ BVH Mode Selection, main paper \S\ref{sec:bvh_mode_selection}). $\mu$ = avg ms/frame (CPU in \textup{k}$=$1000\,ms); ${\pm}20$, ${<}\mu$ in \%. $^{*}$Main paper \S\ref{sec:performance_analysis}.}
  \label{tab:frame_pacing}
  \setlength{\tabcolsep}{1.5pt}%
  \renewcommand{\arraystretch}{0.72}%
  \small
  \begin{tabular}{l rrr rrr rrr rrr}
    \toprule
    & \multicolumn{3}{c}{OptiX}
    & \multicolumn{3}{c}{Embree}
    & \multicolumn{3}{c}{\grcagpu{}}
    & \multicolumn{3}{c}{\grcacpu{}} \\
    \cmidrule(lr){2-4}\cmidrule(lr){5-7}\cmidrule(lr){8-10}\cmidrule(lr){11-13}
    Section & $\mu$ & ${\pm}20$ & ${<}\mu$ & $\mu$ & ${\pm}20$ & ${<}\mu$ & $\mu$ & ${\pm}20$ & ${<}\mu$ & $\mu$ & ${\pm}20$ & ${<}\mu$ \\
    \midrule
    Pose Sensitivity      &  85 & 100 & 49 & 6.8k & 97 & 57 & 52 & 58 & 56 & 1.9k & 35 & 69 \\
    Span Threshold        & --- & --- & --- & --- & --- & --- & 33 & 26 & 61 & 1.4k & 25 & 56 \\
    Co-location           &  79 & 100 & 53 & 3.9k & 73 & 52 & 26 & 25 & 60 & 0.9k & 22 & 61 \\
    Scene~Size            & 0.6 &  27 & 50 & 0.4k & 23 & 54 & 17 & 25 & 58 & 1.2k & 24 & 66 \\
    Mesh~Density$^{*}$    &  65 &  25 & 75 & 1.0k & 12 & 73 & 28 & 21 & 60 & 1.0k & 25 & 61 \\
    Dyn.\ Scale$^{*}$     &  66 &  67 & 50 & 1.5k & 36 & 56 & 31 & 23 & 59 & 1.3k & 25 & 63 \\
    Scale+BAT$^{*}$       & 197 & 100 & 58 & 7.1k & 95 & 59 & 81 & 55 & 52 & 4.2k & 50 & 52 \\
    Range~Var.$^{*}$      &  83 & 100 & 50 & 4.4k & 68 & 53 & 31 & 13 & 59 & 1.1k & 17 & 60 \\
    \bottomrule
  \end{tabular}
\end{table}

%% =============================================================================

\subsubsection{500-Frame Analysis}
\label{sec:frame_pacing_500}

The 50-frame sections deliberately aggregate heterogeneous workloads---Mesh Density (main paper \S\ref{sec:tri_area}) and Scene Size (\S\ref{sec:scene_size}) pool frames across scenes with widely differing per-frame costs, suppressing ${\pm}20\%$ to as low as $12\%$ (Embree) and $21\%$ (\grcagpu{}) by design.
The 500-frame main suite spans all six PC1 scenes; pooling heterogeneous scenes yields ${\pm}20\%$ in a similar range to the 50-frame results (Table~\ref{tab:frame_pacing_500}): OptiX $36$--$42\%$; Embree $28$--$36\%$; \grcagpu{} $24$--$26\%$; \grcacpu{} $20$--$21\%$.
The below-$\mu$ pattern remains consistent: ${\sim}56$--$58\%$ of \grcagpu{} frames and ${\sim}57\%$ of \grcacpu{} frames finish faster than the mean in every condition.

\begin{table}[htb]
  \centering
  \caption{500-frame main suite: frame pacing (PC1).
           ${\pm}20$, ${<}\mu$ in \%.}
  \label{tab:frame_pacing_500}
  \setlength{\tabcolsep}{3pt}%
  \renewcommand{\arraystretch}{0.85}%
  \small
  \begin{tabular}{l rr rr rr rr}
    \toprule
    & \multicolumn{2}{c}{OptiX}
    & \multicolumn{2}{c}{Embree}
    & \multicolumn{2}{c}{\grcagpu{}}
    & \multicolumn{2}{c}{\grcacpu{}} \\
    \cmidrule(lr){2-3}\cmidrule(lr){4-5}\cmidrule(lr){6-7}\cmidrule(lr){8-9}
    Cond. & ${\pm}20$ & ${<}\mu$ & ${\pm}20$ & ${<}\mu$ & ${\pm}20$ & ${<}\mu$ & ${\pm}20$ & ${<}\mu$ \\
    \midrule
    ND  & 42 & 51 & 36 & 53 & 24 & 58 & 20 & 57 \\
    OBD & 36 & 51 & 33 & 55 & 26 & 56 & 20 & 57 \\
    SWD & 39 & 50 & 28 & 55 & 26 & 56 & 21 & 57 \\
    \bottomrule
  \end{tabular}
\end{table}

%% =============================================================================

%% =============================================================================

%% =============================================================================

\subsubsection{Span Threshold Sensitivity}
\label{sec:sat_threshold_extended}
All 36 combinations of $(\gamma_T,\chi_T)\in\{8,16,32,64,96,128\}^2$ were evaluated across
all 6 main suite scenes, all dyn counts, and all $\Omega$ values on PC1
(BAT capacity raised to $4.5{\times}10^6$ to prevent list overflow at small thresholds).
The default $(64,64)$ is within ${\leq}1.3\%$ of the best per scene and condition on the CPU
and within ${\leq}5.1\%$ on the GPU (from the unrounded times behind Table~\ref{tab:sat_threshold_spread}).
The dominant axis is $\chi_T$ (ray-span)---larger $\chi_T$ classifies more rays as SAT,
reducing the deferred BAT workload.
The full spread from worst to best threshold is considerably larger: within the practical
range $\chi_T{\geq}32$, the GPU spread is $5$--$16\%$ across scenes and conditions (median $9\%$);
dropping to $\chi_T{\in}\{8,16\}$ can degrade GPU performance by up to $1.9{\times}$
by over-classifying rays as BATs and overwhelming the deferred queue.
The default $\chi_T{=}64$ already sits near the optimal end of this range.
Table~\ref{tab:sat_threshold_spread} shows the best, default $(64,64)$, and worst avg ms/frame
per scene and mode; compare with Table~\ref{tab:main_suite_full}.

\begin{table}[htb]
  \centering
  \caption{Main suite span-threshold sensitivity (PC1): best (over all 36), default $(64,64)$, and worst (over $\chi_T{\geq}32$) \grca{} avg ms/frame. Spread $=$ (worst$-$best)/best.}
  \label{tab:sat_threshold_spread}
  \resizebox{\columnwidth}{!}{%
  \normalsize\setlength{\tabcolsep}{2pt}%
  \begin{tabular}{llrrrr rrrr}
    \toprule
    & & \multicolumn{4}{c}{\grcacpu{} (ms)} & \multicolumn{4}{c}{\grcagpu{} (ms)} \\
    \cmidrule(lr){3-6}\cmidrule(lr){7-10}
    Scene & Mode & Best & $(64,64)$ & Worst & Spread & Best & $(64,64)$ & Worst & Spread \\
    \midrule
    \multirow{3}{*}{Vokselia Sp.} & ND  & 1154 & 1156 & 1160 & 0.50\% & 12.6 & 12.8 & 14.2 & 12.50\% \\
     & OBD & 1162 & 1163 & 1167 & 0.50\% & 16.6 & 16.7 & 18.0 &  8.70\% \\
     & SWD & 1171 & 1172 & 1177 & 0.50\% & 18.4 & 18.5 & 19.7 &  6.60\% \\
    \midrule
    \multirow{3}{*}{Power Plant} & ND  & 1061 & 1076 & 1113 & 4.90\% & 28.3 & 28.5 & 30.7 &  8.60\% \\
     & OBD & 1066 & 1080 & 1118 & 4.90\% & 34.5 & 34.5 & 36.4 &  5.60\% \\
     & SWD & 1074 & 1088 & 1125 & 4.80\% & 35.5 & 35.6 & 37.6 &  5.90\% \\
    \midrule
    \multirow{3}{*}{Lumberyard} & ND  & 842 & 846 & 856 & 1.70\% & 17.8 & 18.5 & 20.1 & 12.70\% \\
     & OBD & 857 & 862 & 872 & 1.70\% & 24.4 & 25.1 & 26.5 &  8.40\% \\
     & SWD & 876 & 880 & 891 & 1.70\% & 26.6 & 26.9 & 28.4 &  6.70\% \\
    \midrule
    \multirow{3}{*}{Rungholt} & ND  & 2838 & 2840 & 2847 & 0.30\% & 21.7 & 22.2 & 24.0 & 10.40\% \\
     & OBD & 2842 & 2844 & 2851 & 0.30\% & 26.7 & 27.3 & 29.0 &  8.80\% \\
     & SWD & 2847 & 2849 & 2856 & 0.30\% & 28.0 & 28.4 & 30.3 &  8.20\% \\
    \midrule
    \multirow{3}{*}{Emerald Sq.} & ND  & 1226 & 1228 & 1234 & 0.70\% & 17.6 & 18.0 & 19.3 &  9.90\% \\
     & OBD & 1238 & 1240 & 1246 & 0.70\% & 24.2 & 24.2 & 25.7 &  6.30\% \\
     & SWD & 1253 & 1256 & 1262 & 0.70\% & 25.8 & 25.8 & 27.1 &  5.20\% \\
    \midrule
    \multirow{3}{*}{San Miguel} & ND  & 1397 & 1404 & 1420 & 1.70\% & 27.0 & 28.3 & 31.3 & 16.20\% \\
     & OBD & 1643 & 1652 & 1673 & 1.80\% & 55.8 & 57.5 & 60.9 &  9.10\% \\
     & SWD & 1829 & 1839 & 1864 & 2.00\% & 76.8 & 77.9 & 83.3 &  8.50\% \\
    \bottomrule
  \end{tabular}}
\end{table}

\section{Accuracy}
\label{sec:accuracy_detail}

Applied per channel via the two-pass pipeline (\S\ref{sec:pipeline}), the GACP-T intersection defined in \S\ref{sec:gacp_t} is conservative and yields $100\%$ hit-match with OptiX when all performance optimizations are disabled.
With the performance optimizations of \S\ref{sec:accuracy_causes} enabled, hit-distance correctness against OptiX is measured at $0.001\,$m absolute tolerance, sampling $\lfloor N/10 \rfloor$ frames per $N$-frame run; the reported Hit\% floor is the \emph{worst-case per-frame match rate}, and the average floor across configurations is ${\geq}98\%$ (Table~\ref{tab:accuracy_summary}).

\subsection{Identified Causes}
\label{sec:accuracy_causes}

The following are the identified possible causes of below-threshold miss rates.

\begin{enumerate}[label=\textbf{Cause \arabic*:},leftmargin=*,labelsep=0.4em,itemsep=2pt,topsep=3pt]
  \item \textbf{Apparent-area discard ($\varepsilon_A$).} Triangles whose apparent area falls below $\varepsilon_A$ are discarded before any RTIC is attempted; OptiX tests all triangles unconditionally, so these appear as misses even though they subtend sub-sensor-resolution area.
  \item \textbf{$C_{\mathrm{span}}$ binary search initialization.} For obliquely oriented triangles, vertex projections can underestimate $C_{\text{mid}}$, causing the binary search to run in a region with no GACP-T crossings and miss valid intersections.
  \item \textbf{GACP-T early-exit misclassification (both passes).} When all three vertex projections exceed the cone threshold ($\sin\phi_k \ge \sin\phi$), the predicate exits early---correct in exact arithmetic since the cone interior is convex---but near-grazing rounding can push a tangent triangle's vertices just over threshold, producing a false negative in both the early-pass channel search and the late-pass crossing test---both can compound.
  \item \textbf{CW/CCW $R_{\mathrm{span}}$ misprediction.} For BAT triangles spanning the scan seam, the late pass selects CW or CCW via a plane test on arc midpoints; the CCW midpoint is a reflection of the CW midpoint (not the actual buffer ray), which can be wrong for asymmetric scan patterns, and when ambiguous a Möller--Trumbore fallback on the CW midpoint decides---a near-tangent miss routes to the wrong arc.
\end{enumerate}

\subsection{Hit\% Floor}
\label{sec:accuracy_results}

\begin{table}[htb]
\centering
\small
\setlength{\tabcolsep}{2pt}
\renewcommand{\arraystretch}{0.7}
\caption{All cases below the $98\%$ Hit\% floor threshold ($^{\dagger}$).
         All other main-suite and scene-size configurations are $\geq 98\%$.}
\label{tab:accuracy_summary}
\begin{tabular}{c l l c c}
\toprule
\S & Scene & Cond. & \shortstack[c]{Hit\%\\floor} & Causes \\
\midrule
\ref{sec:scene_size}  & San~Miguel     & static & $97.1\%^{\dagger}$ & 1, 2, 3    \\
\ref{sec:scene_size}  & Emerald Square & static & $97.8\%^{\dagger}$ & 1          \\
\ref{sec:main_suite}  & San~Miguel     & ND     & $97.2\%^{\dagger}$ & 1, 2, 3    \\
\ref{sec:main_suite}  & San~Miguel     & OBD    & $96.2\%^{\dagger}$ & 1, 2, 3    \\
\ref{sec:main_suite}  & San~Miguel     & SWD    & $95.8\%^{\dagger}$ & 1, 2, 3, 4 \\
\bottomrule
\multicolumn{5}{l}{\scriptsize $^{\dagger}$Below $98\%$ threshold.}
\end{tabular}
\end{table}

Both backends yield identical floor values.
Only the cases in Table~\ref{tab:accuracy_summary} fall below $98\%$; Emerald Square recovers to ${\geq}98\%$ in the 500-frame main suite.
San~Miguel is the only persistently affected scene (compact scene extent, fine tessellation).
The $95.8\%$ floor is compounded: smallest scene extent ($76\,$m), widest scale range, full SWD simultaneously; scale $[0.001, 5]$ recovers it to ${\geq}97.3\%$ (\S\ref{sec:cross_scene_summary}).
The cross-configuration average floor across all 36 San~Miguel main-suite configurations is $97.98\%$; at scale $[0.001,5]$ it rises to $98.38\%$.
At $\Omega{=}8$, the hit-rate floor is $\mathbf{100.0\%}$ across all conditions and dynamic-mesh counts in Table~\ref{tab:sanmiguel_scale5_perf}---no frame records a miss.

\begin{table}[htb]
  \centering
  \caption{San~Miguel GPU loss mitigation, avg ms/frame, scale $[0.001,5]$, PC1. \textbf{Bold} = best.}
  \label{tab:sanmiguel_scale5_perf}
  \footnotesize
  \setlength{\tabcolsep}{1.5pt}%
  \renewcommand{\arraystretch}{0.70}%
  \resizebox{0.9\columnwidth}{!}{%
  \begin{tabular}{cc cc !{\vrule width 1.2pt} c !{\vrule width 1.2pt} cc !{\vrule width 1.2pt} c !{\vrule width 1.2pt} c}
    \toprule
    & &
      \multicolumn{2}{c}{CPU (ms)} & &
      \multicolumn{2}{c}{GPU (ms)} & & \\
    \cmidrule(lr){3-4}\cmidrule(lr){6-7}
    Dyn & $\Omega$ &
      Embree & \grca{}$_\text{cpu}$ & Sp$_\text{cpu}$ &
      OptiX & \grca{}$_\text{gpu}$ & Sp$_\text{gpu}$ &
      \shortstack[c]{Hit\%\\floor} \\
    \midrule
    \multicolumn{9}{l}{\textit{ND}} \\
    10 & 2 &   708.6 &   334.3 & 2.12 & 34.8 &  7.6 & 4.59 & 98.2 \\
    10 & 4 &   985.2 &   631.2 & 1.56 & 31.5 & 13.4 & 2.35 & 97.5 \\
    10 & 6 & 1{,}248.2 &   902.7 & 1.38 & 31.5 & 16.1 & 1.95 & 98.4 \\
    10 & 8 & 1{,}533.1 & 1{,}235.8 & 1.24 & 31.5 & 23.2 & 1.36 & \textbf{100.0} \\
    20 & 2 & 1{,}197.2 &   364.9 & 3.28 & 43.1 &  9.3 & 4.64 & 98.2 \\
    20 & 4 & 1{,}499.7 &   695.8 & 2.16 & 43.8 & 13.1 & 3.35 & 98.0 \\
    20 & 6 & 1{,}823.0 &   985.4 & 1.85 & 42.7 & 19.5 & 2.19 & 97.6 \\
    20 & 8 & 2{,}099.8 & 1{,}311.2 & 1.60 & 43.5 & 23.9 & 1.82 & \textbf{100.0} \\
    30 & 2 & 1{,}707.4 &   411.8 & \textbf{4.15} & 56.2 &  9.3 & \textbf{6.02} & 97.9 \\
    30 & 4 & 2{,}098.3 &   761.3 & 2.76 & 56.0 & 14.7 & 3.80 & 97.8 \\
    30 & 6 & 2{,}440.5 & 1{,}134.8 & 2.15 & 56.4 & 21.9 & 2.58 & 97.5 \\
    30 & 8 & 2{,}853.4 & 1{,}515.3 & 1.88 & 58.2 & 30.9 & 1.88 & \textbf{100.0} \\
    \midrule
    \multicolumn{9}{l}{\textit{Object-bound Deform}} \\
    10 & 2 &   823.6 &   336.8 & 2.45 & 38.1 &  9.7 & 3.94 & 98.2 \\
    10 & 4 & 1{,}126.9 &   616.9 & 1.83 & 33.4 & 15.0 & 2.22 & 97.5 \\
    10 & 6 & 1{,}461.9 &   912.2 & 1.60 & 33.7 & 22.0 & 1.54 & 98.4 \\
    10 & 8 & 1{,}798.9 & 1{,}248.9 & 1.44 & 34.1 & 29.3 & 1.16 & \textbf{100.0} \\
    20 & 2 & 1{,}374.3 &   373.1 & 3.68 & 47.0 & 10.7 & 4.39 & 98.2 \\
    20 & 4 & 1{,}807.4 &   708.4 & 2.55 & 47.5 & 18.2 & 2.61 & 98.0 \\
    20 & 6 & 2{,}241.5 & 1{,}007.3 & 2.23 & 47.7 & 26.2 & 1.82 & 97.5 \\
    20 & 8 & 2{,}680.5 & 1{,}336.1 & 2.01 & 47.8 & 35.1 & 1.36 & \textbf{100.0} \\
    30 & 2 & 1{,}946.7 &   404.5 & \textbf{4.81} & 61.8 & 12.3 & \textbf{5.01} & 97.7 \\
    30 & 4 & 2{,}482.6 &   748.6 & 3.32 & 62.7 & 21.1 & 2.97 & 97.8 \\
    30 & 6 & 2{,}974.6 & 1{,}109.1 & 2.68 & 61.8 & 30.8 & 2.00 & 97.5 \\
    30 & 8 & 3{,}534.5 & 1{,}488.7 & 2.37 & 63.7 & 36.9 & 1.72 & \textbf{100.0} \\
    \midrule
    \multicolumn{9}{l}{\textit{Scene-wide Deform}} \\
    10 & 2 & 2{,}007.5 &   337.1 & 5.95 & 36.0 & 10.2 & 3.54 & 98.1 \\
    10 & 4 & 3{,}618.8 &   625.4 & 5.79 & 32.5 & 16.1 & 2.02 & 97.4 \\
    10 & 6 & 5{,}230.5 &   923.3 & 5.67 & 33.7 & 23.2 & 1.45 & 98.3 \\
    10 & 8 & 6{,}845.5 & 1{,}227.2 & 5.58 & 35.1 & 30.9 & 1.14 & \textbf{100.0} \\
    20 & 2 & 2{,}972.4 &   381.1 & 7.80 & 43.2 & 12.0 & 3.59 & 98.1 \\
    20 & 4 & 5{,}048.1 &   723.3 & 6.98 & 45.2 & 20.2 & 2.24 & 97.8 \\
    20 & 6 & 7{,}118.1 & 1{,}026.4 & 6.94 & 46.7 & 28.8 & 1.62 & 97.4 \\
    20 & 8 & 9{,}104.8 & 1{,}366.1 & 6.66 & 48.4 & 38.0 & 1.27 & \textbf{100.0} \\
    30 & 2 & 4{,}000.2 &   415.0 & \textbf{9.64} & 55.3 & 14.2 & \textbf{3.89} & 97.7 \\
    30 & 4 & 6{,}717.4 &   770.9 & 8.71 & 58.2 & 24.1 & 2.42 & 97.5 \\
    30 & 6 & 9{,}311.5 & 1{,}143.2 & 8.15 & 61.0 & 34.1 & 1.79 & 97.3 \\
    30 & 8 & 11{,}996.0 & 1{,}533.3 & 7.82 & 63.5 & 39.0 & 1.63 & \textbf{100.0} \\
    \bottomrule
  \end{tabular}}
\end{table}

%% =============================================================================

\section{General Purpose Ray Caster}

\subsection{Ray Noise Simulation}
\label{sec:noise}

Distance noise (range uncertainty) is a post-processing perturbation added after output conversion.
Horizontal angular noise is compatible provided no perturbed ray crosses an adjacent ray; the ray index lookup replaces the nominal index with a three-neighbor argmin over pre-stored perturbed angles~$\theta_k\qual{*}$:
\begin{equation}
  i\qual{*} = \operatorname*{arg\,min}_{k\,\in\,\{i-1,\,i,\,i+1\}}
          \bigl|\theta - \theta_k\qual{*}\bigr|
  \label{eq:noise_search}
\end{equation}
The search is $O(1)$ and no valid hit is excluded.

Per-channel vertical noise---where each channel carries a distinct
pre-stored perturbed elevation $\phi_j\qual{*}$---is also supported under the same
small-perturbation precondition as the horizontal case: no channel offset may
cross an adjacent channel boundary ($|\phi_j\qual{*} - \phi_j| < \Delta\phi$).
Each channel stores its precomputed perturbed elevation $\phi_j\qual{*}$;
the channel assignment uses a three-neighbor argmin identical in form to
Equation~\ref{eq:noise_search}:
\begin{equation}
  j\qual{*} = \operatorname*{arg\,min}_{k\,\in\,\{j-1,\,j,\,j+1\}}
          \bigl|\phi - \phi_k\qual{*}\bigr|
  \label{eq:vnoise_search}
\end{equation}
The GACP-T intersection is evaluated at the nominal channel elevation
$\phi_j$, while the final RTIC uses the actual perturbed ray direction: every reported hit
is exact, but a triangle reached only by a perturbed ray outside the nominal span can be missed.

%% =============================================================================

\subsection{Non-Spinning LiDAR Support}
\label{sec:nonspinning}

The $\gamma_n{\times}\chi_n$ channel--ray grid that natively fits spinning LiDARs also accommodates non-spinning sensors---flash, MEMS, OPA, etc.---whose arbitrary scan patterns map onto it in two ways without modifying the GACP predicate or the two-pass pipeline.

\subsubsection{GAP-only channels}
\label{sec:gap_only}
Each channel can be modeled as a \emph{Geometrically Approximated Plane} (GAP, the $\phi{=}0$ degenerate case of GACP), giving an exact hit-point locus with no conic-arc error.
This fits sensors with ${\leq}180$\textdegree{} FOV---more akin to a camera frustum than a full spinning sweep---where each channel maps cleanly to a single GAP.
For narrow-FOV sensors, all triangles satisfy all-CW (main paper \S\ref{sec:rspan_predict}), so arc disambiguation is skipped and only large-apparent-area triangles reach the late pass.

\subsubsection{Animated ray-mask images}
Arbitrary or sub-frame-varying patterns (MEMS mirrors, OPA steering, Livox rosette) are encoded as an animated 2D image over the $\gamma_n{\times}\chi_n$ grid with two per-cell channels: an \emph{on/off} flag for whether the cell fires this sub-frame, and a stable \emph{ray ID} that survives across frames.
The pipeline skips off cells, so the output buffer looks the same as a native LiDAR output to downstream code.
Static patterns are just the single-frame case.
Per-ray noise composes on top by jittering hits within each active cell, subject to the same conic-locus caveat as GACP channels (\S\ref{sec:gap_only}).
Nothing in the intersection logic changes---arbitrary patterns reduce to authoring the mask.

%% =============================================================================

\section{Additional Data}
\label{sec:additional_data}

Full numeric tables for the Performance Analysis sections whose figures appear in the main body.

Figures~\ref{fig:supp_scene_views} and~\ref{fig:supp_scene_views2} show the main-paper teaser views (Figure~\ref{fig:teaser}) at a larger size, plus hits by camera depth, three scenes per figure.
One $128\times4096$ LiDAR per scene; the five views of a column share one camera.
\textbf{Mesh:} triangles in white, edges in black.
\textbf{Rays tested per triangle:} the colour gives how many rays \grca{} tested against the triangle in this frame, read off the log-scale bar below (right end = the scene's most-tested triangle); dark gray (0) = culled, i.e.\ tested against no ray, and \emph{culled} gives the share of the scene's triangles culled.
\textbf{Hits by LiDAR range:} the colour gives the hit's distance from the LiDAR in metres, read off the bar below, linear between the frame's 2nd and 98th percentile (nearer or farther hits take the end colours); \emph{max} is the LiDAR's maximum range, chosen per scene for legibility.
\textbf{Hits by camera depth:} distance along the viewing camera's axis, like a renderer's depth pass; white = near.
\begin{table}[htb]
  \centering
  \caption{Configurations (PC1) where a BVH
           backend outperforms \grca{}.}
  \label{tab:negative_cases}
  \small
  \setlength{\tabcolsep}{1.7pt}%
  \renewcommand{\arraystretch}{0.7}%
\begin{tabular}{lllccr}
    \toprule
    Scene & Back. & Dyn & $\Omega$ & Cond. & Sp ($\times$) \\
    \midrule
    \multicolumn{6}{l}{\textit{CPU: Embree $>$ \grcacpu{}}} \\
    Rungholt     & CPU & 10    & 2--8 & ND, OBD               & 0.32--0.66 \\
    Rungholt     & CPU & 20    & 4--8 & ND, OBD               & 0.44--0.71 \\
    Rungholt     & CPU & 30    & 4--8 & ND, OBD               & 0.55--0.94 \\[2pt]
    Power~Plant  & CPU & 10    & 4--8 & ND; OBD ($\Omega{\geq}6$)  & 0.64--0.91 \\
    Power~Plant  & CPU & 20    & 8    & ND                & 0.99       \\
    Emerald Sq.  & CPU & 10    & 4--8 & ND; OBD ($\Omega{\geq}6$)  & 0.76--0.99 \\
    Vokselia Sp. & CPU & 10    & 6--8 & ND, OBD               & 0.80--0.98 \\
    \midrule
    \multicolumn{6}{l}{\textit{GPU: OptiX $>$ \grcagpu{}}} \\
    San~Miguel   & GPU & 10--30 & 4--8 & SWD; OBD ($\Omega{\geq}6$) & 0.52--0.92 \\[2pt]
    Lumberyard   & GPU & 10     & 8    & All                     & 0.85--0.99 \\
    Rungholt     & GPU & 10     & 8    & OBD, SWD                & 0.88--0.91 \\
    Vokselia Sp. & GPU & 10     & 8    & SWD                     & 0.98       \\
    \bottomrule
  \end{tabular}
\end{table}

\begin{table}[htb]
  \centering
  \caption{Avg ms/frame vs subdivision level and $\Omega$, scale\,1.0, ND, PC1.
           \textbf{Bold} = min ms per row.}
  \label{tab:subdiv}
  \setlength{\tabcolsep}{1.5pt}%
  \renewcommand{\arraystretch}{0.7}%
  \footnotesize
  \begin{tabular}{lcccc}
    \toprule
    Configuration (Avg BAT) & Embree & \grcacpu{} & OptiX & \grcagpu{} \\
    \midrule
    \multicolumn{5}{l}{\textit{Original ($1{\times}$, 300\,603 dyn.\ tris, 3.56\,cm$^2$/tri)}} \\
    \quad $\Omega{=}$2\ (47{,}776)  &\phantom{0,}126.3 & \phantom{0,}335.4 & 43.5  & \phantom{0}\textbf{8.8}  \\
    \quad $\Omega{=}$4\ (104{,}484) &\phantom{0,}243.1 & \phantom{0,}721.1 & 42.3  & \textbf{19.0}            \\
    \quad $\Omega{=}$6\ (158{,}002) &\phantom{0,}379.4 & 1{,}218.3         & 42.1  & \textbf{31.7}            \\
    \quad $\Omega{=}$8\ (197{,}991) &\phantom{0,}501.3 & 1{,}543.9         & 43.7  & \textbf{45.2}            \\
    \midrule
    \multicolumn{5}{l}{\textit{Sub$\div$1 ($4{\times}$, 1\,202\,368 dyn.\ tris, 0.89\,cm$^2$/tri)}} \\
    \quad $\Omega{=}$2  &\phantom{0,}215.6 & \phantom{0,}339.1 & 45.0  & \phantom{0}\textbf{9.1}  \\
\bottomrule
  \end{tabular}
\end{table}

\begin{table}[htb]
  \centering
  \caption{Avg ms/frame across mesh scale ranges (Power~Plant, avg 10/20/30~dyn, avg $\Omega{=}2$--$8$, ND, PC1).}
  \label{tab:scale_variation}
  \setlength{\tabcolsep}{1.5pt}%
  \small
  \resizebox{\columnwidth}{!}{%
  \begin{tabular}{l rrr c rrrr c r}
    \toprule
    & \multicolumn{3}{c}{CPU (ms)} & & \multicolumn{4}{c}{GPU (ms)} & & \\
    \cmidrule(lr){2-4}\cmidrule(lr){6-9}
    Scale range & Embree & \grcacpu{} & Sp($\times$) & & OptiX & \grcagpu{} & Sp($\times$) & BAT avg & & \shortstack[c]{Hit\,(\%)\\floor} \\
    \midrule
    $[0.001,\,0.1]$ & 1\,326 & 1\,096 & 1.21          & & 64.1 & 29.2 & 2.20          & 132\,K & & 99.0 \\
    $[0.01,\,1]$    & 1\,302 & 1\,097 & 1.19          & & 65.2 & 28.6 & 2.28          & 132\,K & & 99.0 \\
    $[0.1,\,10]$    & 1\,242 & 1\,123 & 1.11          & & 65.5 & 28.3 & \textbf{2.31} & 133\,K & & 99.0 \\
    $[0.3,\,30]$    & 1\,304 & 1\,268 & 1.03          & & 65.7 & 30.0 & 2.19          & 136\,K & & 99.0 \\
    $[0.001,\,30]$  & 1\,367 & 1\,138 & 1.20          & & 67.0 & 29.2 & 2.30          & 133\,K & & 99.0 \\
    $[0.6,\,60]$    & 1\,612 & 1\,544 & 1.04          & & 66.3 & 33.7 & 1.97          & 147\,K & & 98.8 \\
    $[1,\,100]$     & 2\,196 & 1\,914 & 1.15          & & 67.4 & 39.8 & 1.69          & 167\,K & & 98.7 \\
    $[0.001,\,100]$ & 1\,992 & 1\,271 & \textbf{1.57} & & 68.5 & 31.2 & 2.20          & 141\,K & & 99.0 \\
    \bottomrule
  \end{tabular}}
\end{table}

\begin{table}[htb]
  \centering
  \caption{\grca{} speedup over rebuild-only per range. Power~Plant, 30~dyn, $\Omega{=}8$, PC1. \textbf{Bold} = best per column.}
  \label{tab:range_variation}
  \small
  \renewcommand{\arraystretch}{0.8}%
  \setlength{\tabcolsep}{1.5pt}%
  \resizebox{\columnwidth}{!}{%
  \begin{tabular}{l cc !{\vrule width 1.2pt} ccc !{\vrule width 1.2pt} c cc !{\vrule width 1.2pt} ccc !{\vrule width 1.2pt}}
    \toprule
    & \multicolumn{5}{c}{CPU} & & \multicolumn{5}{c}{GPU} \\
    \cmidrule(lr){2-6}\cmidrule(lr){8-12}
    & \multicolumn{2}{c}{Avg ms} & \multicolumn{3}{c}{Sp ($\times$)} & & \multicolumn{2}{c}{Avg ms} & \multicolumn{3}{c}{Sp ($\times$)} \\
    \cmidrule(lr){2-3}\cmidrule(lr){4-6}\cmidrule(lr){8-9}\cmidrule(lr){10-12}
    Rng (m) & Embree & \grcacpu{} & ND & OBD & SWD & & OptiX & \grcagpu{} & ND & OBD & SWD \\
    \midrule
       5 &   2{,}136 &   \textbf{507} & $\mathbf{3.59}$ & $\mathbf{3.83}$ & $5.22$ & &  80 &  \textbf{15} & $\mathbf{7.19}$ & $\mathbf{5.31}$ & $\mathbf{4.04}$ \\
      10 &   2{,}274 &   517 & $3.57$ & $3.82$ & $5.80$ & &  80 &  16 & $7.02$ & $5.15$ & $3.94$ \\
      15 &   2{,}409 &   551 & $3.39$ & $3.64$ & $6.08$ & &  80 &  16 & $6.73$ & $4.99$ & $3.84$ \\
      25 &   2{,}666 &   659 & $2.89$ & $3.10$ & $6.14$ & &  81 &  17 & $6.12$ & $4.68$ & $3.66$ \\
      35 &   2{,}919 &   714 & $2.70$ & $2.93$ & $6.62$ & &  81 &  19 & $5.62$ & $4.29$ & $3.43$ \\
\bottomrule
  \end{tabular}}
\end{table}

\clearpage
\begin{table*}[!htbp]
  \centering
  \renewcommand{\arraystretch}{0.7}
  \scriptsize
  \caption{Full main suite: avg ms/frame, all $\Omega{=}2/4/6/8$, dyn\,10/20/30, ND/OBD/SWD, scale $[0.001,30]$.
           PC1: all scenes; PC2--PC5: worst case (Power~Plant, 30~dyn, $\Omega{=}8$).
           \textbf{Bold} = best speedup per dyn group.}
  \label{tab:main_suite_full}
  \setlength{\tabcolsep}{2pt}
  \resizebox{\textwidth}{!}{%
  \begin{tabular}{rrrr c@{\hspace{1pt}}c@{\hspace{2pt}}c@{\hspace{1pt}}c@{\hspace{2pt}}c@{\hspace{1pt}}c !{\vrule width 1.2pt} @{\hspace{0.3em}}c@{\hspace{0.3em}}c@{\hspace{0.3em}}c@{\hspace{0.3em}} !{\vrule width 1.2pt} c@{\hspace{1pt}}c@{\hspace{2pt}}c@{\hspace{1pt}}c@{\hspace{2pt}}c@{\hspace{1pt}}c !{\vrule width 1.2pt} @{\hspace{0.3em}}c@{\hspace{0.3em}}c@{\hspace{0.3em}}c@{\hspace{0.3em}} !{\vrule width 1.2pt}}
    \toprule
    & & & &
      \multicolumn{6}{c}{CPU ms/frame} &
      \multicolumn{3}{c}{SpeedUp ($\times$)} &
      \multicolumn{6}{c}{GPU ms/frame} &
      \multicolumn{3}{c}{SpeedUp ($\times$)} \\
    \cmidrule(lr){5-10}\cmidrule(lr){11-13}\cmidrule(lr){14-19}\cmidrule(lr){20-22}
    & & & &
      \multicolumn{2}{c}{ND} & \multicolumn{2}{c}{OBD} & \multicolumn{2}{c}{SWD} &
      & & &
      \multicolumn{2}{c}{ND} & \multicolumn{2}{c}{OBD} & \multicolumn{2}{c}{SWD} &
      & & \\
    \cmidrule(lr){5-6}\cmidrule(lr){7-8}\cmidrule(lr){9-10}
    \cmidrule(lr){14-15}\cmidrule(lr){16-17}\cmidrule(lr){18-19}
    Dyn & Total tris & $\Omega$ & Rays/f
      & Embree & \grcacpu{} & Embree & \grcacpu{} & Embree & \grcacpu{}
      & ND & OBD & SWD
      & OptiX & \grcagpu{} & OptiX & \grcagpu{} & OptiX & \grcagpu{}
      & ND & OBD & SWD \\
    \midrule
    \multicolumn{22}{l}{\textbf{PC1}} \\
    \multicolumn{22}{l}{\textit{\quad Vokselia Spawn (1{,}875{,}632 static tris)}} \\
    10 & 4{,}881{,}662 & 2 & 1{,}048{,}576 & 687.8 & 451.1 & 759.4 & 451.7 & 2047.4 & 454.0 & \textbf{1.52} & \textbf{1.68} & \textbf{4.51} & 18.59 & 4.33 & 19.95 & 5.21 & 18.08 & 5.91 & \textbf{4.30} & \textbf{3.83} & \textbf{3.06} \\
    10 & 4{,}881{,}662 & 4 & 2{,}097{,}152 & 932.7 & 896.9 & 1040.0 & 880.8 & 3681.6 & 903.9 & 1.04 & 1.18 & 4.07 & 18.44 & 8.67 & 20.19 & 10.13 & 19.24 & 11.09 & 2.13 & 1.99 & 1.74 \\
    10 & 4{,}881{,}662 & 6 & 3{,}145{,}728 & 1171.3 & 1333.9 & 1318.5 & 1339.2 & 5350.8 & 1344.2 & 0.88 & 0.98 & 3.98 & 18.62 & 10.93 & 20.46 & 15.27 & 20.43 & 16.46 & 1.70 & 1.34 & 1.24 \\
\bottomrule
  \end{tabular}}
\end{table*}

\clearpage
\begin{table*}[!htbp]
  \centering
  \renewcommand{\arraystretch}{0.7}
  \scriptsize
    \caption{Range optimization ($D\qual{\max}{=}100$\,m): avg ms/frame, PC1 all scenes, $\Omega{=}2/4/6/8$, dyn\,10/20/30, ND/OBD/SWD, scale $[0.001,30]$. SpeedUp\,=\,Embree/OptiX\,$\div$\,\grcacpu{}/\grcagpu{}. \textbf{Bold}\,=\,best speedup per dyn group.}
  \label{tab:range_suite}
  \setlength{\tabcolsep}{2pt}
  \resizebox{\textwidth}{!}{%
\begin{tabular}{rrrr c@{\hspace{1pt}}c@{\hspace{2pt}}c@{\hspace{1pt}}c@{\hspace{2pt}}c@{\hspace{1pt}}c !{\vrule width 1.2pt} @{\hspace{0.3em}}c@{\hspace{0.3em}}c@{\hspace{0.3em}}c@{\hspace{0.3em}} !{\vrule width 1.2pt} c@{\hspace{1pt}}c@{\hspace{2pt}}c@{\hspace{1pt}}c@{\hspace{2pt}}c@{\hspace{1pt}}c !{\vrule width 1.2pt} @{\hspace{0.3em}}c@{\hspace{0.3em}}c@{\hspace{0.3em}}c@{\hspace{0.3em}} !{\vrule width 1.2pt}}
    \toprule
    & & & &
      \multicolumn{6}{c}{CPU ms/frame} &
      \multicolumn{3}{c}{SpeedUp ($\times$)} &
      \multicolumn{6}{c}{GPU ms/frame} &
      \multicolumn{3}{c}{SpeedUp ($\times$)} \\
    \cmidrule(lr){5-10}\cmidrule(lr){11-13}\cmidrule(lr){14-19}\cmidrule(lr){20-22}
    & & & &
      \multicolumn{2}{c}{ND} & \multicolumn{2}{c}{OBD} & \multicolumn{2}{c}{SWD} &
      & & &
      \multicolumn{2}{c}{ND} & \multicolumn{2}{c}{OBD} & \multicolumn{2}{c}{SWD} &
      & & \\
    \cmidrule(lr){5-6}\cmidrule(lr){7-8}\cmidrule(lr){9-10}
    \cmidrule(lr){14-15}\cmidrule(lr){16-17}\cmidrule(lr){18-19}
    Dyn & Total tris & $\Omega$ & Rays/f
      & Embree & \grcacpu{} & Embree & \grcacpu{} & Embree & \grcacpu{}
      & ND & OBD & SWD
      & OptiX & \grcagpu{} & OptiX & \grcagpu{} & OptiX & \grcagpu{}
      & ND & OBD & SWD \\
    \midrule
    \multicolumn{22}{l}{\textbf{PC1}} \\
    \multicolumn{22}{l}{\textit{\quad Vokselia Spawn (1{,}875{,}632 static tris)}} \\
    10 & 4{,}881{,}662 & 2 & 1{,}048{,}576 & 670.8 & 255.2 & 733.3 & 256.1 & 1787.0 & 258.1 & \textbf{2.63} & \textbf{2.86} & \textbf{6.92} & 18.3 & 3.42 & 19.8 & 3.77 & 17.8 & 4.51 & \textbf{5.36} & \textbf{5.26} & \textbf{3.96} \\
    10 & 4{,}881{,}662 & 4 & 2{,}097{,}152 & 897.7 & 481.2 & 991.5 & 484.0 & 3157.7 & 488.0 & 1.87 & 2.05 & 6.47 & 18.3 & 5.41 & 20.1 & 6.47 & 18.8 & 7.56 & 3.38 & 3.10 & 2.49 \\
    10 & 4{,}881{,}662 & 6 & 3{,}145{,}728 & 1117.9 & 741.9 & 1244.5 & 716.3 & 4560.7 & 722.6 & 1.51 & 1.74 & 6.31 & 18.5 & 7.72 & 20.3 & 9.35 & 19.8 & 10.7 & 2.40 & 2.17 & 1.85 \\
\bottomrule
  \end{tabular}}
\end{table*}

\clearpage
\begin{table*}[!htbp]
  \centering
  \renewcommand{\arraystretch}{0.7}
  \scriptsize
    \caption{Hybrid pipeline ($D\qual{\max}{=}1000$\,m, static OptiX/Embree\,+\,dynamic \grca{}): avg ms/frame, PC1 all scenes, $\Omega{=}2/4/6/8$, dyn\,10/20/30, ND/OBD/SWD, scale $[0.001,30]$. SpeedUp\,=\,Embree/OptiX\,$\div$\,Hybrid. \textbf{Bold}\,=\,best speedup per dyn group;}
  \label{tab:hybrid_suite}
  \setlength{\tabcolsep}{2pt}
  \resizebox{\textwidth}{!}{%
\begin{tabular}{rrrr c@{\hspace{1pt}}c@{\hspace{2pt}}c@{\hspace{1pt}}c@{\hspace{2pt}}c@{\hspace{1pt}}c !{\vrule width 1.2pt} @{\hspace{0.3em}}c@{\hspace{0.3em}}c@{\hspace{0.3em}}c@{\hspace{0.3em}} !{\vrule width 1.2pt} c@{\hspace{1pt}}c@{\hspace{2pt}}c@{\hspace{1pt}}c@{\hspace{2pt}}c@{\hspace{1pt}}c !{\vrule width 1.2pt} @{\hspace{0.3em}}c@{\hspace{0.3em}}c@{\hspace{0.3em}}c@{\hspace{0.3em}} !{\vrule width 1.2pt}}
    \toprule
    & & & &
      \multicolumn{6}{c}{CPU ms/frame} &
      \multicolumn{3}{c}{SpeedUp ($\times$)} &
      \multicolumn{6}{c}{GPU ms/frame} &
      \multicolumn{3}{c}{SpeedUp ($\times$)} \\
    \cmidrule(lr){5-10}\cmidrule(lr){11-13}\cmidrule(lr){14-19}\cmidrule(lr){20-22}
    & & & &
      \multicolumn{2}{c}{ND} & \multicolumn{2}{c}{OBD} & \multicolumn{2}{c}{SWD} &
      & & &
      \multicolumn{2}{c}{ND} & \multicolumn{2}{c}{OBD} & \multicolumn{2}{c}{SWD} &
      & & \\
    \cmidrule(lr){5-6}\cmidrule(lr){7-8}\cmidrule(lr){9-10}
    \cmidrule(lr){14-15}\cmidrule(lr){16-17}\cmidrule(lr){18-19}
    Dyn & Total tris & $\Omega$ & Rays/f
      & Embree & Hyb-CPU & Embree & Hyb-CPU & Embree & Hyb-CPU
      & ND & OBD & SWD
      & OptiX & Hyb-GPU & OptiX & Hyb-GPU & OptiX & Hyb-GPU
      & ND & OBD & SWD \\
    \midrule
    \multicolumn{22}{l}{\textbf{PC1}} \\
    \multicolumn{22}{l}{\textit{\quad Vokselia Spawn (1{,}875{,}632 static tris)}} \\
    10 & 4{,}881{,}662 & 2 & 1{,}048{,}576 & 687.8 & 239.4 & 759.4 & 211.0 & 2047.4 & 213.6 & \textbf{2.87} & \textbf{3.60} & \textbf{9.59} & 18.6 & 1.63 & 19.9 & 2.24 & 18.1 & 3.06 & \textbf{11.41} & \textbf{8.91} & \textbf{5.91} \\
    10 & 4{,}881{,}662 & 4 & 2{,}097{,}152 & 932.7 & 456.6 & 1040.0 & 460.0 & 3681.6 & 464.8 & 2.04 & 2.26 & 7.92 & 18.4 & 2.20 & 20.2 & 3.78 & 19.2 & 4.93 & 8.37 & 5.35 & 3.90 \\
    10 & 4{,}881{,}662 & 6 & 3{,}145{,}728 & 1171.3 & 677.0 & 1318.5 & 681.0 & 5350.8 & 687.9 & 1.73 & 1.94 & 7.78 & 18.6 & 3.04 & 20.5 & 5.34 & 20.4 & 6.76 & 6.13 & 3.83 & 3.02 \\
\bottomrule
  \end{tabular}}
\end{table*}

\clearpage
\begin{table*}[!htbp]
  \centering
  \renewcommand{\arraystretch}{0.7}
  \scriptsize
    \caption{Range+Hybrid ($D\qual{\max}{=}100$\,m, static OptiX/Embree\,+\,dynamic \grca{}): avg ms/frame, PC1 all scenes, $\Omega{=}2/4/6/8$, dyn\,10/20/30, ND/OBD/SWD, scale $[0.001,30]$. SpeedUp\,=\,Embree/OptiX(100\,m)\,$\div$\,Hybrid(100\,m). \textbf{Bold}\,=\,best speedup per dyn group.}
  \label{tab:range_hybrid_suite}
  \setlength{\tabcolsep}{2pt}
  \resizebox{\textwidth}{!}{%
\begin{tabular}{rrrr c@{\hspace{1pt}}c@{\hspace{2pt}}c@{\hspace{1pt}}c@{\hspace{2pt}}c@{\hspace{1pt}}c !{\vrule width 1.2pt} @{\hspace{0.3em}}c@{\hspace{0.3em}}c@{\hspace{0.3em}}c@{\hspace{0.3em}} !{\vrule width 1.2pt} c@{\hspace{1pt}}c@{\hspace{2pt}}c@{\hspace{1pt}}c@{\hspace{2pt}}c@{\hspace{1pt}}c !{\vrule width 1.2pt} @{\hspace{0.3em}}c@{\hspace{0.3em}}c@{\hspace{0.3em}}c@{\hspace{0.3em}} !{\vrule width 1.2pt}}
    \toprule
    & & & &
      \multicolumn{6}{c}{CPU ms/frame} &
      \multicolumn{3}{c}{SpeedUp ($\times$)} &
      \multicolumn{6}{c}{GPU ms/frame} &
      \multicolumn{3}{c}{SpeedUp ($\times$)} \\
    \cmidrule(lr){5-10}\cmidrule(lr){11-13}\cmidrule(lr){14-19}\cmidrule(lr){20-22}
    & & & &
      \multicolumn{2}{c}{ND} & \multicolumn{2}{c}{OBD} & \multicolumn{2}{c}{SWD} &
      & & &
      \multicolumn{2}{c}{ND} & \multicolumn{2}{c}{OBD} & \multicolumn{2}{c}{SWD} &
      & & \\
    \cmidrule(lr){5-6}\cmidrule(lr){7-8}\cmidrule(lr){9-10}
    \cmidrule(lr){14-15}\cmidrule(lr){16-17}\cmidrule(lr){18-19}
    Dyn & Total tris & $\Omega$ & Rays/f
      & Embree & Hyb-CPU & Embree & Hyb-CPU & Embree & Hyb-CPU
      & ND & OBD & SWD
      & OptiX & Hyb-GPU & OptiX & Hyb-GPU & OptiX & Hyb-GPU
      & ND & OBD & SWD \\
    \midrule
    \multicolumn{22}{l}{\textbf{PC1}} \\
    \multicolumn{22}{l}{\textit{\quad Vokselia Spawn (1{,}875{,}632 static tris)}} \\
    10 & 4{,}881{,}662 & 2 & 1{,}048{,}576 & 670.8 & 225.7 & 733.3 & 197.4 & 1787.0 & 199.8 & \textbf{2.97} & \textbf{3.72} & \textbf{8.94} & 18.3 & 1.48 & 19.8 & 1.98 & 17.8 & 2.86 & \textbf{12.42} & \textbf{10.01} & \textbf{6.25} \\
    10 & 4{,}881{,}662 & 4 & 2{,}097{,}152 & 897.7 & 429.6 & 991.5 & 432.6 & 3157.7 & 437.1 & 2.09 & 2.29 & 7.22 & 18.3 & 2.04 & 20.1 & 3.21 & 18.8 & 4.48 & 8.99 & 6.25 & 4.21 \\
    10 & 4{,}881{,}662 & 6 & 3{,}145{,}728 & 1117.9 & 637.7 & 1244.5 & 641.4 & 4560.7 & 648.3 & 1.75 & 1.94 & 7.03 & 18.5 & 2.65 & 20.3 & 4.47 & 19.8 & 6.02 & 6.99 & 4.54 & 3.29 \\
\bottomrule
  \end{tabular}}
\end{table*}

%% Teaser gallery last: \clearpage flushes every pending table first, so it lands after all tables
%% (standalone supplementaries: right before the references; preprint: last page).
\clearpage
%% All but the render column rendered with GRCA-CUDA by GRCA/presets/benchmark_config_vis_teaser_<Scene>.json.
%% TODO: realistic renders. The render column shows a placeholder until Teaser_<Scene>_render.png exists.
% Figures S5 and S6: the main-paper teaser views enlarged, three scenes per page (scenes as columns, views as rows),
% each image as wide as the page height allows. Under the heat and rainbow rows, a colour bar per scene; values from
% the captures' _scale.txt / _depth.txt and the run logs.
\newlength{\svw}%
% Non-image height per page (scene names, colour bars, row gaps, a two-line caption) is about 75pt; 90pt leaves slack.
\setlength{\svw}{\dimexpr(\textheight-90pt)*4/15\relax}%
% Width cap: three images plus the row-label column (1.2em) and three 3pt gaps fill the text width exactly.
\newlength{\svwmax}\setlength{\svwmax}{\dimexpr(\textwidth-1.2em-9pt)/3\relax}%
\ifdim\svw>\svwmax\setlength{\svw}{\svwmax}\fi
\typeout{supp scene views: image width \the\svw}%
\newcommand{\svgap}{\hspace{3pt}}%
\newcommand{\svlab}[1]{\parbox[b][0.75\svw][c]{1.2em}{\centering\rotatebox{90}{\scriptsize #1}}}%
\newcommand{\svnolab}{\makebox[1.2em]{}}%
\newcommand{\svrender}[1]{\IfFileExists{#1}{\includegraphics[width=\svw]{#1}}{{\setlength{\fboxsep}{0pt}%
  \colorbox{black!12}{\parbox[b][0.75\svw][c]{\svw}{\centering\scriptsize TODO: render}}}}}%
\newcommand{\svimg}[2]{\includegraphics[width=\svw]{Teaser_#1_#2.png}}%
% Colour bars, \svw wide; y in pt (not em) so the geometry is the same in every document class.
% \svheat{rays at the 4 ticks}{% culled}; \svrange{distance at the 5 ticks}{max range}
\newcommand{\svheat}[5]{\begin{tikzpicture}[x=\svw,y=-1pt,inner sep=0pt,font=\tiny]
  \path (0,0) rectangle (1,13.5);
  \fill[black!80] (0,0) rectangle (0.05,5);
  \node[anchor=north west] at (0.08,0) {\includegraphics[width=0.56\svw,height=5pt]{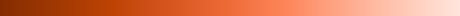}};
  \foreach \x in {0.22,0.36,0.50,0.64} \draw[line width=0.3pt] (\x,5) -- (\x,6.5);
  \node[anchor=north] at (0.025,7.5) {0};
  \node[anchor=north] at (0.22,7.5) {#1}; \node[anchor=north] at (0.36,7.5) {#2};
  \node[anchor=north] at (0.50,7.5) {#3}; \node[anchor=north] at (0.64,7.5) {#4};
  \node[anchor=east] at (1,2.5) {#5\% culled};
\end{tikzpicture}}%
\newcommand{\svrange}[6]{\begin{tikzpicture}[x=\svw,y=-1pt,inner sep=0pt,font=\tiny]
  \path (0,0) rectangle (1,13.5);
  \node[anchor=north west] at (0.04,0) {\includegraphics[width=0.62\svw,height=5pt]{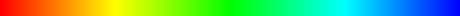}};
  \foreach \x in {0.04,0.195,0.35,0.505,0.66} \draw[line width=0.3pt] (\x,5) -- (\x,6.5);
  \node[anchor=north] at (0.04,7.5) {#1}; \node[anchor=north] at (0.195,7.5) {#2};
  \node[anchor=north] at (0.35,7.5) {#3}; \node[anchor=north] at (0.505,7.5) {#4};
  \node[anchor=north] at (0.66,7.5) {#5};
  \node[anchor=east] at (1,2.5) {max #6\,m};
\end{tikzpicture}}%
% \svscene{file key}{name}{{heat ticks}{% culled}}{{range ticks}{max range}}
\newcommand{\svscene}[4]{\expandafter\def\csname svname#1\endcsname{#2}%
  \expandafter\def\csname svheat#1\endcsname{\svheat#3}%
  \expandafter\def\csname svrange#1\endcsname{\svrange#4}}%
\svscene{Vokselia}{Vokselia Spawn}{{14}{230}{3.5k}{53k}{90.3}}{{2.3}{32.5}{62.8}{93.1}{123}{200}}%
\svscene{Bistro}{Lumberyard Bistro}{{14}{230}{3.5k}{53k}{96.3}}{{2.2}{10.1}{18.0}{25.8}{33.7}{50}}%
\svscene{SanMiguel}{San Miguel}{{15}{242}{3.8k}{59k}{96.2}}{{1.1}{3.1}{5.1}{7.1}{9.1}{10}}%
\svscene{Rungholt}{Rungholt}{{12}{178}{2.4k}{32k}{94.8}}{{1.7}{30.9}{60.2}{89.4}{119}{200}}%
\svscene{EmeraldSquare}{Emerald Square}{{20}{425}{8.8k}{181k}{97.6}}{{2.3}{12.0}{21.7}{31.4}{41.1}{50}}%
\svscene{PowerPlant}{Power Plant}{{21}{471}{10k}{223k}{96.2}}{{1.5}{14.2}{26.8}{39.5}{52.1}{200}}%
\newcommand{\svrowof}[2]{\svlab{#1}\svgap#2}%
\newcommand{\svpage}[3]{%
  \svnolab\svgap\makebox[\svw]{\scriptsize\csname svname#1\endcsname}\svgap
  \makebox[\svw]{\scriptsize\csname svname#2\endcsname}\svgap\makebox[\svw]{\scriptsize\csname svname#3\endcsname}\\[1pt]
  \svrowof{Render}{\svrender{Teaser_#1_render.png}\svgap\svrender{Teaser_#2_render.png}\svgap\svrender{Teaser_#3_render.png}}\\[2pt]
  \svrowof{Mesh}{\svimg{#1}{plain}\svgap\svimg{#2}{plain}\svgap\svimg{#3}{plain}}\\[2pt]
  \svrowof{Rays tested per triangle}{\svimg{#1}{heat}\svgap\svimg{#2}{heat}\svgap\svimg{#3}{heat}}\\[1pt]
  \svnolab\svgap\csname svheat#1\endcsname\svgap\csname svheat#2\endcsname\svgap\csname svheat#3\endcsname\\[2pt]
  \svrowof{Hits by LiDAR range}{\svimg{#1}{rainbow}\svgap\svimg{#2}{rainbow}\svgap\svimg{#3}{rainbow}}\\[1pt]
  \svnolab\svgap\csname svrange#1\endcsname\svgap\csname svrange#2\endcsname\svgap\csname svrange#3\endcsname\\[2pt]
  \svrowof{Hits by camera depth}{\svimg{#1}{gray}\svgap\svimg{#2}{gray}\svgap\svimg{#3}{gray}}}%
\begin{figure*}[p]
  \centering
  \svpage{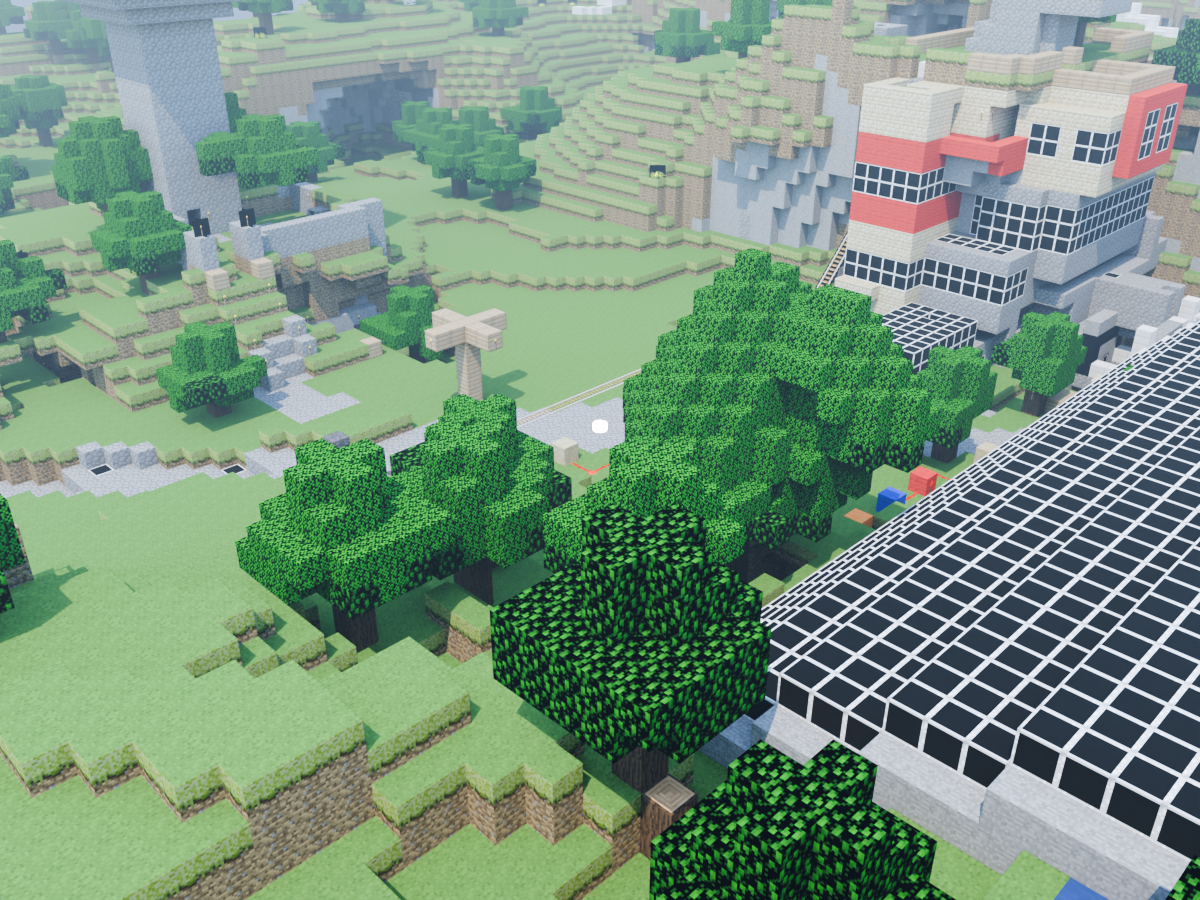}{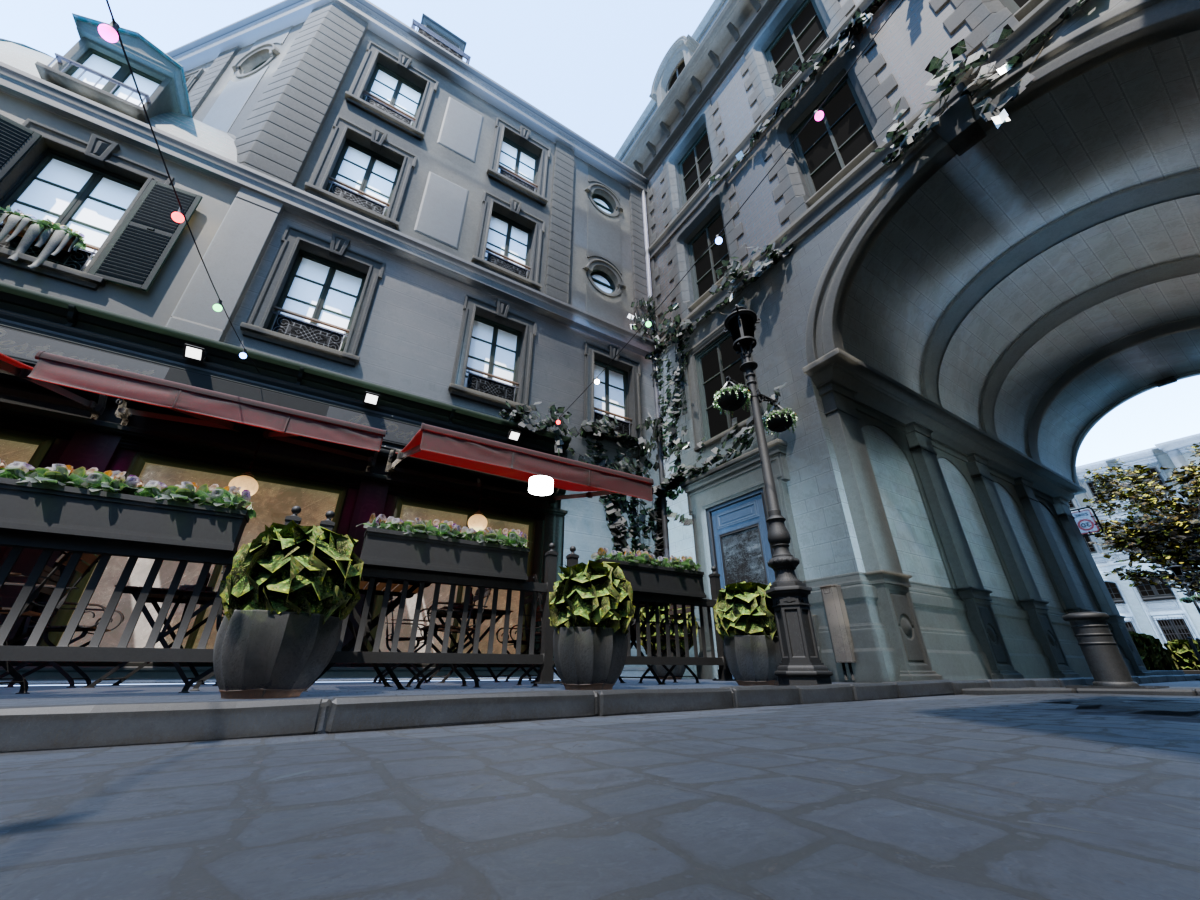}{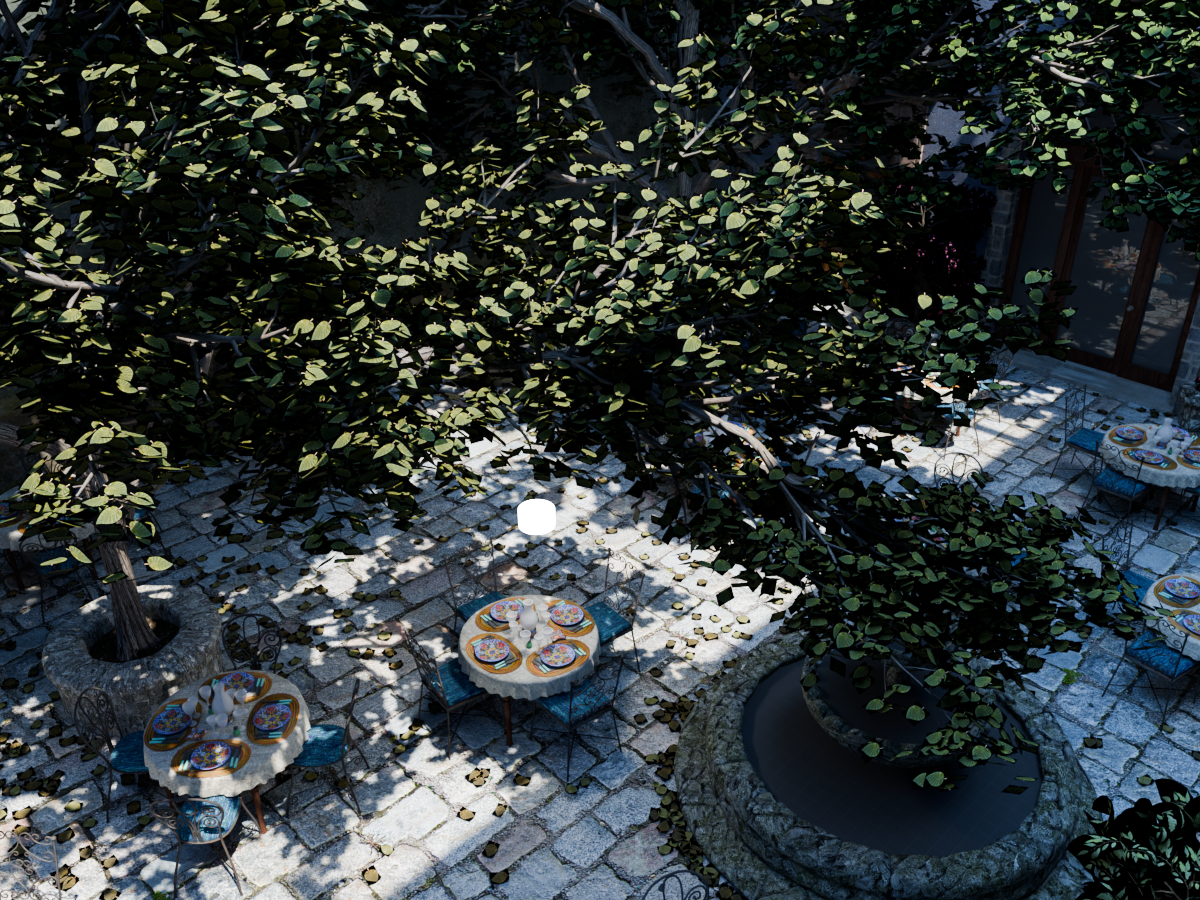}
  \caption{Main-paper teaser views, enlarged, plus hits by camera depth, for Vokselia Spawn, Lumberyard Bistro and San
    Miguel (the other three scenes: Figure~\ref{fig:supp_scene_views2}); views and colour bars as described in
    Section~\ref{sec:additional_data}.}
  \Description{A grid of three columns, one per scene (Vokselia Spawn, Lumberyard Bistro, San Miguel), and five rows: a
    rendering of the scene, its triangle mesh in white with black edges, a heat map of rays tested per triangle with
    culled triangles in gray, the LiDAR hit points colored red to blue by distance, and the hit points as a grayscale
    depth pass. Under the heat-map and LiDAR-range rows, a colour bar per scene gives its ray counts or hit distances
    along the ramp, the share of triangles culled, and the LiDAR's maximum range.}
  \label{fig:supp_scene_views}
\end{figure*}
\begin{figure*}[p]
  \centering
  \svpage{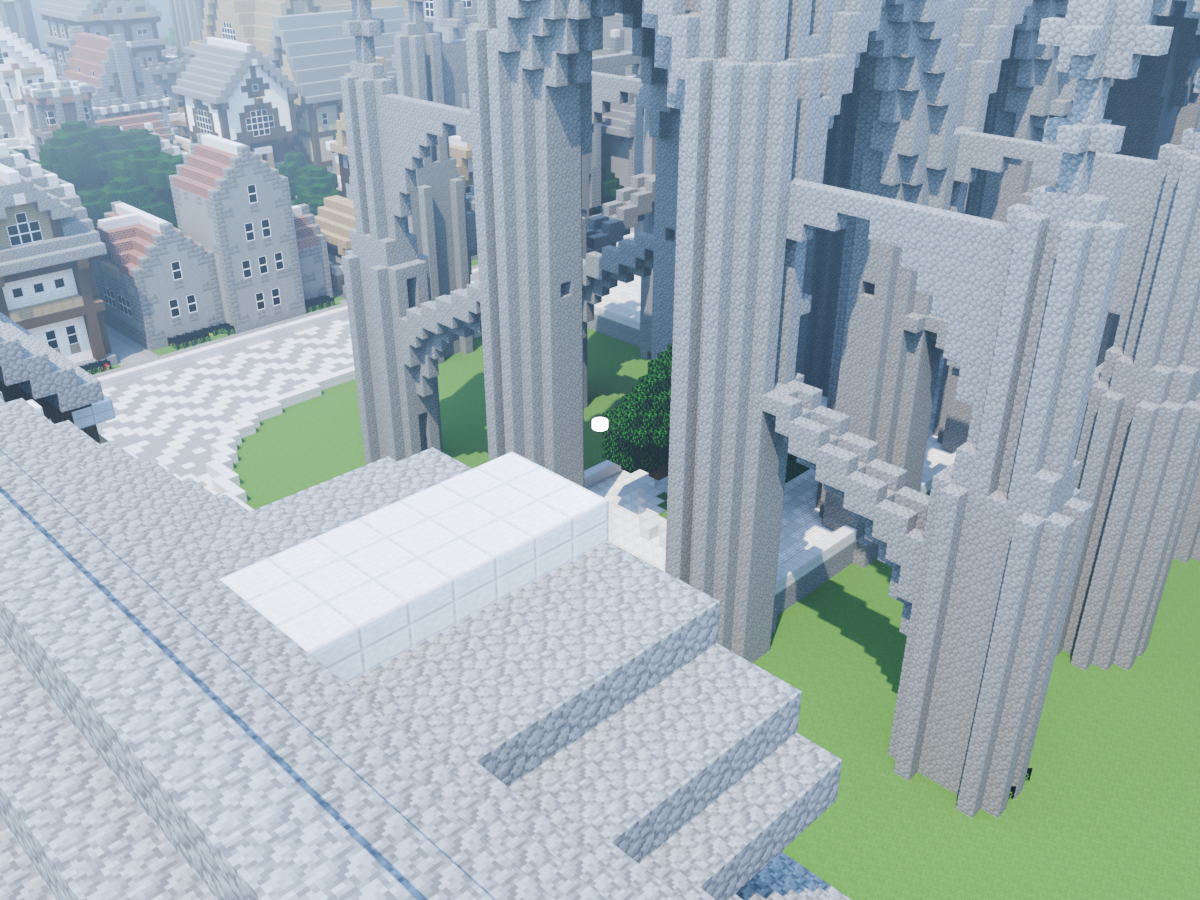}{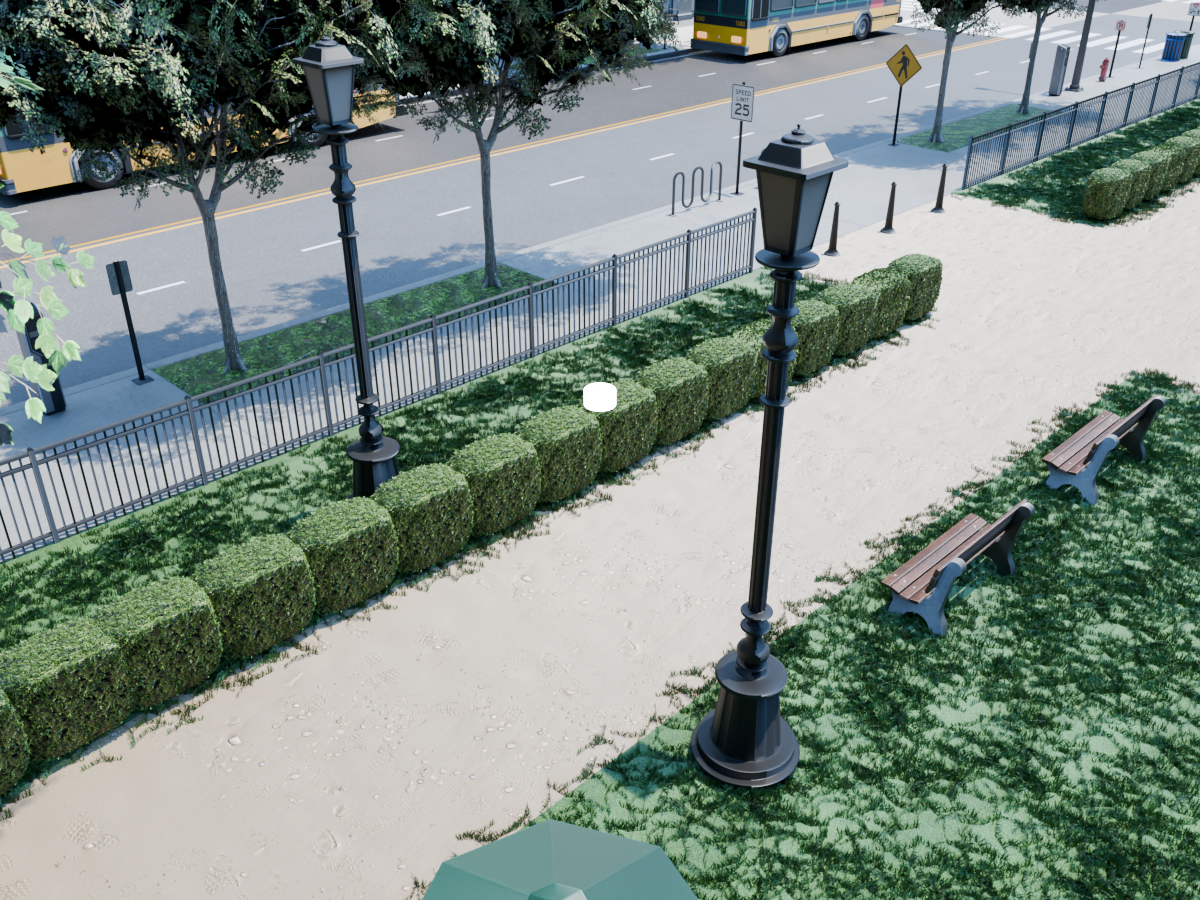}{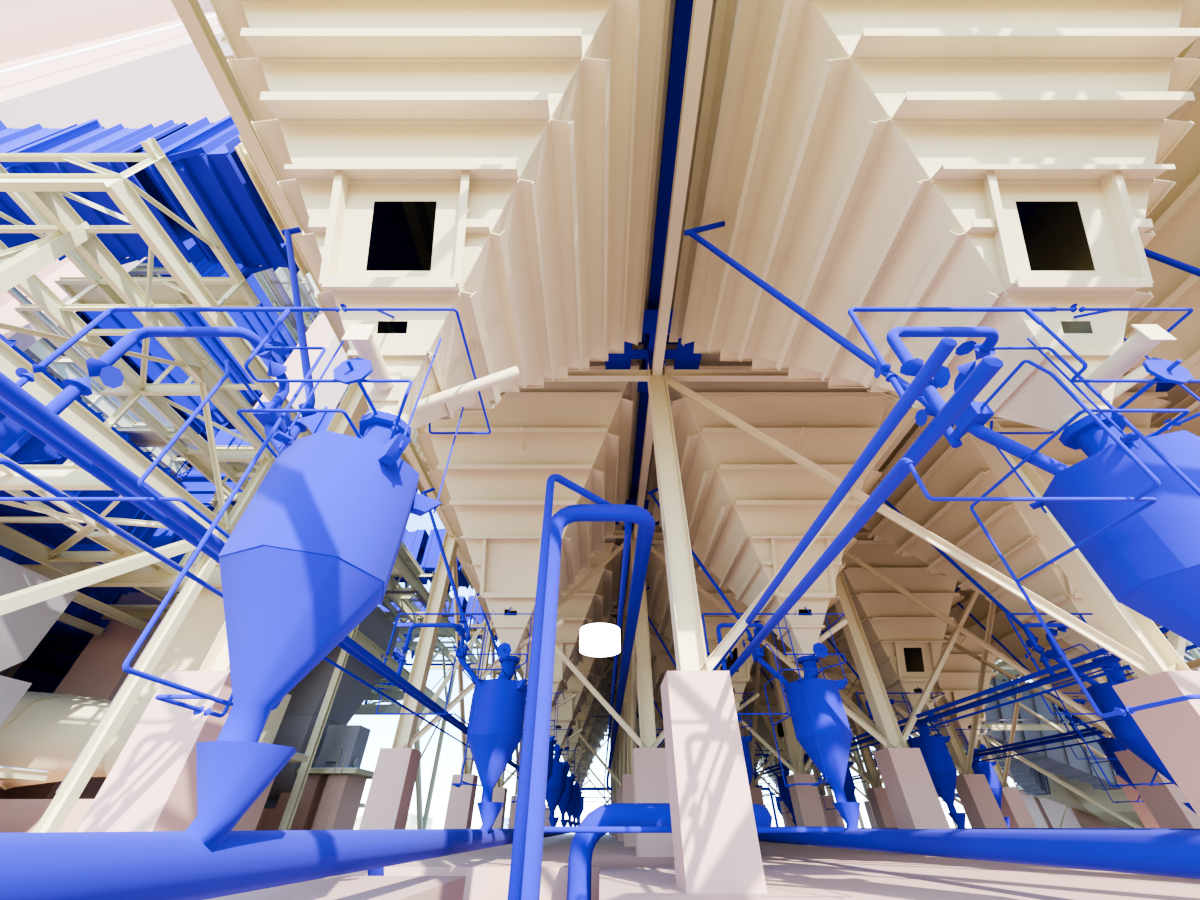}
  \caption{Figure~\ref{fig:supp_scene_views}, continued, for Rungholt, Emerald Square and Power Plant: same views,
    colours and colour bars.}
  \Description{The same grid as the previous figure for Rungholt, Emerald Square and Power Plant.}
  \label{fig:supp_scene_views2}
\end{figure*}

\end{document}